# BUNDLEP: Prioritizing Conflict Free Regions in Multi-Threaded Programs to Improve Cache Reuse

Corey Tessler
Wayne State University
corey.tessler@wayne.edu

Nathan Fisher
Wayne State University
fishern@wayne.edu

***Abstract*—In "BUNDLE: Real-Time Multi-Threaded Scheduling to Reduce Cache Contention", Tessler and Fisher propose a scheduling mechanism and combined worst-case execution time calculation method that treats the instruction cache as a beneficial resource shared between threads. Object analysis produces a worst-case execution time bound and separates code segments into regions. Threads are dynamically placed in bundles associated with regions at run time by the BUNDLE scheduling algorithm where they benefit from shared cache values.**

**In the evaluation of the previous work, tasks were created with a predetermined worst-case execution time path through the control flow graph. Apriori knowledge of the worst-case path is an impractical restriction on any analysis. At the time, the only other solution available was an all-paths search of the graph, which is an equally impractical approach due to its complexity.**

**The primary focus of this work is to build upon BUNDLE, expanding its applicability beyond a proof of concept. We present a complete a worst-case execution time calculation method that includes thread level context switch costs, operating on real programs, with representative architecture parameters, and compare our results to those produced by Heptane's state of the art method. To these ends, we propose a modification to the BUNDLE scheduling algorithm called BUNDLEP. Bundles are assigned priorities that enforce an ordered flow of threads through the control flow graph – avoiding the need for multiple all-paths searches through the graph. In many cases, our evaluation shows a run-time and analytical benefit for BUNLDEP compared to serialized thread execution and state of the art WCET analysis.**

## I. Introduction

For hard real-time systems, cache memory complicates the calculation of a task's worst-case execution time (WCET) bound [1]–[4] . An architecture that includes an instruction cache creates the possibility of (at least) two execution times for every instruction. Executing an instruction from the cache, a cache hit, typically takes less time than an instruction that must fetched from main memory (a cache miss). For hierarchical caches, the multiple loading times from one level to another increases the diversity of execution times per instruction. Previous work preceding BUNDLE [5] accounts for variations in these execution times due to tasks sharing a single cache by extending execution times of the preempted or preempting task [2], [6]–[8].

BUNDLE moves away from the classical (negative) perspective of caches, by treating the cache as a benefit to multi-threaded hard real-time tasks. It is comprised of two parts, a worst-case execution time with cache overhead (WCETO) analysis and scheduling algorithm. The analysis leverages BUNDLE's thread-level scheduling algorithm in calculating a bound which includes the cache benefit between threads. In addition, the analysis identifies conflict free regions used to make scheduling decisions.

As a first work, the evaluation in BUNDLE [5] served as a proof of concept, demonstrating the potential benefit in WCETO and run-time savings. For WCETO calculations, it described an all-paths walk with complexity $\mathbb{O}((|V|!)^m)$ for $V$ conflict free regions, and $m$ threads. The evaluation subverted this bound by construction of synthetic programs with a known set of worst-case execution paths.

Synthetic tasks, all-paths walks, and prior knowledge of their worst-case paths are barriers to BUNDLE's practical application and acceptance. The primary goal of this work is to provide a suitable WCETO calculation method and run-time evaluation for BUNDLE scheduling that can be compared to the classical approach. The following contributions are made to reach these goals:

- A method for calculating conflict free regions (CFRs) where instructions participate in exactly one CFR.
- A BUNDLE-based scheduling algorithm that prioritizes conflict free regions named BUNDLEP.
- A suitable WCETO calculation method for BUNDLEP which incorporates **context switch costs**.
- A complete evaluation and simulation environment based on the Heptane[1] package available for download [9].

The remainder of this work is structured as follows. Section II sets BUNDLE and BUNDLEP in context of the related work. Section III summarizes the core concepts of BUNDLE and background required for the extensions. Section IV gives a brief overview of BUNDLEP's approach in contrast to BUNDLE. Section V outlines the creation of conflict free regions. Section VI is divided into three subsections, describing the thread bottlenecks, priority assignment, and the BUNDLEP scheduling algorithm. Section VII details the WCETO method for BUNDLEP. Section VIII describes the evaluation method as a compliment to Heptane and results. Our work concludes with final remarks and potential extensions in Section IX.

## II. Related Work

Other efforts have been made to mitigate or manage the cache impact of concurrent tasks. Memory-Centric Schedul-ing [10] is influenced by the cache impact of each task. However, it only supports PREM-compliant [11] tasks that have been divided into load and execution phases. Loading phases are isolated from one another preventing an inter-thread cache benefit between them. Another approach to predictable cache behavior is taken by [12], using management techniques such as cache coloring and blocking. Cache reuse is increased for a single task, but the method does not accommodate cached values being shared between distinct threads or tasks. These are representative examples of the classical (negative) perspective of caches, in contrast to BUNDLE's positive view.

[1] See https://team.inria.fr/alf/software/heptane

Aside from BUNDLE, the only works we are currently aware of taking a positive perspective on caches with respect to schedulability are related to Persistent Cache Blocks (PCBS) [13], [14] or cache spread [15]. PCBs are cache blocks that remain in the cache after a job completes to be reused with the next release. However, PCBs are limited to a single task (or thread), and the analytical benefits is limited to subsequent jobs. Calandrino's [15] examination of cache spread is limited to empirical analysis with a more coarse grained approach than BUNDLE.

## III. Summary of BUNDLE

The motivation for BUNDLE stems from a positive perspective of caches in the setting of multi-threaded tasks on a shared processor. From this perspective, there is an *inter-thread cache benefit* when a thread encounters an unexpected cache hit due to the previous execution of a thread from the same task. Other concurrent approaches [16], [17] do not account for the benefit in their analysis due to their focus on finding the worst-case cache interleavings. Conversely, BUNDLE schedules threads to avoid the worst-case and creates a quantifiable benefit of cache reuse between threads.

The sporadic model [18] is extended to support BUNDLE's thread-level scheduling algorithm and analysis. Each task $\tau_i$ in the set of tasks $\tau$ is represented by a tuple of minimum inter-arrival time, relative deadline, and initial ribbon: $\tau_i = \langle p_i, d_i, R_i \rangle$. A *ribbon* is the set of reachable instructions from a single entry instruction, described by a conflict free region graph $R_i$. The initial ribbon of a task is a starting point for the first *threads* of execution released with each job.

Due to the complexity of intra-task scheduling, BUNDLE's scheduling algorithm and analysis is limited to one processor, a single task[2], with one ribbon, releasing $m$ threads per job. Execution on the shared processor is aided by a single level direct-mapped cache with $l$ lines. Loading a block from main memory to the cache takes $\mathbb{B}$ cycles, with a uniform number of clock cycles per instruction (CPI) denoted $\mathbb{I}$. This paper is the first, implementation-based step towards our larger goal of bringing BUNDLE to a fully preemptive system-wide scheduler, with multiple tasks, hierarchical caches, on many cores.

To quantify the inter-thread cache benefit, BUNDLE schedules threads in a manner cognizant of the program's structure as well as potential cache conflicts within and between threads. Central to BUNDLE's scheduling decisions and WCETO analysis are conflict free regions.

Conflict free regions are created from the control flow graph [19] (CFG) of a ribbon. A *control flow graph* is a weakly connected directed graph representing the flows of execution through an executable object. Described by a triple $G = (N, E, h)$ of nodes, edges, and entry instruction. Typically the nodes $n \in N$ of a CFG are basic blocks, for simplicity of presentation nodes of CFGs within this work are single instructions. The directed edges $(u, v) \in E \wedge u, v \in N$ define the possible paths through the program, with $h \in N$ at the root of all paths, ending in a single terminal node.

A *conflict free region* (CFR) is a subset of nodes and edges from the CFG of a ribbon. The subset is determined by including instructions which, when executed, would not cause an eviction of each other. When a CFR is extracted from the CFG $G$ the structure of the program is maintained. An extracted CFR is itself a CFG, denoted $F = (N, E, h)$ with the following properties:

1) No two instructions (outside of the same block) map to the same cache line.
2) All instructions of $F$ are weakly connected to the entry instruction $h$.
3) For any two instructions in $(u, v) \in F$, if there was an edge between them in $G$ then $(u, v) \in E$ (of $F$).

The set of CFRs of a ribbon's CFG are collected in the ribbon's *conflict free region graph* (CFRG). A CFRG $R = (N, E, h)$ is a CFG where the nodes are CFRs. Connectivity between CFRs from the CFG is preserved in the edges of the CFRG. For an edge $(n_1, n_2)$ in the CFG, if $n_1$ and $n_2$ are placed in distinct CFRs, then the CFRG must contain an edge between the CFRs. Figure 1 illustrates the relationships between the CFG, CFRs, and CFRG.

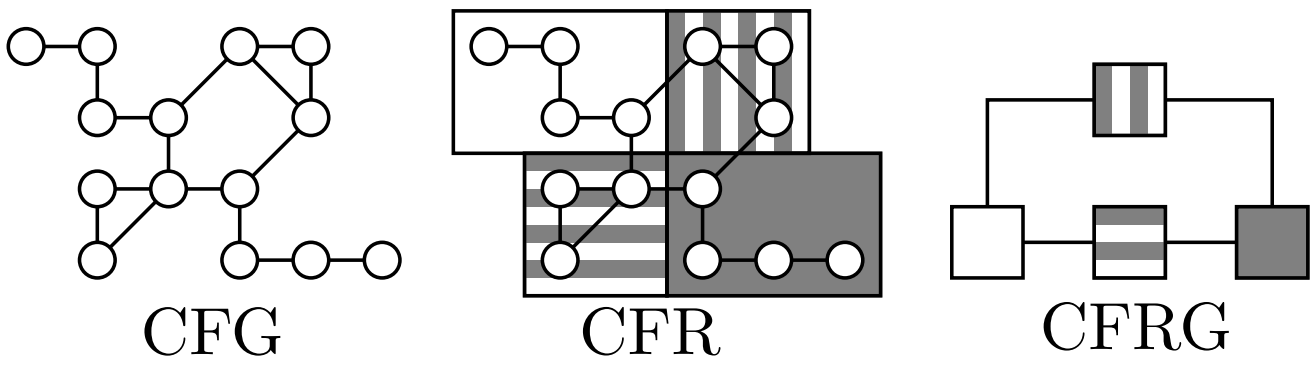

Fig. 1: CFG, CFR, and CFRG of a ribbon

It is the CFRG which drives the scheduling decisions of BUNDLE's algorithm. For each CFR within the CFRG, the scheduling algorithm creates a container for threads called a *bundle*. Only one bundle is *active* at any time, and only threads of the active bundle are allowed to execute. If a thread of the active bundle attempts to execute an instruction outside of the active CFR, the thread is blocked. After being blocked, the thread is placed in the bundle of the CFR it attempted to enter. After all threads of the active bundle have blocked, the bundle is *depleted* and another bundle is selected as active.

*Graphical Notation*: Execution under BUNDLE is illustrated in Figure 2. Annotation of the CFRG $R = (N, E, h)$ will remain consistent with other figures. Nodes $n_i \in N$ are CFRs,

[2]Without modification BUNDLE's techniques could be applied to a non-preemptive task-level scheduler.

$n_i = (N_i, E_i, h_i)$. The entry instruction $h_i$ of CFR $n_i$ has a main memory address, which is denoted $a_i$. When presented as nodes in graphs, CFRs will be squares and instructions circles. When needed, the CFR shaded light gray indicates its bundle is active. Small black squares adjacent to or within a CFR are the threads of the associated bundle.

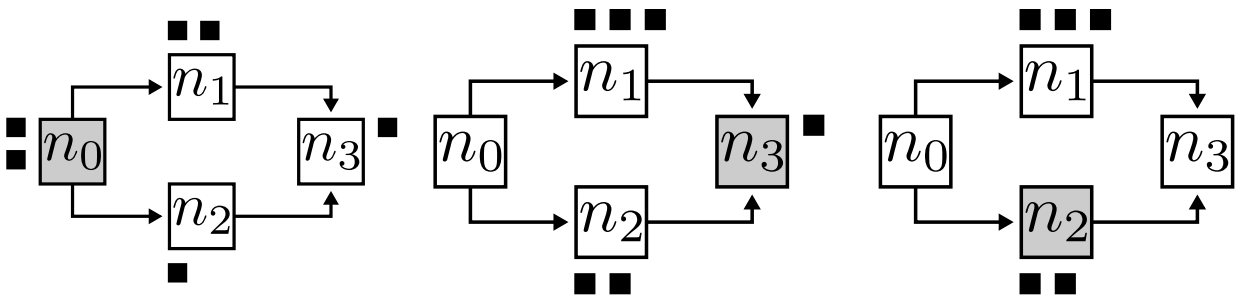

(a) $n_0$ active (b) $n_3$ active (c) $n_2$ active

Fig. 2: BUNDLE Execution

In Figure 2a, $n_0$ is active. Accordingly, BUNDLE executes each of $n_0$'s threads until they leave the CFR, entering $n_1$ and $n_2$. The next active bundle selected is $n_3$ in Figure 2b, its one thread executes until termination. This process repeats until all threads have terminated, and the job has been completed.

By restricting execution by CFR (bundle), the execution time of threads are lowered by sharing cached values where they otherwise would not (e.g. letting each thread complete the execution of the entire CFG before switching to the next thread). In BUNDLE [5], the selection of which bundle to activate is arbitrary. Selecting in such a manner can reduce the inter-thread cache benefit, which can be observed in Figure 2. Had $n_1$ and $n_2$ been activated before $n_3$, more threads would have received the benefit when $n_3$ was activated.

## IV. Overview of BUNDLEP

Arbitrary selection of bundle activation has the deleterious effect of increasing the complexity of WCETO analysis. BUNDLEP addresses both issues of increasing the inter-thread cache benefit, and simplifying WCETO calculation with a single solution: assign priorities to the CFRs of the CFRG. At run-time, a bundle inherits the priority of its associated CFR, and the bundle with the best priority is always activated.

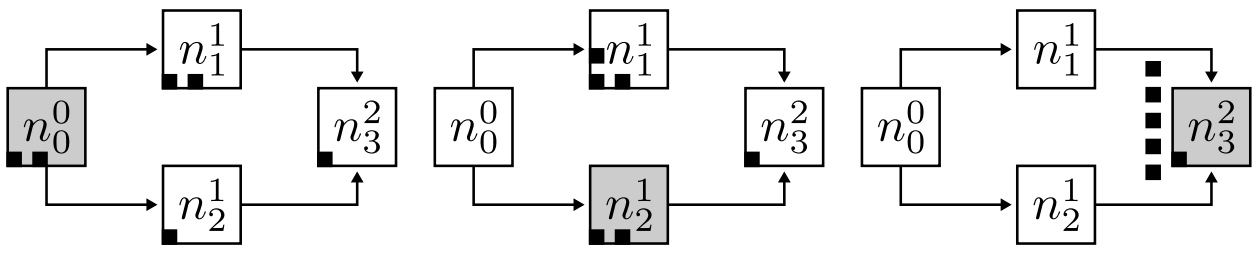

(a) (b) (c)

Fig. 3: BUNDLEP Execution

Priorities are assigned to CFRs to guarantee the minimum number of activations, which maximizes the inter-thread cache benefit per activation. Priority assignments are based on the intuition that the terminal node of the CFRG should have the lowest priority. Those nodes immediately preceding the terminal node are given the second lowest, etc. This intuition is illustrated by example in Figure 3.

CFR's have their priority listed as superscripts, where the lowest value has the best priority, ie. node $n_2^1$ has priority 1 which is worse than $n_0^0$. Figure 3 revisits the CFRG of Figure 2 to highlight the benefit of prioritizing CFRs. Starting with 3a, the best priority CFR is selected as active ($n_0$) and the bundle is depleted. BUNDLEP's next activation is a choice between $n_1$ and $n_2$, they have equal priority, either would be a valid choice. The selection of $n_2$ is made before $n_1$. Figure 3b illustrates only the first choice, omitting depicting the activation of $n_1$ in favor of $n_3$. Figure 3c shows the clear benefit of priority assignment. The bundle for $n_3$ is activated only once for six threads, this maximizes the inter-thread cache benefit which is equivalent to minimizing the number of activations. Compared to Figure 2 where $n_3$ could be activated up to four times, possibly quadrupling the cache load penalty for the CFR at run-time and during WCETO analysis.

The priority assignment which minimizes the number of activations is shown in Theorem 1 as the longest path value for each node of the CFRG from the initial node given positive edge weights. This influences the creation of CFRs and the treatment of loops in the CFRG. As such, the CFRG must be a DAG, where the assignment of priorities to CFRs is performed in polynomial time with the added benefit of enabling a tractable WCETO analysis.

**Theorem 1** (Maximum Bundle Activations). *For a CFRG $R = (N, E, h)$, which is a DAG, where each node $n_i \in N$ has priority $\pi_i$ equal to the length of the longest path from $h$ to $n_i$ the bundle $n_i$ will be activated at most once per job using BUNDLEP.*

*Proof.* To illustrate a contradiction assume a CFR $n_i$ is activated more than once. Then there must exist a node $n_j$ with a worse priority (greater value) $\pi_j > \pi_i$ on a valid path $\langle h, ..., n_j, ..., n_i, ...\rangle$. Given that $R$ is a DAG, there can be no path from $n_i$ to $n_j$. Since priorities are assigned equal to the longest path from $h$ to a node, then $\pi_j < \pi_i$ contradicting $\pi_j > \pi_i$. Therefore, $n_i$ can be activated at most once. □

## V. Conflict Free Region Creation

To ensure the CFRG of a ribbon is a DAG[3], the process of converting a ribbon's executable object into a CFG, then CFRs, and finally a CFRG must avoid introducing ambiguity or loops into the CFG or CFRG. To do so, the process is divided into two stages: 1.) create an expanded CFG 2.) create and link CFRs. In Subsection V-A the motivation and definition of expanded CFGs is provided. Subsection V-B details the assignment of individual instructions to CFRs and their compilation into a CFRG.

### A. Expanded Control Flow Graphs

Typically, for a CFG $G = (N, E, h)$, a node $n \in N$ is a basic block identified by the memory address of the first instruction of the block. In this work, nodes are individual instructions. However, nodes are not identified by their address. They are identified by their address and call stack. This prevents loops from being introduced into the CFG.

Common to other hard real-time programs, ribbons are restricted from including infinite loops, function pointers,

[3]CFRG's cannot be a DAG in the presence of user defined loops, to maintain the DAG structure loops are collapsed – which is described in Section VI-C.

long jmp's, or unbounded recursions. With these restrictions in place, it is still possible to create loops and ambiguity in the CFG of the ribbon. As illustrated by Figure 4.

```
1    procedure a(x, y)
2      c = x
3      x = b(y)
4      y = b(2 * c)
5      return x + y
6    end procedure
```

Listing 1: Procedure a()

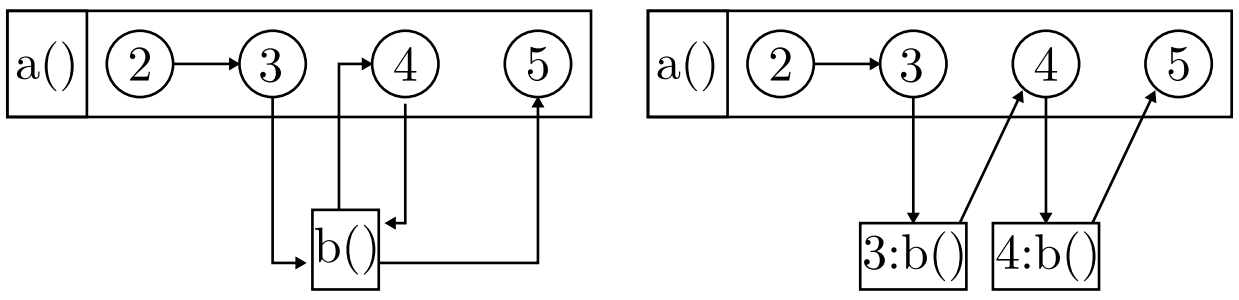

a: Introduced Loop    b: Breaking the Loop

Fig. 4: Preventing Introduced Loops

In Figure 4, Listing 1 is used to generate the two CFGs 4a and 4b. Numbers adjacent to lines in the pseudocode are the in memory address of the statements. There are no cycles in the procedure, however the CFG in 4a contains a cycle with 4 and b(). This is due to nodes of the CFG being identified by their address coupled with two calls to b(). The loop is broken in Figure 4b by identifying each node by the instruction's address and complete call stack. Identifying instructions in this way preserves the structure of the program, without introducing loops into the CFG.

Throughout the remainder of this work, all CFG operations take place over the *expanded CFG* of ribbons. Where a node $n \in N$ is identified by its address $a$ and callstack $s$ of depth $k$, $s = \langle n_1, n_2, ..., n_k \rangle$. Each entry of the callstack is a node in $N$, where the first entry is the top of the stack – the node calling $n$'s function. In the case of the first instruction $h$ (and all other nodes reachable without a function call), the callstack has one element $\varnothing$ indicating no parenting call.

Creating an expanded CFG from a ribbon is a straightforward modification of common CFG [20], [21] program analysis. As such, a detailed description of expanded CFG creation is omitted.

### *B. Conflict Free Region Assignment*

A CFG serves as the basis of construction of conflict free regions and subsequently the conflict free region graph. In the previous work [5], a single node of the CFG could belong to multiple CFRs. Nodes of the CFG are excluded from participating in multiple CFRs under BUNDLEP. If nodes from the CFG participated in multiple CFRs, then loops may be introduced into the CFRG. Additionally, WCETO calculation and scheduling decisions developed in this work, rely upon nodes being assigned to exactly one CFR.

Section III listed the three requirements of CFRs from [5]. A fourth requirement is added to accommodate BUNDLEP scheduling and WCETO calculation for a CFG $G = (N, E, h)$ and CFRs $(F_1, F_2, ..., F_f)$ where $F_i = (N_i, E_i, h_i)$:

4) A node $n \in N$ is present in at most one CFR: $n \in N_i \implies \forall_{k \neq i}\ n \notin N_k$

Placing a node of the CFG in a single CFR is referred to as *assignment*. The assignment process relies upon nodes of the expanded CFG being annotated with their call stack and inner-most loop head. A single node may participate in multiple loops (loops embedded within another). All loops have a head; a starting instruction that determines if the loop will repeat. The inner-most loop head of an instruction is the loop head closest to the node in the hierarchy of embedded loops it belongs to, identified by any suitable algorithm [22].

Assignment begins with a bi-level depth first search (DFS). The top-level DFS marks nodes of $G$ as CFR entry points. The bottom-level DFS marks nodes with their CFR while handling the special cases that could create loops in the CFRG. Both use conflicts to bound their searches and return the set of conflicting nodes as successor nodes to continue the search. The coordinated result of the bi-level searches complete the first stage of assignment, called *CFR entry tagging*. The top-level DFS procedure TAGCFRS() is presented as pseudocode in Algorithm 1. It makes use of a simulated cache $C$ to identify conflicts, with three notable methods: $C$.insert($a$) caches $a$'s block, $C$.clear() removes all blocks, and $C$.conflicts($a$) returns true if $C$.insert($a$) would evict a cached block.

**Algorithm 1** TAGCFRS()

```
G = (N, E, h)                              ▷ Expanded CFG G
C                                          ▷ Simulated Cache
procedure TAGCFRS
    s.clear()                              ▷ Local stack
    v.clear()                              ▷ Visited node array
    s.push(h)                              ▷ Starting condition
    while not s.empty() do
        n ← s.pop()                        ▷ Take a node
        v[n] ← true                        ▷ Mark the node as visited
        C.clear()                          ▷ Reset the cache
        X ← LABELNODES(n)                  ▷ Label CFR nodes
        for x ∈ X do
            s.push(x) if not v[x]  ▷ Conflict begins a CFR
        end for
    end while  ▷ v[n] = true indicates n is a CFR entry.
end procedure
```

The TAGCFRS() procedure is responsible for tracking the entry instructions of CFRs. It resembles a typical DFS, marking nodes as visited, and adding them to the search list when they have not been visited. Where TAGCFRS() differs is the selection of subsequent nodes. In a typical DFS, the subsequent nodes are the immediate successors of the current node, in TAGCFRS() the subsequent nodes are the entry instructions of successive CFRs.

Those entry instructions are determined by the LABELNODES() method. Since no CFR may contain a cache conflict, any instruction that would conflict must be the entry instruction of a subsequent CFR. Figure 5 provides an example call to LABELNODES($n_3$), $n_3$ is the entry instruction of the current CFR and LABELNODES($n_3$) returns the set of entry instructions for subsequent CFRs $\{n_7, n_8, n_9\}$. Placed below each node in the figure is the cache block it maps to.

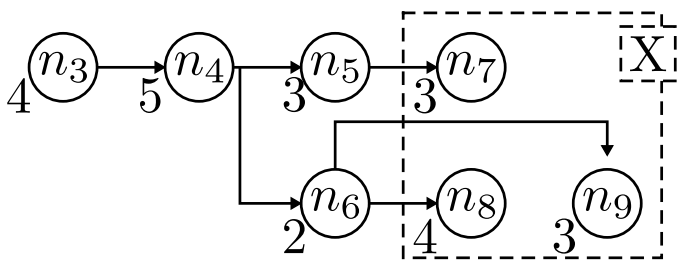

Fig. 5: Call to LABELNODES($n_3$) Returning X

In addition to returning the entry instructions of subsequent CFRs, the LABELNODES() procedure is responsible for marking nodes of the CFG with the CFR they belong to. In Figure 5 the nodes $\{n_3, n_4, n_5, n_6\}$ are labeled with their CFR $n_3$. The pseudocode for LABELNODES() is given in Algorithm 2.

**Algorithm 2** LABELNODES()

```
G = (N, E, h)               ▷ CFG G, shared with TAGCFRS()
C                           ▷ Simulated cache, shared with TAGCFRS()
procedure LABELNODES(n)
    s, x        ▷ Local stacks (not shared with TAGCFRS())
    v ▷ Local visited array (not shared with TAGCFRS())
    if n.label ≠ ∅ then
        ℓ ← n.label                 ▷ Breaking an existing CFR
    end if
    s.push(n)
    while not s.empty() do
        u ← s.pop()
        if (n.isHead() ∧ not n.inLoop(u)) ∨
                                        ▷ Case 1, Loop Exit
            (u.isHead() ∧ u ≠ n) ∨
                                        ▷ Case 2, Loop Head
            (u.label ≠ ∅ ∧  u ≠ ℓ) ∨
                                        ▷ Case 3, Already Assigned
            (C.conflicts(u.a))          ▷ Case 4, Cache Conflict
             then
            x.push(u)                   ▷ Push the Conflict
            v[u] ← true                 ▷ Skip u's successors.
        end if
        next while if v[u]              ▷ Already visited
        v[u] ← true
                                        ▷ Case 5, Add to CFR n
        u.label ← n                     ▷ Label u with CFR n
        C.insert(u.a)                   ▷ Insert u into the Cache
        for y ∈ G.succ(u) do
            if not v[y] then
                s.push(y)
            end if
        end for
    end while
    return x
end procedure
```

During each iteration of the DFS within LABELNODES(), a candidate node $u$ is deemed within the CFR or an entry instruction of a subsequent CFR. If $u$ is within the CFR, the node is labeled with the CFR's initial instruction $n$, and $u$'s successors are added to the search list $s$. If $u$ is an entry instruction it is placed in the set of conflicts $x$, and no successors of $u$ are added to $s$. When the search list is empty, the set of conflicts $x$ is returned to the caller.

There are four cases in LABELNODES when the node $u$ may be deemed a conflict. The simplest is Case 4, when $u$ conflicts with another value present in the cache. Since CFRs must contain no conflicts, $u$ cannot be added to the CFR $n$.

Case 1 and 2 are related to loops. If $u$ falls under Case 1, then $n$ is a loop-head to which $u$ does not belong. Since $u$ was reachable from some instruction in the CFR $n$, it must be an exit point of the loop and will not be permitted as part of the CFR. If $u$ falls into Case 2, then $u$ is a loop-head. Loops are collapsed (described in Section VI) to ensure the CFRG is a DAG. To permit collapsing, loop-heads must start CFRs and CFRs must only contain instructions of the same loop.

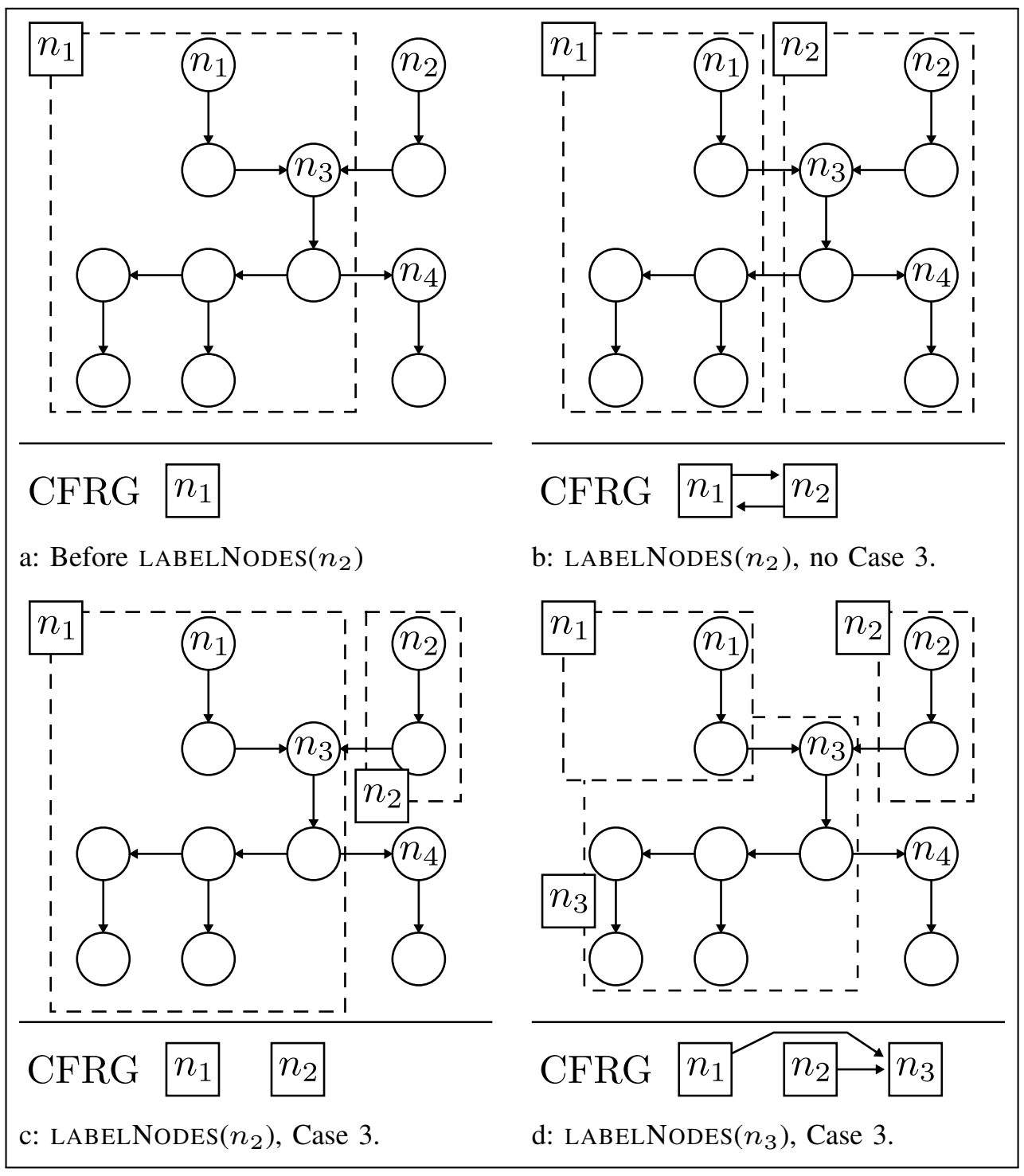

Fig. 6: Case 3 Protection

Case 3 provides two necessary forms of protection. The first is against loops being added to the CFRG (ie. the fourth CFR requirement). The second is to prevent the nodes of a CFR becoming disconnected. For the WCETO method each CFR must have a single WCET path, which will not exist if the CFR is not weakly connected.

Figure 6 compares the result if Case 3 was not present versus it being present. Divided into four parts, Figure 6a is a snapshot of the CFRG from the TAGCFRS() procedure's perspective after calling LABELNODES($n_1$) before calling

LABELNODES($n_2$); the top half is the CFG, the bottom half is the resulting CFRG. Figure 6b shows the result of calling LABELNODES($n_2$) without Case 3 in place, creating both a disconnected CFR $n_1$ and a loop in the CFRG between $n_1$ and $n_2$. Figure 6c is the result of calling LABELNODES($n_2$) with Case 3 protection. With the protection in place when LABELNODES($n_2$) returns to TAGCFRS(), it returns $\{n_3\}$ as the set of entry points for subsequent CFRs. There is a an issue to be resolved with respect to $n_3$ in Figure 6c. Before the call $n_3$ was previously assigned to the CFR $n_1$, but is now an entry node to a new CFR. As previously stated, a node must reside in exactly one CFR. Case 3 resolves this issue as well in Figure 6d, the result of calling LABELNODES($n_3$).

When the TAGCFRS() procedure returns, tagging has been completed. All nodes of the CFG have been assigned to CFRs given by their $n$.label. What remains to complete the CFRG is to add edges between CFRs. This final stage is referred to as *linking*. Pseudocode for linking is ommitted due to the simplicity of the operation; a DFS of the CFG where unique labels are added to the CFRG as nodes, and edges added when edges in the CFG have differing labels at the end points.

The result is a CFRG $R = (N, E, h)$, where $N$ is the set of CFRs, $E$ the edges between CFRs, and $h$ the entry CFR. For consistency and clarity, a CFR $F_i$ is identified by its entry node in the CFG $n_i$, so is the corresponding node in the CFRG. For example, in Figure 6 the node in the CFRG and the CFR $n_3$ is labeled so because its entry instruction was $n_3$ from the CFG.

## VI. BUNDLEP

After converting a CFG to a CFRG, priorities are assigned to CFRs to minimize the number of activations of each bundle. At run-time, the BUNDLEP scheduling algorithm relies on some hardware mechanism that halts threads attempting to execute the entry instruction of an inactive CFR. Such a mechanism was proposed in [23]; it relied on cache evictions rather than thread execution, making it unsuitable for BUNDLEP. This section proposes a new conflict interrupt mechanism suitable for halting threads. The XFLICT interrupt behavior closely matches that of hardware breakpoints [24].

### A. Hardware Support

The XFLICT interrupt represents the attempted execution of an instruction that may result in a cache conflict. Since the execution may or may not result in a conflict, the conclusion that an interrupt should be raised cannot be made without additional information. That additional information is encoded in the XFLICT TABLE. When the program counter is set to a value present in the XFLICT TABLE, the proposed hardware mechanism halts the CPU before executing the instruction and raises an XFLICT interrupt carrying ancillary data encoding the address of the instruction that raised it.

To illustrate how the interrupt, table, and scheduling algorithm work together the process of activating $n_2$ is illustrated below. In Figure 7a, $n_1$ has been depleted and its two threads have moved to $n_3$ where they are blocked. The BUNDLEP scheduling algorithm will now select $n_2$ as active and begin executing threads. Before $n_2$ is activated, the XFLICT TABLE (abbreviated **X** in the figure) is cleared, and the values are replaced with the addresses of the entry instructions of the subsequent CFRs of $n_2$: $(a_3, a_4)$ – recall every node of the CFG is identified by it's address $a$ and stack $s$, CFRs are identified by their entry nodes from the CFG, in this case $n_3$ and $n_4$. In Figure 7b, $n_2$ has been activated and threads are permitted to execute. When any thread attempts to execute either $a_3$ or $a_4$ the XFLICT interrupt is raised before the thread can execute it. The BUNDLE scheduling algorithm receives the interrupt, blocks the thread, and places it into the correct bundle. Figure 7c provides a snapshot of execution after the first two threads of $n_2$ have exited the bundle and are blocked awaiting the activation of $n_3$ or $n_4$.

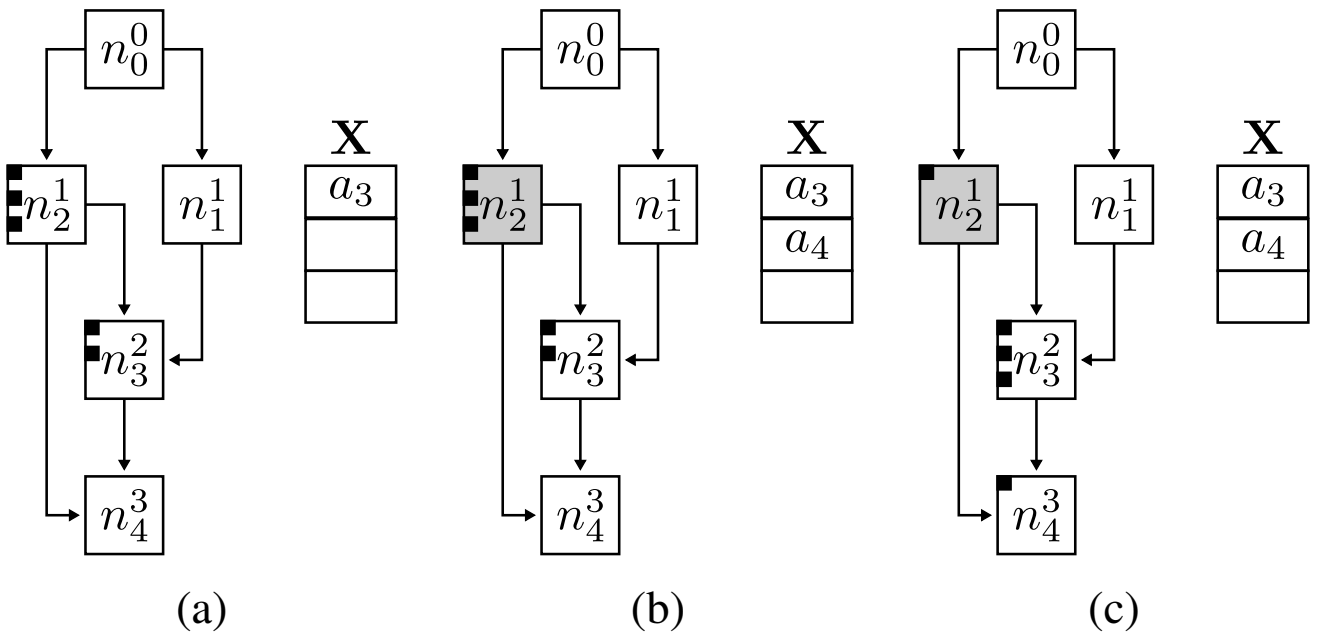

Fig. 7: XFLICT Interrupts and BUNDLEP

### B. BUNDLEP's Scheduling Algorithm

Incorporation of the XFLICT interrupt and priorities into BUNDLE scheduling is given by Algorithm 3, the pseudocode for BUNDLEP. Bundles are stored in the array $B$, indexed by the nodes of the CFRG $R$. A single bundle $b$ has four members: a starting address $b.a$, node from the CFRG $b.n$, priority $b.\pi$, and set of threads $b.t$. When a bundle is ready, it is placed in the priority queue $P$.

Lines 6-8 are responsible for adding the threads of the job to the initial bundle and adding the bundle to the priority queue. Each iteration of the outer while loop starting on line 9 removes the best priority bundle from the queue and activates it.

Lines 10-17 manage the XFLICT TABLE and a local mapping from address to CFRG node. Addresses of successors to $b$ are added to the XFLICT TABLE so that interrupts will be delivered when threads exit $b$. However, those addresses may map to more than one CFR. To avoid ambiguity, the $S$ array maps addresses of successor CFRs to their proper node in the CFRG.

The inner for loop starting on line 18, executes each thread of the active bundle. When an XFLICT interrupt is raised, the thread is placed in the successor's bundle and added to the priority queue as a candidate for activation. If the interrupt is not raised, the thread has terminated and belongs to no bundle.

It is appropriate to mention the data structures used for bundles and threads at this point. Threads of bundles are removed

**Algorithm 3** BUNDLEP Scheduling Algorithm

1: $T$ ▷ Set of Threads
2: $R = (N, E, h)$ ▷ Conflict Free Region Graph
3: $P$ ▷ Priority Queue of Ready Bundles
4: $B$ ▷ Array of bundles indexed by their node $n \in N$
5: **procedure** BUNDLEP
6: $b \leftarrow B[h]$
7: $b.t$.add($T$)
8: $P$.insert($b$, $b.\pi$)
9: **while** $b \leftarrow P$.removeMax() **do** ▷ Best Bundle
10: $S \leftarrow \varnothing$ ▷ Clear the successor array
11: XFLICT_CLEAR() ▷ Clear the XFLICT table
12: ▷ Create the mapping of address to node
13: **for** $s \in R$.succs($b.n$) **do**
14: $b_s \leftarrow B[s]$
15: $S[b_s.a] \leftarrow b_s$
16: XFLICT_ADD($b_s.a$)
17: **end for**
18: **for** $t \in b.t$ **do**
19: **try** {
20: RUN($t$)
21: } **catch** (XFLICT $x$) {
22: $b_{next} \leftarrow S[x.a]$ ▷ Get the next bundle
23: $b_{next}.t$.add($t$)
24: P.insert($b_{next}$, $b_{next}.\pi$)
25: next **for** ▷ $t$ has not terminated
26: }
27: **end for**
28: **end while**
29: **end procedure**

on line 18, the order is irrelevant requiring nothing more than an array with $\mathbb{O}(1)$ time for extraction. Bundle activation order is important, being removed from a priority queue on line 9. An efficient implementation using a Fibonacci heap has complexity $\mathbb{O}(1)$ for insertions and amortized complexity $\mathbb{O}(\log n)$ for removeMax. Thus each iteration of the while loop is dominated by the removeMax operation, which will be accounted for in the WCETO analysis of Section VII.

### C. Priority Assignment

Priorities are assigned to CFRs during the offline object analysis based on their longest path value from the entry CFR $h$. The priority assignment method requires the CFRG to be a DAG. This requirement is impossible to meet in all circumstances. Although the analysis does not introduce loops, the program structure may contain them. These *user defined* loops may be contained within a single CFR, which would ease the analysis, or they may span multiple CFRs creating a necessary loop in the CFRG.

The structure of CFRs from user defined loops allows them to be collapsed into a single *false* node. Every loop has a head, members, and at least one exit CFR. The loop head identifies all members of the loop, which are are *collapsed* into a single false node during priority assignment and WCETO calculation. Priorities are first assigned to the interior nodes, then assigned the false node in the greater *scope*.

Figure 8 illustrates two scopes of loop collapse. In Figure 8a, at the greatest scope there is a single loop which is collapsed into a single false node $n_1$. As part of being collapsed, the interior nodes are assigned priorities corresponding to their longest path from the loop head. The false node is given a priority according to the longest path to reach it from the entry node, with two caveats. First, the interior *real* (ie., not false) node that is the loop head must have a priority worse than all other members of the loop. Second, each false node is given a unique priority within its scope, this guarantees only members of the loop execute until all threads exit the loop. This is why the false node $n_1$ denoted by a hexagon has priority 4, which is inherited by the loop head.

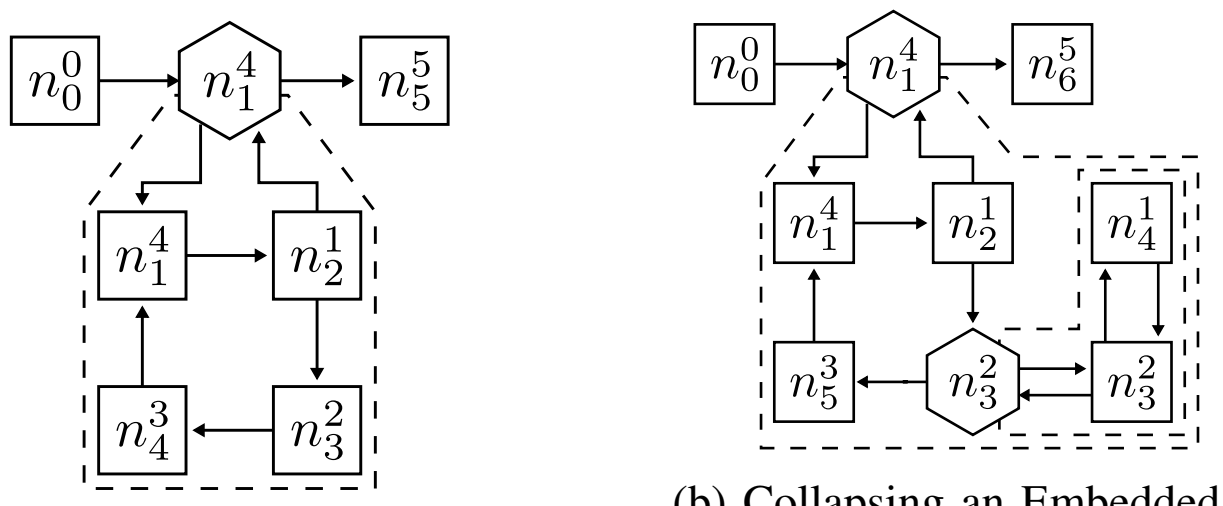

(a) Collapsing a Loop

(b) Collapsing an Embedded Loop

Fig. 8: Loop Collapse and False Nodes

Figure 8b illustrates a smaller scope of embedded loop collapse. The inner loop is processed first, which affects the priority values of members within the outer loop.

By repeatedly collapsing loops into false nodes, all loops are removed and the CFRG is converted to a DAG. Additionally, at each scope within a collapsed node the graph structure is also DAG (when excluding the incoming edges to the loop head). Assigning priorities to interior nodes of the same scope equal to their longest path from the loop head (with the worst priority) guarantees each CFR is activated at most once per iteration shown in Theorem 2.

**Theorem 2** (Maximum Bundle Activations per Iteration)**.** *For a graph of the collapsed node $G = (N, E, n_0)$ with loop head $n_0$, set of in-scope nodes $N$ and edges $E$, where each node $n_i \in N$ has priority $\pi_i$ equal to the longest path from $n_0$ to $n_i$, and $n_0$ has priority worse than all others $\{\pi_0 \mid \forall_{n_j \in N}\ \pi_0 > \pi_j\}$ the bundle $n_i$ will be activated at most once per iteration of $n_0$.*

*Proof. Observation I:* For a node $n_\ell$ in scope of $n_0$ that is a loop head, $n_\ell$ will be collapsed with all other nodes that have $n_\ell$ as their inner-most loop head. Only the collapsed node $n_\ell$ will be in scope of $n_0$. Therefore, for any collapsed node, there is exactly one loop in scope with head $n_0$.

*Observation II:* A single iteration of the loop contained within a collapsed node is defined as the series of activations that begins with the activation of $n_0$ and ends just before $n_0$ would be selected as active once again. Since $n_0$ has the worst priority among all in-scope nodes, all other nodes must have been depleted before $n_0$ could be activated again.

*Observation III:* All other threads not in the loop of $n_0$ must be in bundles with worst priority than any bundle in the current scope. (Otherwise, bundles in $n_0$'s scope would not be scheduled). Thus, the current loop (and embedded loops) will complete all iterations before any out-of-scope bundle is executed.

Consider the graph $G$ where the incoming edges to $n_0$ have been removed, removing the cycle, i.e., $E = \{(u, v) | (u, v \neq n_0) \in E\}$ as a graph $G = (N, E, n_0)$. By Observation I, $G$ is now a DAG of CFRs. Treating a single iteration as a job release and applying Theorem 1 to $G$, each $n \in N$ is activated at most once per iteration for all threads executing the loop. □

## VII. BUNDLEP WCETO Calculation

As a practical effort, the focus of this work is on the calculation of an effective, safe WCETO bound. To that goal, the bound calculation is formulated as an integer linear program (ILP) the number of variables grow at $\mathbb{O}(V + E)$. This section is devoted to describing the transformation of a CFRG into a set of constraints and an objective function.

Assigning priorities to nodes of the CFRG and collapsing loops (as described in Section VI) guarantees each node is activated at most once. As such, the contributions of individual nodes may be considered in isolation. What determines a node's individual contribution is the number of threads assigned to it. The maximization problem becomes finding the greatest sum of contributions of individual nodes for a valid assignment of threads. Figure 9 illustrates the relationship between the CFRG, WCETO of individual nodes $\omega_n(t_n)$, and objective function $\Omega = \sum_{n \in N} \omega_n(t_n)$. Refer to Figure 13 in the appendix for a more detailed example.

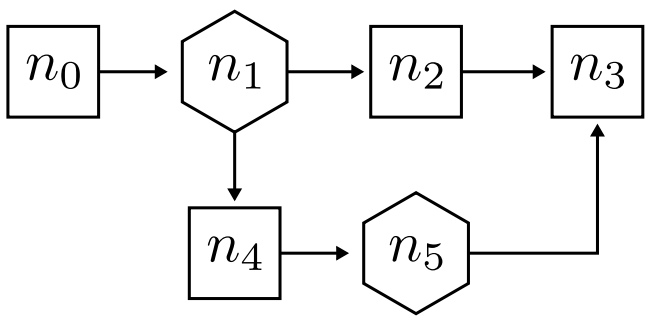

| $i$ | $n_0$ | $n_1$ | $n_2$ | $n_3$ | $n_4$ | $n_5$ | |
|---|---|---|---|---|---|---|---|
| $t_i$ | 4 | 4 | 2 | 4 | 2 | 2 | $\Omega$ |
| $\omega_i$ | 40 | 20 | 70 | 30 | 50 | 10 | 220 |

Fig. 9: CFRG Individual Nodes and ILP Objective

The WCETO of a node $\omega_n(t_n)$ depends on the number of threads assigned to it $t_n$. For real nodes, the function takes the form of Equation 1. We assume a timing-compositional architecture [25]; the number of cycles required to complete a single node is divided into two parts: the *memory demand* and the *execution demand*. The memory demand of a node $n$ is the product of the set of ECBs found in the CFR $ECB_n$ and the block reload time $\mathbb{B}$, $\gamma_n = |ECB_n| \cdot \mathbb{B}$. The execution demand is the product of the worst-case execution time of a single thread over the node $c_n$ and the number of threads assigned $t_n$. Two context switch costs are included to reflect the penalty of BUNDLEP scheduling, $\mathbb{X}_b$ is the number of cycles required to switch to a new active bundle, and $\mathbb{X}_t$ is the cost of selecting a thread from the active bundle. The costs $\mathbb{X}_b$ and $\mathbb{X}_t$ are directly related to lines 9 and 18 of Algorithm 3.

$$\omega_n(t_n) = \begin{cases} 0, & t_n = 0 \\ c_n \cdot t_n \cdot \mathbb{X}_t + \mathbb{X}_b + \gamma_n, & t_n \geq 1 \end{cases} \tag{1}$$

The WCETO for false nodes is given in Equation 2. It depends on $\mathbb{I}_n$ the maximum number of iterations of the loop collapsed into the false node $n$, and $inscope(n)$. The set of nodes returned by $inscope(n)$ are the interior nodes of the false node $n$, which may include other false nodes. For example in Figure 8b, $inscope(n_1) = \{n_1, n_2, n_3, n_5\}$ where $n_1$ is a real node and $n_3$ a false node.

$$\omega_n(t_n) = \begin{cases} 0, & t_n = 0 \\ \mathring{\gamma}_n + \mathbb{I}_n \cdot \sum_{i \in inscope(n)} \mathring{\omega}_i(t_i), & t_n \geq 1 \end{cases} \tag{2}$$

The memory and execution demand of a false node are not entirely separable. Individual nodes within scope of $n$ have their per-iteration contribution bounded by $\mathring{\omega}_i(t_i)$, described later. An initial memory demand for the false node $n$ is calculated as $\mathring{\gamma}_n$, it represents the number of cycles required to cache all blocks of nodes within the collapsed node regardless of scope. The set of nodes $allscope(n)$ includes any real node that has been collapsed under $n$. Using Figure 8b, $allscope(n_1) = \{n_1, n_2, n_3, n_4, n_5\}$ where no node is false.

Using the set of all nodes collapsed under $n$, the multi-set union of their ECBs is formed and labeled $\mathring{ECBs}_n = \bigcup_{i \in allscope(n)} ECBs_i$. The product of cardinality of the ECB multiset and the block reload time produces $\mathring{\gamma}_n = \mathbb{B} \cdot |\mathring{ECBs}_n|$. This value accounts for all of the cycles required to load every cache block found in the collapsed node.

For a false node $i$ collapsed under a false node $n$, its WCETO contribution $\mathring{\omega}_i(t_i)$ is defined by Equation 2. For a real node $i$ under a false node $n$, its WCETO contribution $\mathring{\omega}_i(t_i)$ is given by Equation 3. It includes context switch costs, execution demand, and the worst-case memory demand. A method similar to the ECB-Union CRPD approach [2] is employed to calculate the memory demand from the perspective of the affected node $i$. The worst-case occurs when another real node collapsed under $n$ evicts the ECBs of $i$, forcing the blocks $ECBs_i$ to be loaded when $i$ is activated. The number of evictions can be bounded by the ECBs of all loop members, specifically those that occur more than once in the loop. The set of ECBs found more than once under the collapsed node are given by $\overset{2}{ECBs}_n = \{\bigcup_{u \cdot k} \mid u \cdot k \in \mathring{ECBs}_n \wedge k \geq 2\}$. Thus, the memory demand bound for $i$ is $\overset{2}{\gamma}_i = |\overset{2}{ECBs}_n \cap ECBs_i| \cdot \mathbb{B}$. Incorporating per-iteration context switches, execution and memory demand into the bound for $i$ yields Equation 3.

$$\mathring{\omega}_i(t_i) = \begin{cases} 0, & t_n = 0 \\ \mathbb{X}_b + \mathbb{X}_t \cdot t_i \cdot c_i + \overset{2}{\gamma}_i, & t_n \geq 1 \end{cases} \tag{3}$$

A valid assignment of threads takes into account the structure of the CFRG. To reflect the structure, threads are treated as flow traversing the edges of nodes. The entry node is treated as the source of flow, providing a total $m$ threads on its outgoing edges. All threads much reach the terminal node. At each node the sum of threads along incoming edges and outgoing edges must be equal (except the entry and terminal nodes).

The ILP finds the assignment of threads according to the flow of the CFRG which maximizes the number of cycles required to complete $m$ threads according to BUNDLEP scheduling, thus bounding the WCETO of a job. To conserve space in the main body of the work, the details of transforming the formulae to ILP constraints is placed in Appendix A.

## VIII. Evaluation

The evaluation takes the approach of comparing BUNDLEP's thread-level scheduling algorithm to a naive algorithm which executes threads one after another (serially). Individual benchmarks from the Mälardalen [26] MRTC suite are treated as ribbons releasing $m$ threads per job. The WCET of each job is analyzed twice, once for a single multi-threaded task scheduled by BUNDLEP, and again for $m$ serial threads by Heptane. Similarly, the run-time behavior is collected for each benchmark under BUNDLEP and serial execution. A fully functional virtual machine with the tools and source is available for download to recreate these results or expand upon them [9].

Ideally, BUNDLEP would also be compared with BUNDLE. However, the evaluation in [5] used synthetic programs rather than compiled source (for any architecture). WCETO analysis for BUNDLE is also intractable with complexity $\mathbb{O}((|N|!)^m)$. This is due to the nature of the algorithm, it does not restrict the flow of threads through the CFRG, which demands all-paths be repeatedly searched. A novel BUNDLE WCETO implementation of an intractable solution, which is known to be dominated by BUNDLEP is not compelling, as such it is omitted from the evaluation.

The target platform for WCETO analysis and execution is a MIPS 74K processor with a direct mapped single level instruction cache. Cache blocks are restricted to 32 bytes. The CPI $\mathbb{I}$, block reload time $\mathbb{B}$, and number of cache blocks $\ell$ vary based on Table I. Additionally, the number of threads per job $m$ vary from 1 to 16 by powers of two. Jobs are executed on a MIPS simulator provided by Heptane and modified to execute BUNDLEP scheduling or a serial batch of threads.

| CPI ($\mathbb{I}$) | BRT ($\mathbb{B}$) | $\ell$ | $m$ |
|---|---|---|---|
| 1 | 100 | {8, 16, 32} | {2, 4, 8 ,16} |
| 10 | 100 | {8, 16, 32} | {2, 4, 8 ,16} |

TABLE I: MIPS 74K Architecture Parameters

Of the 27 MRTC benchmarks, 18 were evaluated. The selection is limited by Heptane's ability to perform WCET analysis using the `lp_solve` ILP solver and the 12 gigabytes of RAM available (the complete results are available in the technical report [27]). For each benchmark, Heptane produces a single WCET value for the execution of one thread through the ribbon denoted $c_H$. To compare Heptane's WCET to BUNDLEP's WCETO $\Omega$, the number of threads and context switch costs are incorporated and quantified as a difference $\Delta_\omega = m \cdot (c_H + \mathbb{X}_b) - \Omega$. Similarly, the number of cycles required to execute on the simulator serially is denoted $E_H$, cycles required for BUNDLEP execution denoted $E_B$, and the comparison $\Delta_B = E_H - E_B$. A positive $\Delta$ value indicates the BUNDLEP approach provides a benefit.

Context switch costs are encapsulated in BUNDLEP's WCETO $\Omega$, they are not incorporated into $c_H$. Serial execution models threads as jobs of distinct tasks, switching between jobs incurs a task-level context switch cost. A job switch includes more heavyweight operations than a bundle-level context switch, such as exchanging task control blocks instead of thread control blocks. To favor the classical approach, the bundle-level context switch cost $\mathbb{X}_b$ is also used as the task-level context switch cost.

There are two context switch costs for BUNDLEP: between bundles $\mathbb{X}_b$ and between threads $\mathbb{X}_t$. To find representative values for both costs, sample programs were written for the target architecture and analyzed using Heptane. Selecting a thread from an array and jumping to a new instruction address took less than 10 cycles. For $\mathbb{X}_b$ a precise value would require the implementation of a priority queue, supported by an optimized heap, that could be analyzed by Heptane. The implementation of such a priority queue is beyond the scope of this work. However, a limited set of queue operations were analyzed for removing one of two items taking less than 55 cycles. Since removeMax grows at $\log_2(m)$ for bundle selection, the context switch costs are set to $\mathbb{X}_b = 55 \cdot \log_2(m)$ and $\mathbb{X}_t = 10$.

Figures 10a and 10b summarize the results of the evaluation. The y-axis represents the number of benchmarks where BUNDLEP benefits the task. Along the x-axis, the groups separate the architecture parameters which are enumerated by their "($\mathbb{B}$:$\mathbb{I}$, $\ell$)". For each group the result is tallied by the number of threads per job, from 1 to 16.

There are several interesting observations to be made in Figure 10a. Though BUNDLEP analysis provides a benefit in the majority of cases, it does not always. As the cache size is reduced the number of benchmarks that benefit increases. Similarly, as the number of threads per job increases so do the number of benchmarks that benefit. These trends are due to the number of misses (typically) increasing as the cache size is reduced, or the number of threads increased. BUNDLEP avoids these conflicts or converts them to cache hits. Surprisingly, for a single thread per job BUNDLEP may provide a lower bound – this is likely due to the use of the expanded CFG instead of the conventional CFG used by Heptane's analysis.

The run-time benefit summary in Figure 10b more heavily favors BUNDLEP, with unsurprising trends. For a single thread per job, BUNDLEP provides no benefit since there is no reason to block and incur context switch costs. As the number of threads increases so does the run-time performance. As the cache size decreases, the number of benchmarks that see a run-time performance increases. When compared to the WCETO benefit, more benchmarks benefit from the run-time behavior than the analysis would suggest. This implies further refinements of the analysis are possible.

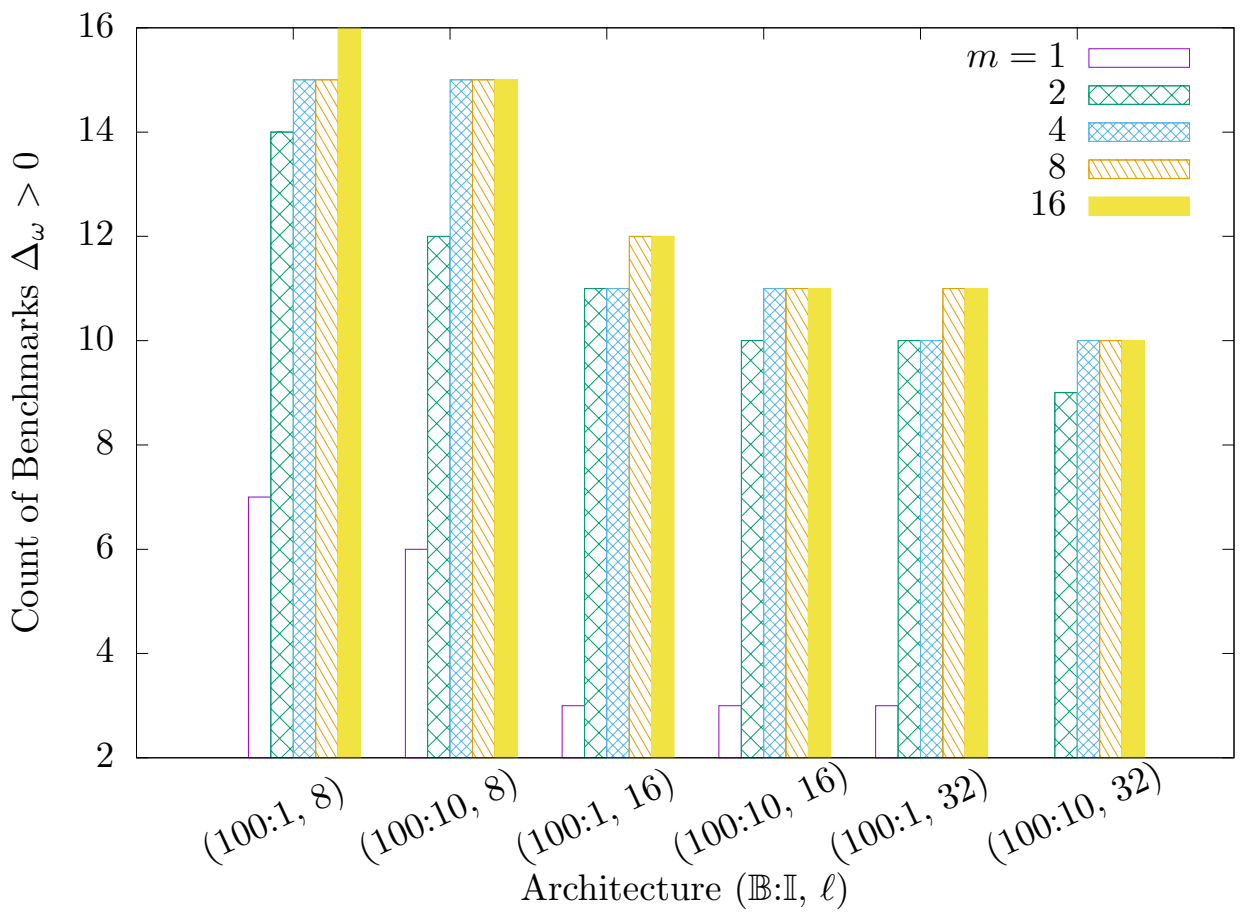

(a) Analytical Benefit of `BUNDLEP`

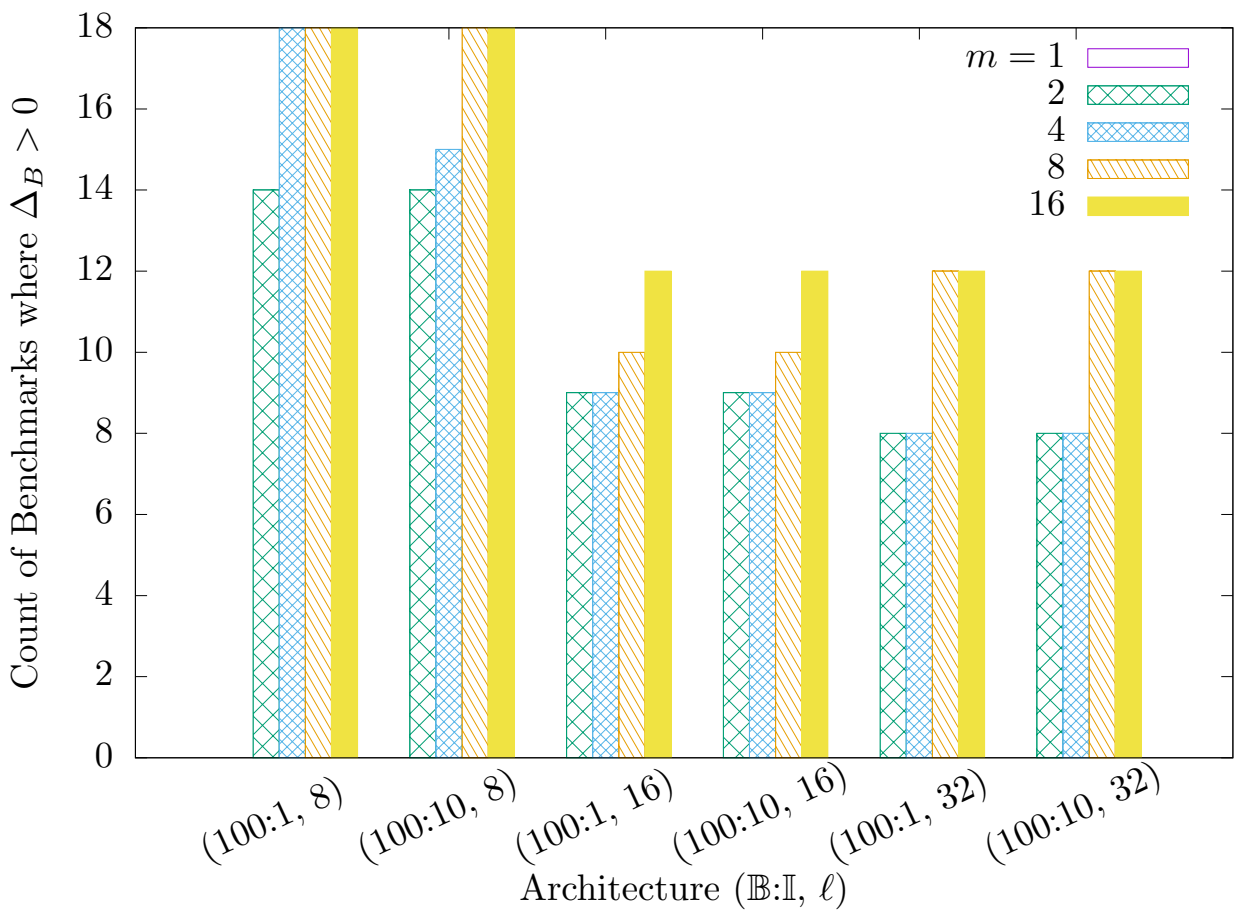

(b) Run-Time Benefit of `BUNDLEP`

Fig. 10: Benefits of `BUNDLEP`

Across the four dimensions of the evaluation (cache size, BRT, CPI, and number of threads per job), the expectation of `BUNDLEP`'s benefit will increase as the cache size decreases, increase as the BRT increases, decrease as the CPI increases, and increase as the number of threads per job increase. Many of the benchmarks match these expectations, such as the results for `ud` found in Figure 11.

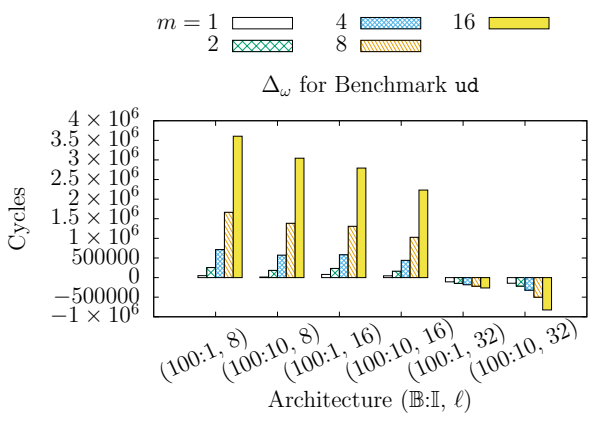

(a) Analytical Benefit

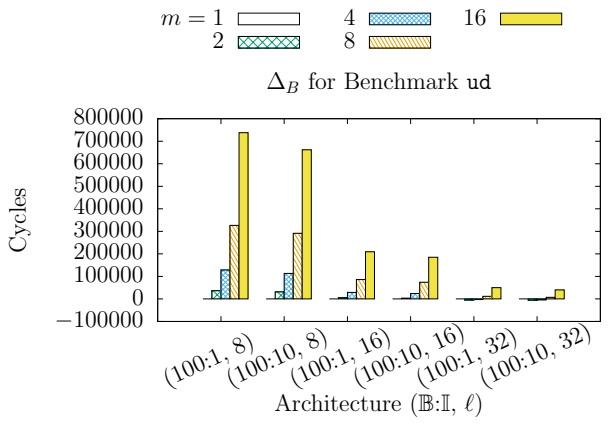

(b) Run-Time Benefit

Fig. 11: Results for the `ud` Benchmark

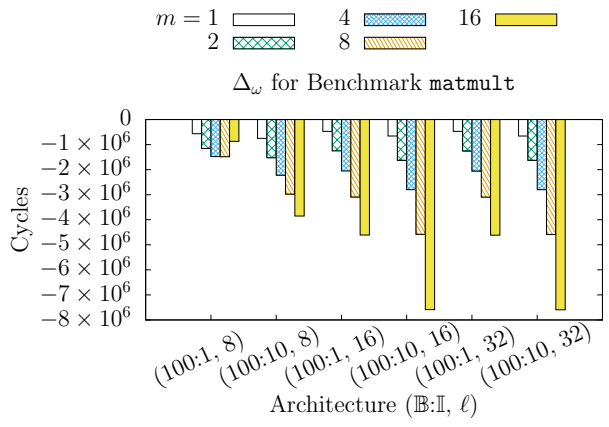

(a) Analytical Benefit

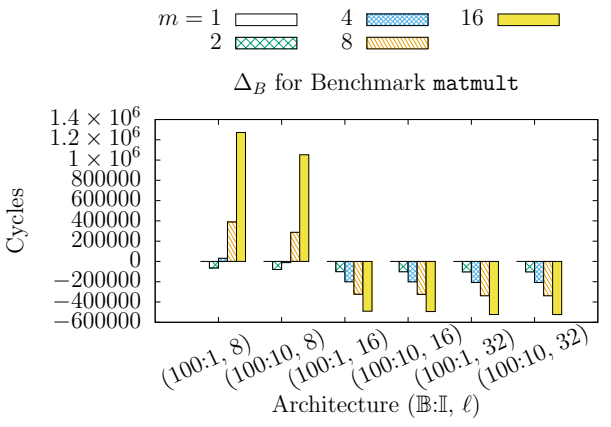

(b) Run-Time Benefit

Fig. 12: Results for the `matmult` Benchmark

The anomalies provide insights into the circumstances where `BUNDLEP` may be improved. Using the `matmult` benchmark as an example, `BUNDLEP` never produces an analytical benefit, and rarely a run-time benefit (Figure 12). Counter-intuitively as the number of threads increase the analytical result worsens compared to the serial bound. This is due to the structure of the CFRG, which has several small CFRs contained within multiply embedded loops. The number of bundle-level context switches with cost $\mathbb{X}_b$ outweighs the benefit of sharing cache values.

## IX. Conclusion

This work expands upon the foundation set in `BUNDLE` [5]. The central principles of treating the cache as a benefit to execution times by scheduling threads in a cache cognizant manner are refined and improved. In [5] permitting instructions to reside in multiple CFRs created ambiguity in scheduling decisions. The methods of [5] to create CFRs and CFRGs introduced unnecessary loops.

Refining the creation of CFRs removes ambiguity from scheduling decisions and prevents loops from being added to the CFRG. The preclusion of additional loops, and the assignment of priorities to CFRs reduces the complexity of the WCETO calculation. Additionally, restricting the flow of threads through the CFRG tightens the WCETO bound.

These improvements are demonstrated in the toolkit available for download and reuse for future use and expansion, demonstrating `BUNDLEP`'s practical applicability for representative compiled programs. The evaluation shows a benefit to `BUNDLEP` scheduling in terms of analysis and run-time behavior for specific programs and architecture parameters.

The evaluation provides further motivation for future improvements in the extraction of CFRs, scheduling mechanism, and bound calculation. Benchmarks with anomalous results highlight the cost of `BUNDLEP` scheduling, the greater number of context switch costs. This cost must be balanced against the inter-thread cache benefit, which is not always the case.

Future work seeks to find a balance in two ways: 1.) ensuring CFRs are of greatest size 2.) developing criteria to permit some cache conflicts when an imbalance exists. These efforts coincide with our ongoing development of a multi-task version of `BUNDLEP` integrated with the evaluation toolkit.

## X. Acknowledgments

We would like to express our gratitude to Isabelle Puaut and Damien Hardy for their personal assistance with Heptane and MIPS simulator. Without a freely available, extensible, and reusable platform this work would not have been possible, thank you!

## Appendix A
## ILP Constraints from WCETO Contributions

This appendix is dedicated to describing the transformation of equations 1, 2, 3 and the supply of threads into the constraints of the ILP for WCETO calculation. For a CFRG $R = (N, E, h)$, the objective of the ILP is to maximize:

$$\Omega = \sum_{n \in N} \omega_n(t_n)$$

Several variables are added to the ILP which are not present in the formulae. A binary selector variable $b_n \in \{0, 1\}$ is added for each node, when the value is 1 the node has at least one thread assigned to it. For every edge $(u, v) \in E$, the variable $t_{(u,v)}$ represents the number of threads passed from node $u$ to $v$. The terminal node of the CFRG is identified as $z \in N$, having out-degree zero.

Two functions are defined for each node. The successor and predecessor functions return the sets given by their names. Both of these functions properly obey the scope of the provided node $n$. If a false node is provided, the nodes collapsed within it will not be included in the set. If a real node collapsed within a false node is provided the set will include only nodes found within the collapsed node (real or false).

*Functions*

$preds(n) \triangleq \{u | (u, n) \in E\}$

Set of immediate predecessors of $n \in N$.

$succs(n) \triangleq \{v | (n, v) \in E\}$

Set of immediate successors of $n \in N$.

What follows are the individual constraints generated for each node. To clarify, a top-most false node contributes its WCETO directly to the objective, being a part of $n \in N$. Nodes collapsed within the false contribution to the objective indirectly. Nodes hidden by collapse have their WCETO reflected by their containing false node's $\omega$ value.

*Node Constraints*

$t_n \in \{0, m\}$

Number of threads assigned to node $n$.

$b_n \in \{0, 1\} \leq t_n$

Binary selector for $n$, indicating that $n$ has at least one thread assigned.

$t_n \triangleq \sum_{u \in preds(n)} t_{(u,n)}$

Number of threads assigned to $n$ must be equal to the sum of all entering $n$.

$t_n \triangleq \sum_{v \in succs(n)} t_{(n,v)}$

Number of threads assigned to $n$ must be equal to the sum of all leaving $n$.

$\omega_n \triangleq c_n \cdot t_n \cdot \mathbb{X}_t + \mathbb{X}_b \cdot b_n + \gamma_n \cdot b_n$

WCETO of a regular $n$, see Equation 1.

$\omega_n \triangleq (\mathring{\gamma}_n \cdot b_n) + \mathbb{I}_n \left( \sum_{i \in inscope(n)} \mathring{\omega}_i \right)$

WCETO for a false node $n$, see Equation 2.

$\mathring{\omega}_n \triangleq c_n \cdot t_n \cdot \mathbb{X}_t + (\mathbb{X}_b + \overset{2}{\gamma}_n) \cdot b_n$

WCETO for a real node $n$ within a collapsed node.

*Special Case Constraints*

$t_h \triangleq m$

The initial node $h$ must have all $m$ threads assigned.

$t_z \triangleq m$

The terminal node $z$ must have all $m$ threads assigned.

## Appendix B
## WCETO Example

The ILP objective function $\Omega$, is the sum of the contributions of the CFRs of the CFRG given an assignment of threads per node. Figure 13 illustrates the source of each node's contribution for four threads ($m = 4$). It reuses the structure of Figure 9 with detailed memory and execution demand values that are closer to those found in the evaluation.

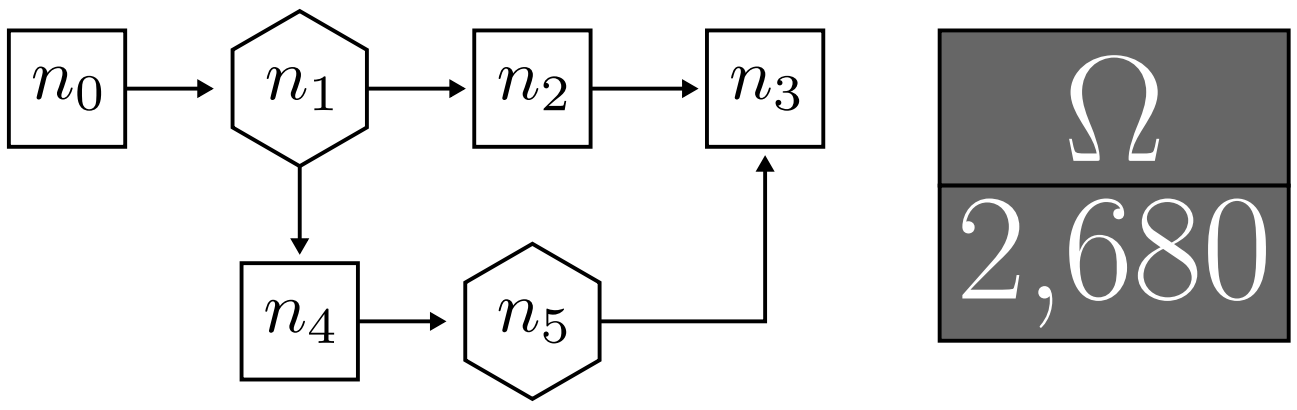

| $i$ | $n_0$ | $n_2$ | $n_3$ | $n_4$ |
|---|---|---|---|---|
| $\gamma_i$ | 400 | 500 | 100 | 300 |
| $c_i$ | 40 | 110 | 5 | 100 |
| $t_i$ | 4 | 2 | 4 | 2 |
| $\omega_i$ | 560 | 720 | 120 | 500 |

| $i$ | $n_1$ | $n_5$ |
|---|---|---|
| $\mathring{\gamma}_i$ | 200 | 300 |
| $\mathbb{I}_{n_i}$ | 8 | 2 |
| $\sum \mathring{\omega}$ | 30 | 20 |
| $t_i$ | 4 | 2 |
| $\omega_i$ | 440 | 340 |

Fig. 13: CFRG Individual Nodes and ILP Objective

When completed, the ILP determines the WCETO bound is 2,680 cycles. Understanding how the bound is calculated is made easier by considering the memory demand indepedently of the execution demand. For $n_0$, there is no decision four threads are assigned. The memory demand for $n_0$ is 400 cycles, and 40 cycles per thread for 560 cycles total.

There are no decisions to be made for $n_1$ or $n_3$, the number of threads assigned to them are determined by the structure of the graph. For threads to be assigned to $n_4$ and $n_5$, their combined execution and memory demand must be compared to $n_2$. For one thread, $n_2$ has a total demand of 610 cycles. For one thread, the combined demand for $n_4$ and $n_5$ is 710 cycles (the demand for interior nodes of $n_5$ is 10 cycles per thread. Though not explicitly listed in the figure, this is the reason $\sum \mathring{\omega} = 20 = 2 \cdot 10$).

The execution demand for a second thread (or third) of $n_2$ is 110 cycles, and the combined execution demand for a second thread of $n_4$ and $n_5$ is also 110 cycles. Any assignment where $t_2$, $t_4$, and $t_5$ are greater than or equal to one will result in the same WCETO value. The assignment in Figure 13 has balanced the threads across paths.

Table 14 lists the benchmarks evaluated

## Appendix C
## Technical Report on WCETO Results

From the MRTC benchmark suite, 18 of the 27 tests were evaluated for their WCET and run-time performance. Benchmarks are treated as one code segment with $m$ requests for execution per job release. The classical perspective treats each execution as a job of a distinct task, while `BUNDLEP` views them as threads.

From the classical perspective jobs are scheduled serially, one after another. This is compared to `BUNDLEP`'s thread level scheduling using two metrics. The first $\Delta_\omega$ is the difference between the WCET of $m$ serial threads and `BUNDLEP`'s WCETO. When $\Delta_\omega$ is positive, `BUNDLEP` provides an analytical benefit. The second metric is $\Delta_B$, the difference in the number of cycles required to complete $m$ threads serially versus `BUNDLEP`'s scheduling mechanism. A positive $\Delta_B$ value indicates a shorter run-time under `BUNDLEP`.

Names of each of the benchmarks evaluated are listed in Figure 14. Starting on the next page are the graphical results. Two graphs are given for each benchmark, one to illustrate the WCETO benefit ($\Delta_\omega$), and a second for the run-time benefit ($\Delta_B$).

| | |
|---|---|
| bs | bsort100 |
| crc | expint |
| fft | insertsort |
| jfdctint | lcdnum |
| matmult | minver |
| ns | nsichneu |
| qurt | select |
| simple | sqrt |
| statemate | ud |

Fig. 14: Benchmarks of MRTC

For each benchmark, altering the architecture parameters has the following anticipated effects on the performance of `BUNDLEP`. In general, as the number of cache lines is increased, the analytical and run-time benefit of `BUNDLEP` is expected to decrease; since the shared resource `BUNDLEP` benefits from becomes less scarce. As the number of threads increase, the analytical and run-time benefit of `BUNDLE` will increase since the use of the cache increases. As the ratio of cache block reload time to instruction execution time $\mathbb{B} : \mathbb{I}$ increases so should `BUNDLEP`'s performance.

For many of the benchmarks `BUNDLEP` performs as expected when varying cache sizes, threads, and CPI. Examples include: `bs`, `bsort100`, `crc`, `insertsort`, `select`, `qurt`, and `ud`. A highlight of the analysis and run-time benefit is the `select` benchmark. In the best case `BUNDLEP` reduces the WCETO by roughly fifty percent, from the serial bound of nearly four million close to two million. At run-time, the observed execution time is roughly two-thirds of the serial version.

For a few of the benchmarks: `matmult`, `expint`, and `ns`, `BUNDLEP` analysis is worse for almost every architecture and thread configuration. For each of these benchmarks the run-time behavior favors `BUNDLEP` only slightly better than the WCETO analysis. Examining the structure of the CFRG for these benchmarks we find embedded loops with small CFRs. Thus, the benefit of cache sharing is reduced and the penalty of bundle-level and thread-level context switch costs are relatively high.

In terms of analysis, the `minver` benchmark has the most inconsistent results. For instance, comparing the configurations of (100:1, 16) and (100:10, 16) as the number of threads increases the performance of `BUNDLEP` increases for the former. Yet, in the latter performance decreases, then increases. This benchmark lies on the cusp of analytical benefit and will receive further investigation, in part motivated by the consistent run-time benefit.

We find merit in `BUNDLEP`'s scheduling and analytical approach. Under the best circumstances (100:1, 8) and $m = 16$, `BUNDLEP` provides an analytical benefit for 16 of the 18 benchmarks, and a run-time benefit for 18. For the least favorable configuration (100:10, 32) and $m = 2$, `BUNDLEP` provides an analytical benefit for 9 of 18, and a run-time benefit for 8 of the 18 benchmarks.

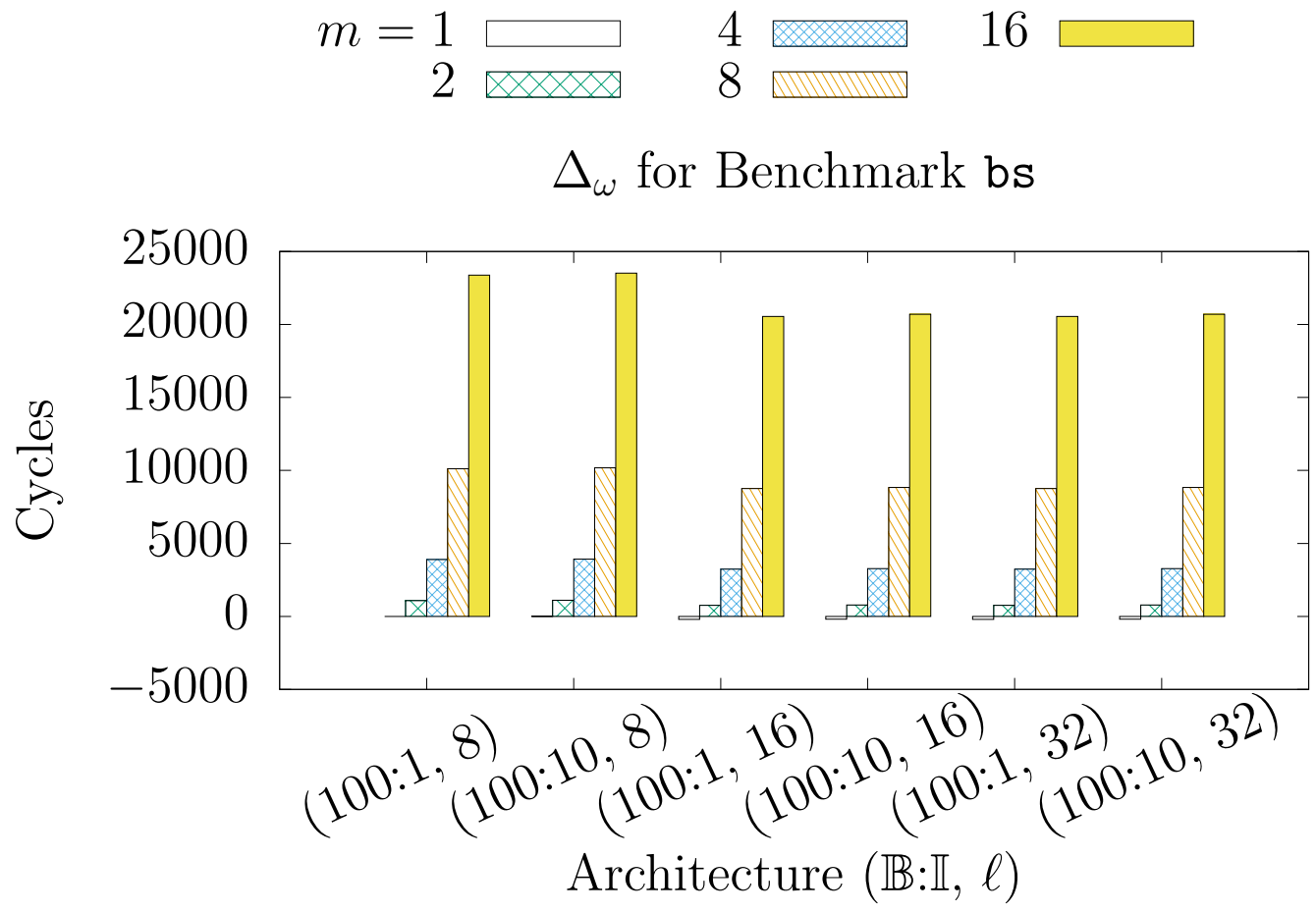

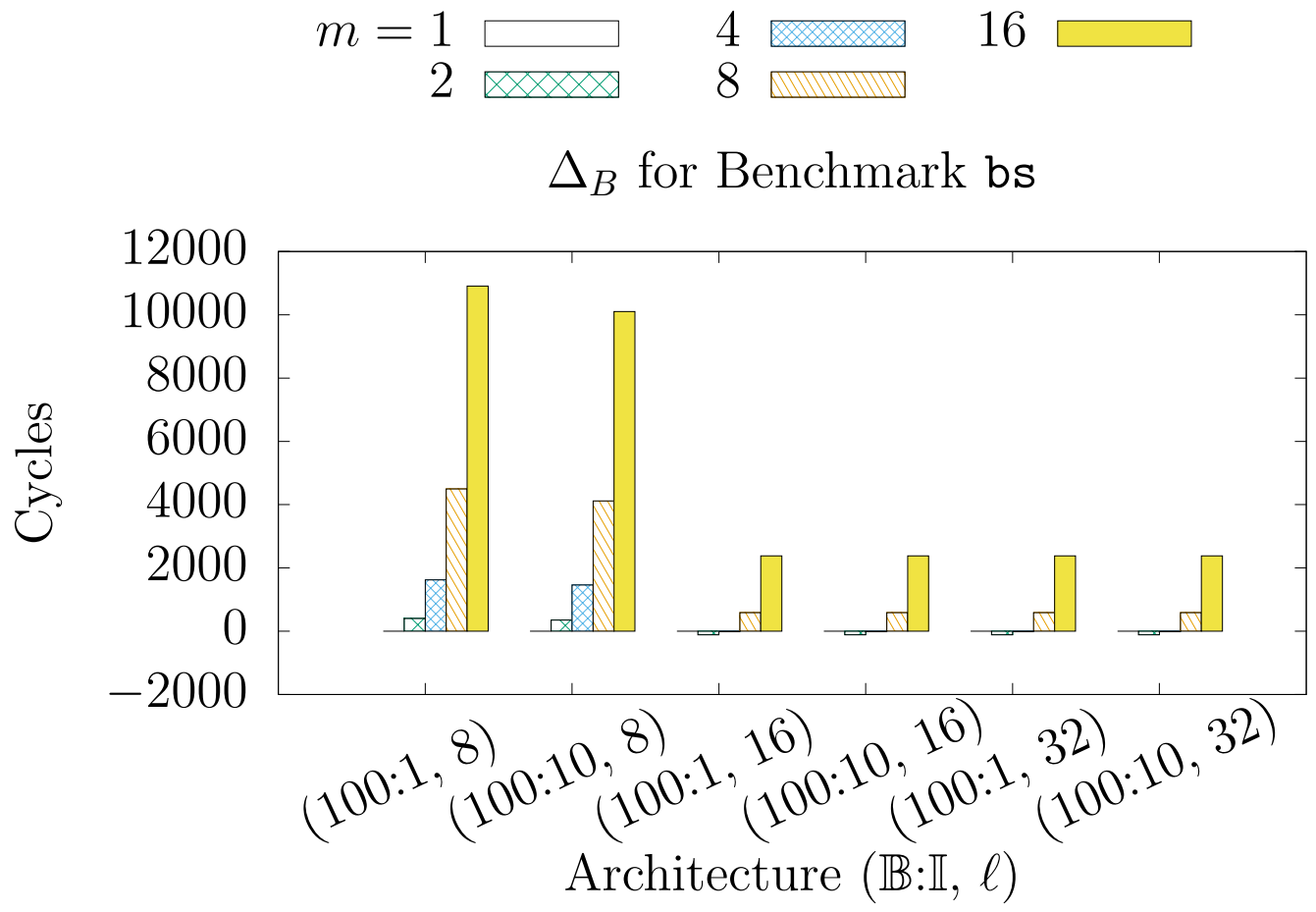

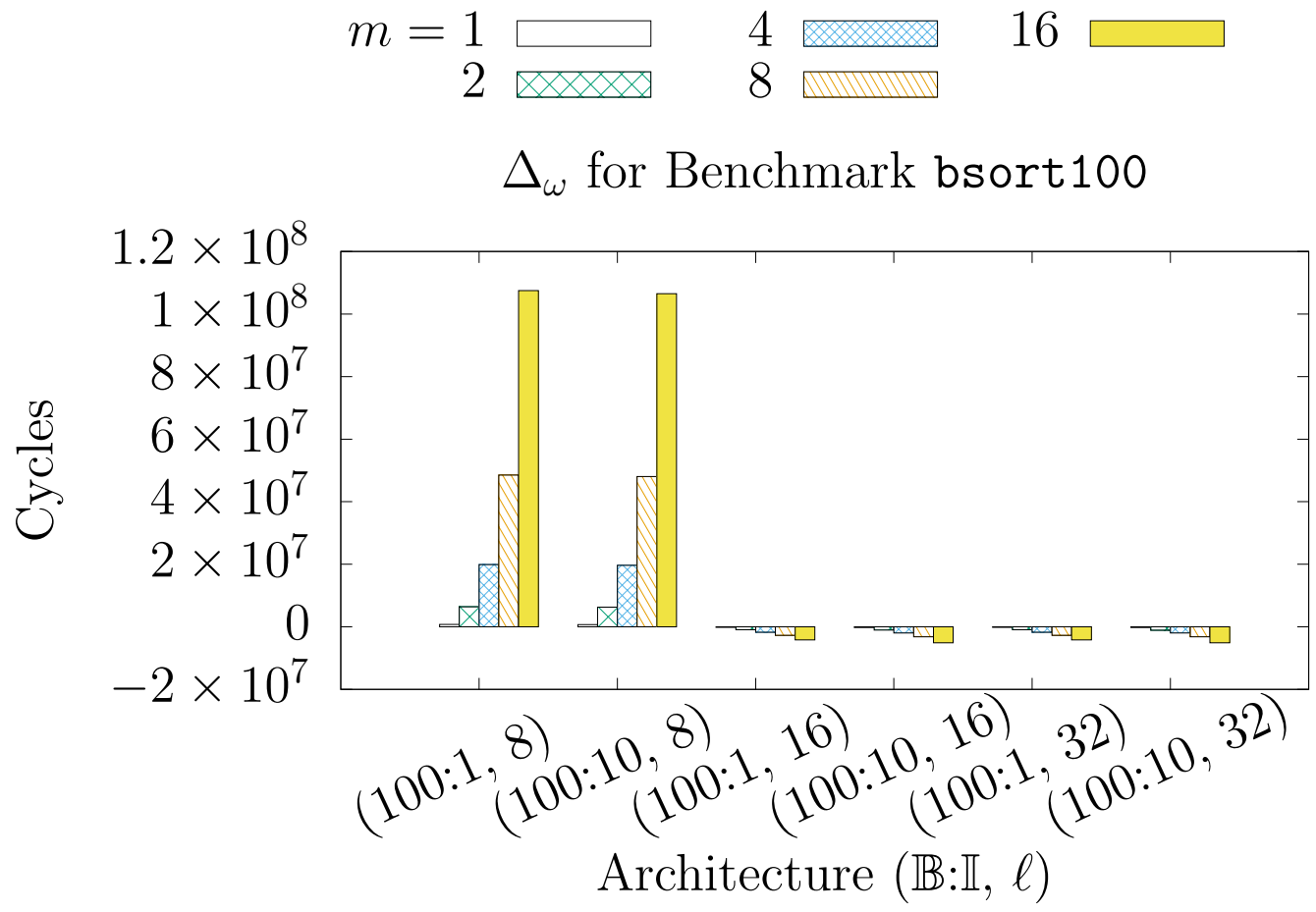

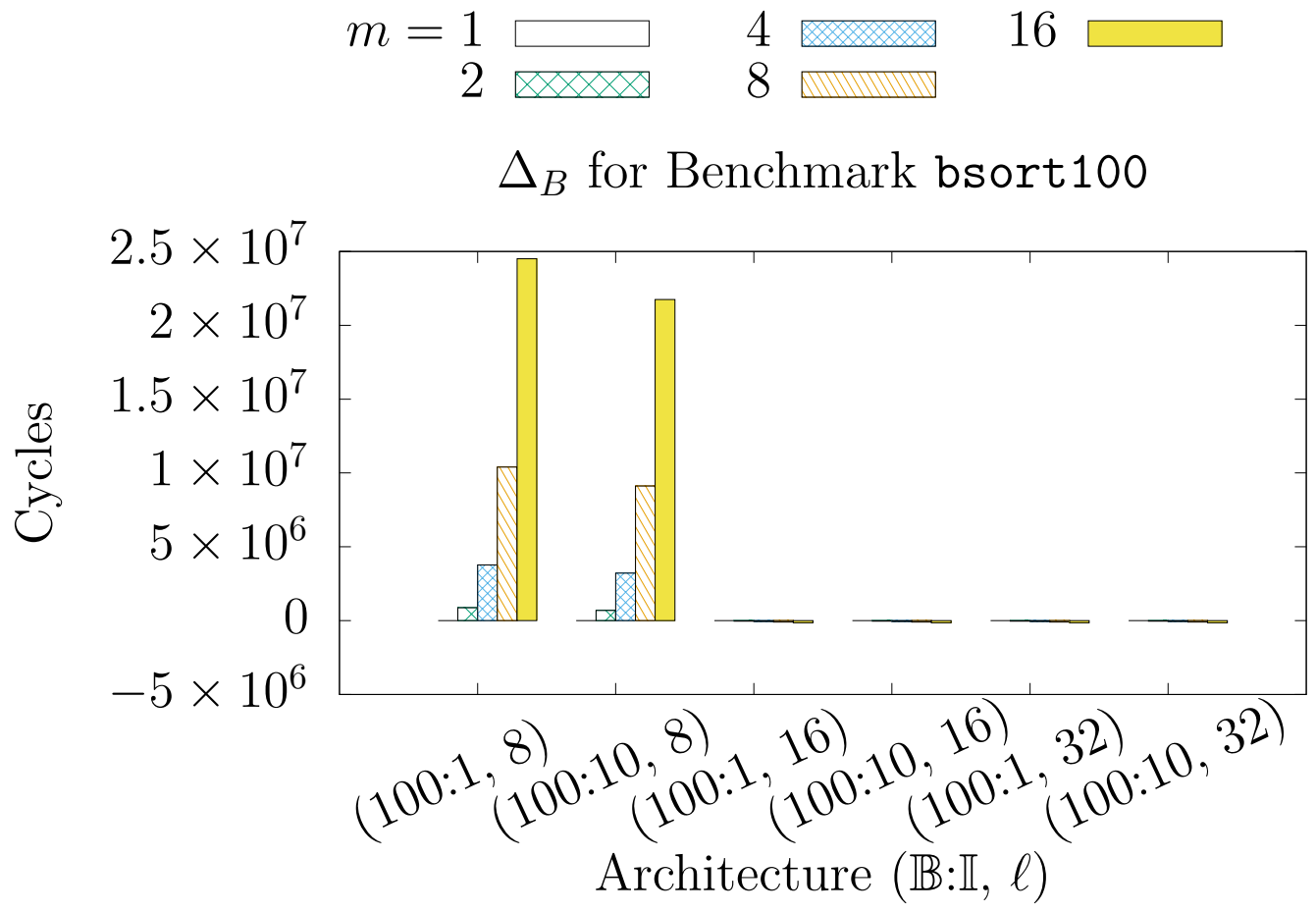

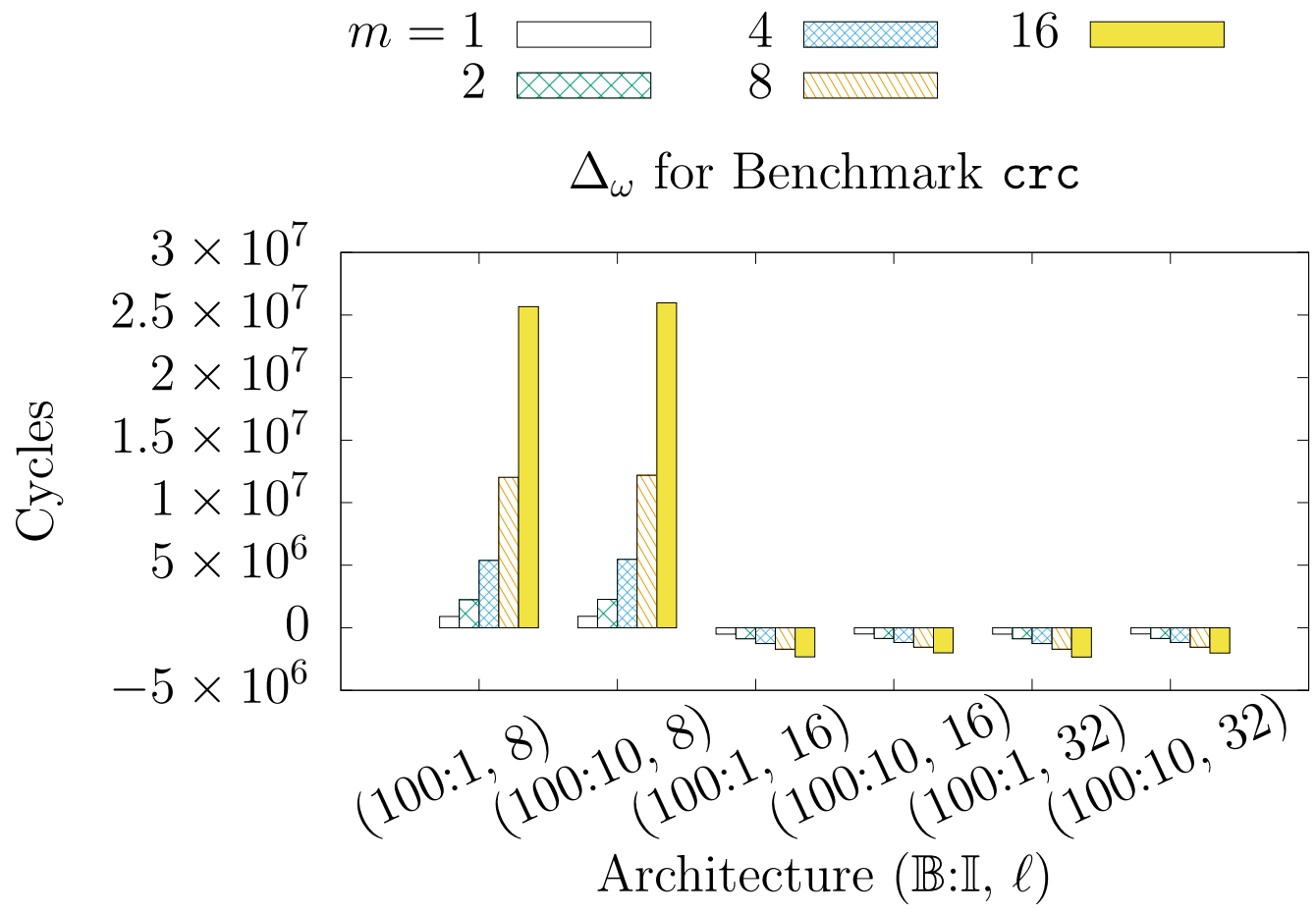

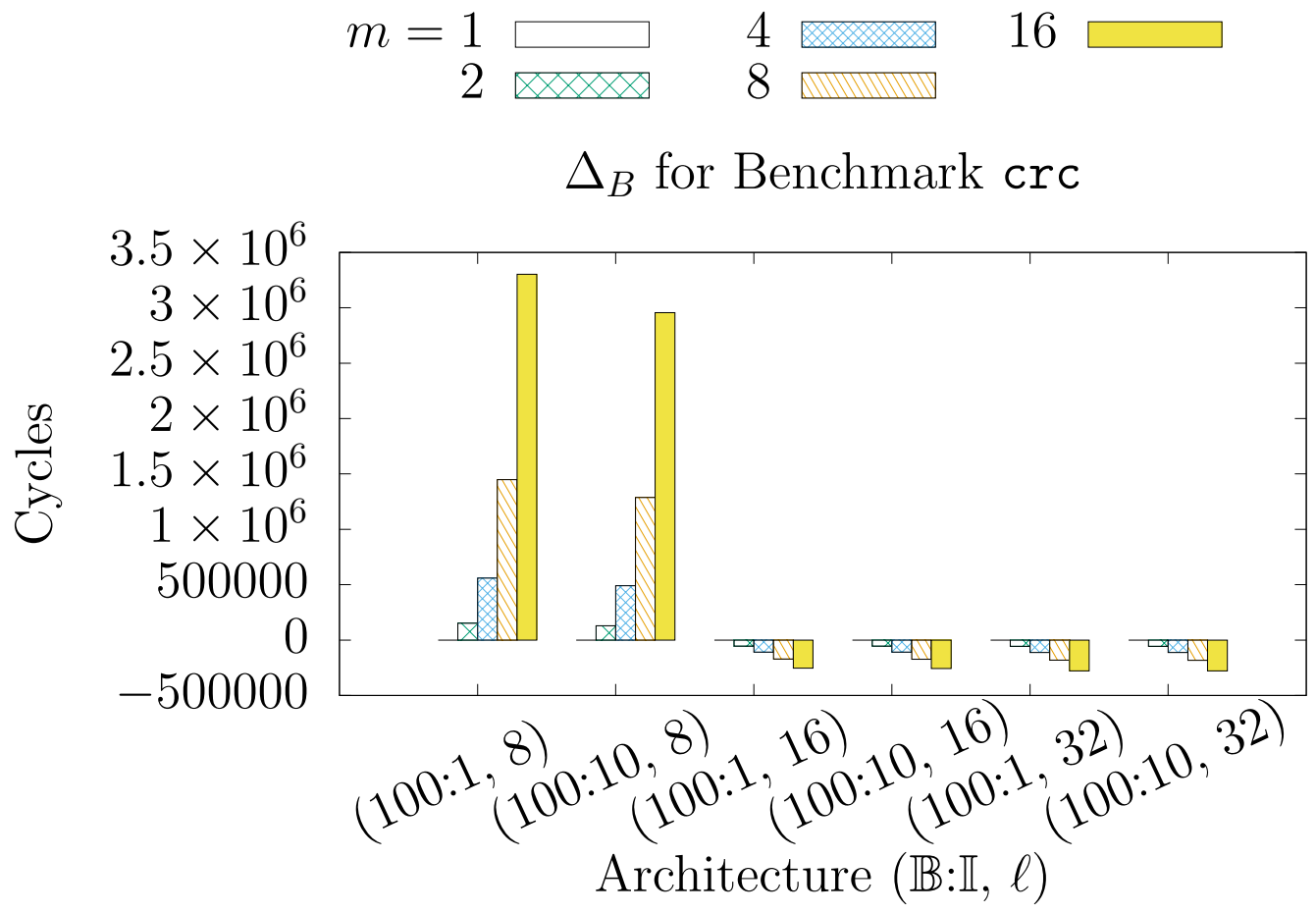

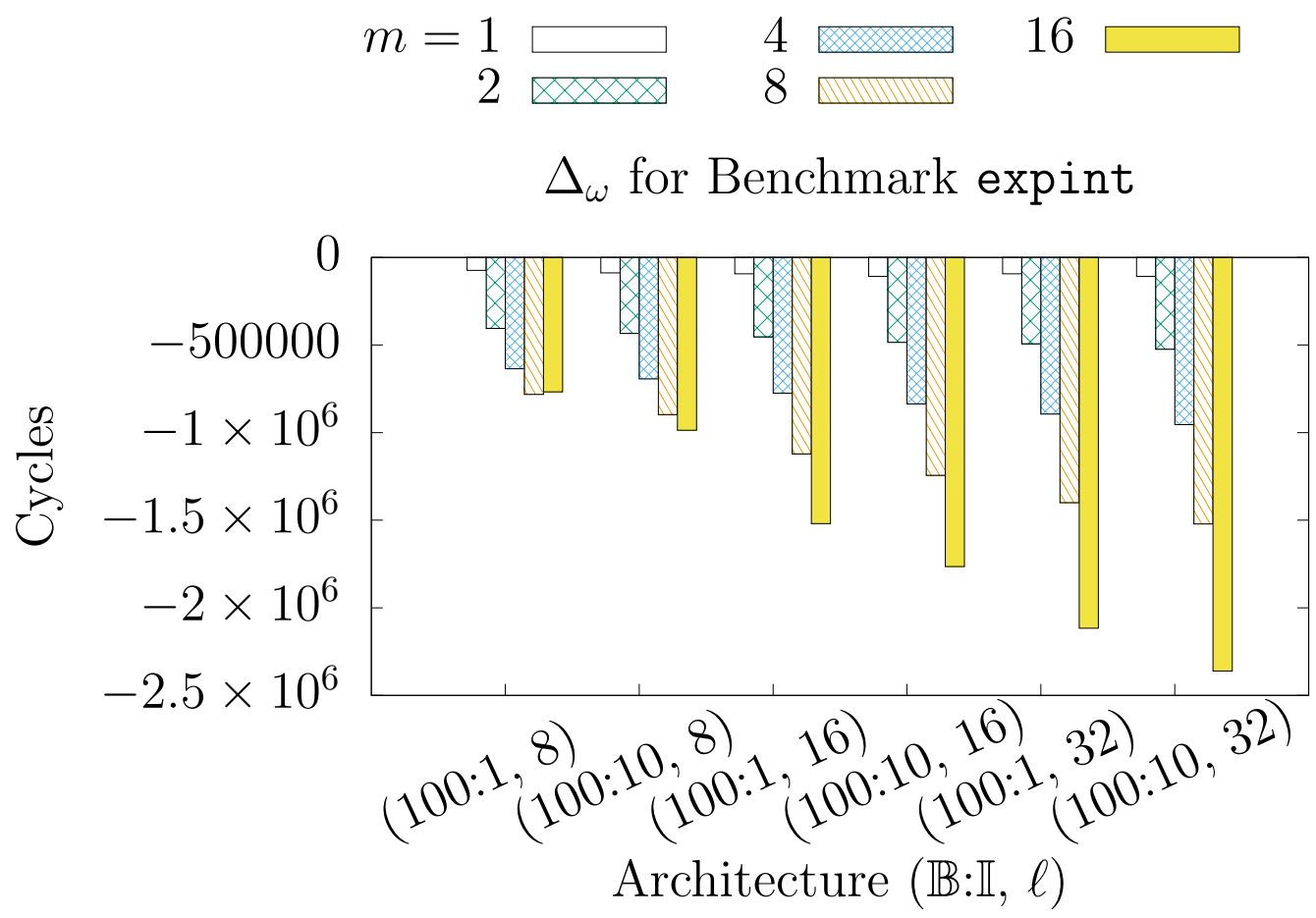
$m = 1$ 2 4 8 16
$\Delta_\omega$ for Benchmark expint
Cycles
0
−500000
$-1 \times 10^6$
$-1.5 \times 10^6$
$-2 \times 10^6$
$-2.5 \times 10^6$
(100:1, 8) (100:10, 8) (100:1, 16) (100:10, 16) (100:1, 32) (100:10, 32)
Architecture ($\mathbb{B}$:$\mathbb{I}$, $\ell$)

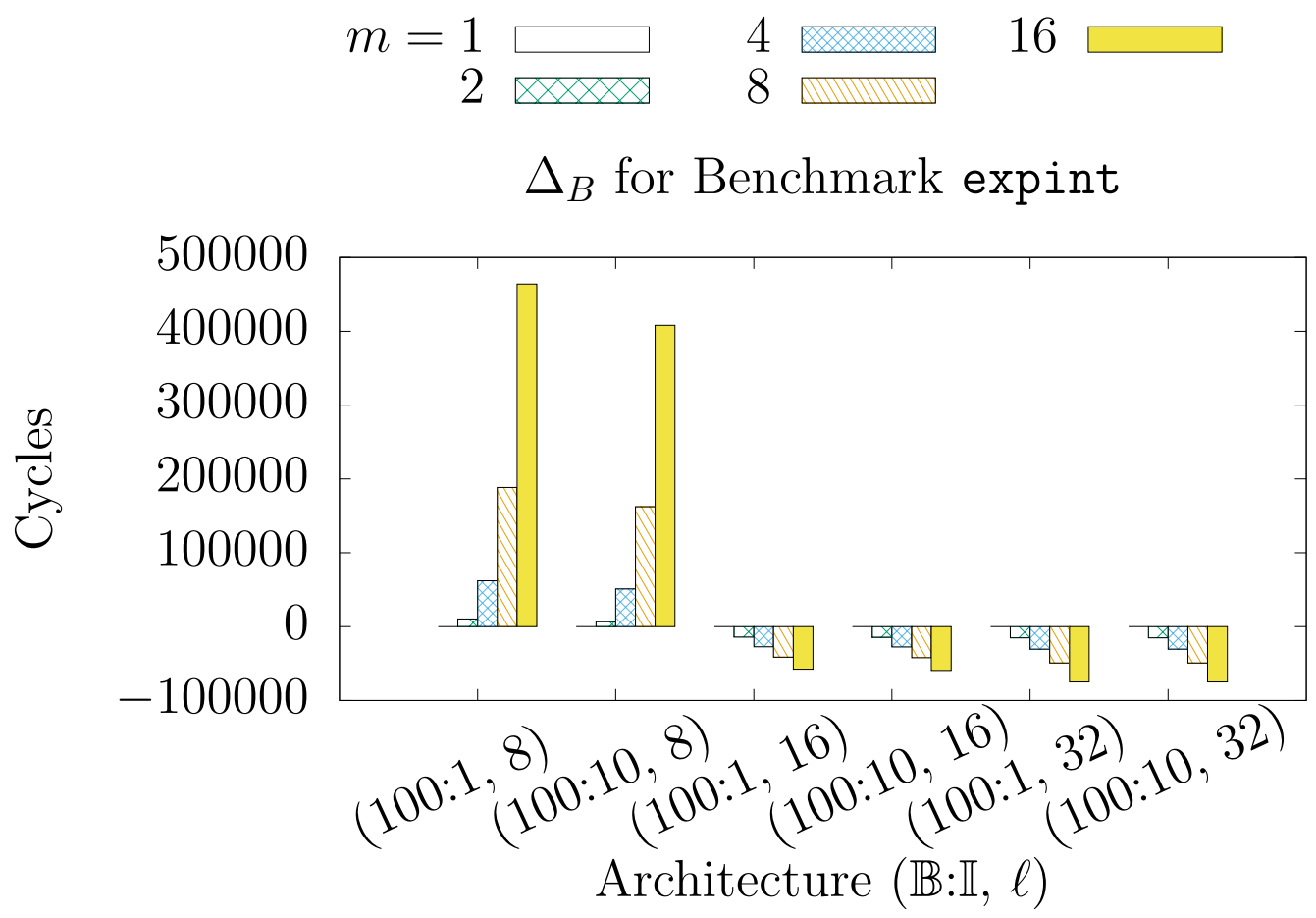
$m = 1$ 2 4 8 16
$\Delta_B$ for Benchmark expint
Cycles
500000
400000
300000
200000
100000
0
−100000
(100:1, 8) (100:10, 8) (100:1, 16) (100:10, 16) (100:1, 32) (100:10, 32)
Architecture ($\mathbb{B}$:$\mathbb{I}$, $\ell$)

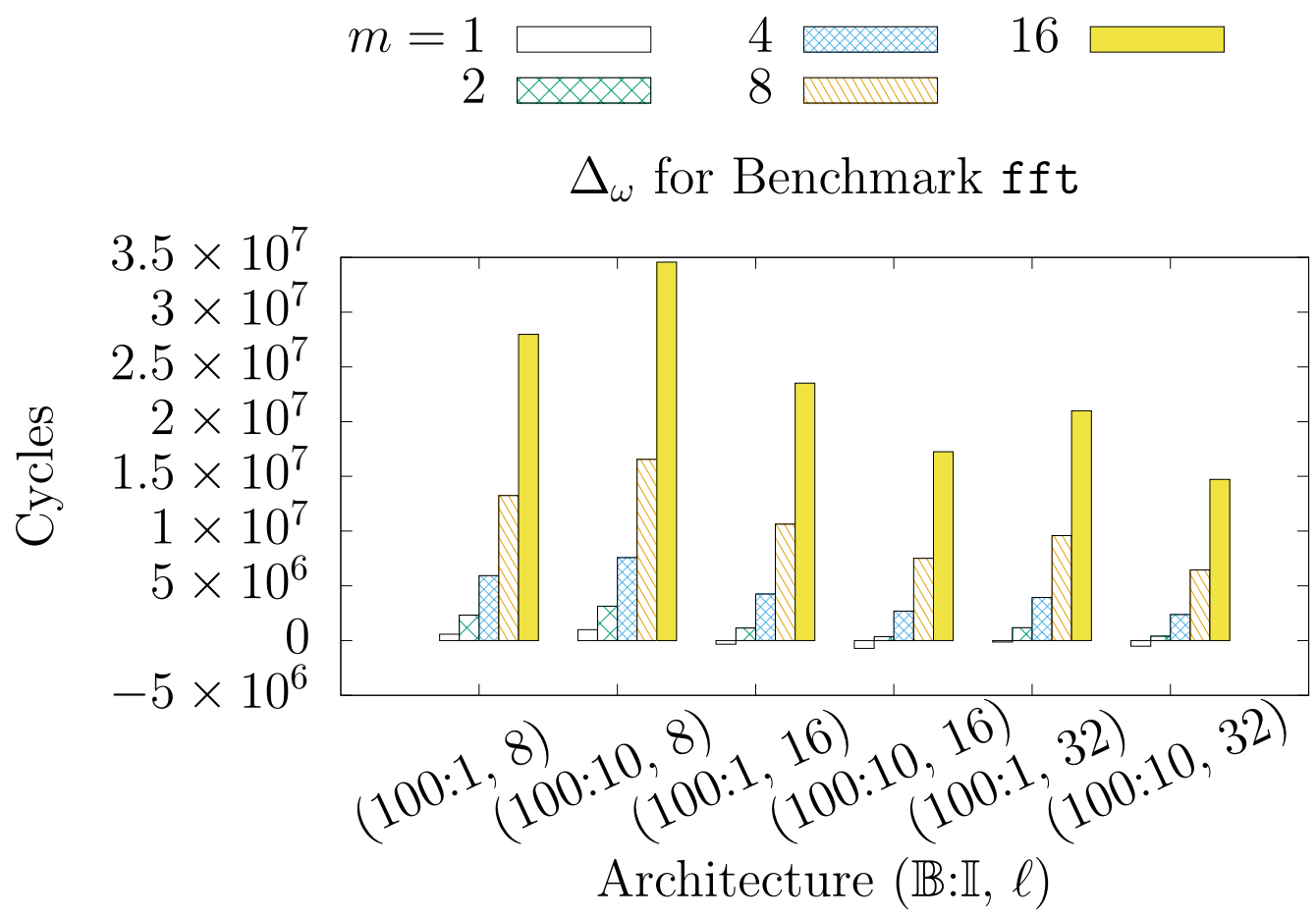
$m = 1$ 2 4 8 16
$\Delta_\omega$ for Benchmark fft
Cycles
$3.5 \times 10^7$
$3 \times 10^7$
$2.5 \times 10^7$
$2 \times 10^7$
$1.5 \times 10^7$
$1 \times 10^7$
$5 \times 10^6$
0
$-5 \times 10^6$
(100:1, 8) (100:10, 8) (100:1, 16) (100:10, 16) (100:1, 32) (100:10, 32)
Architecture ($\mathbb{B}$:$\mathbb{I}$, $\ell$)

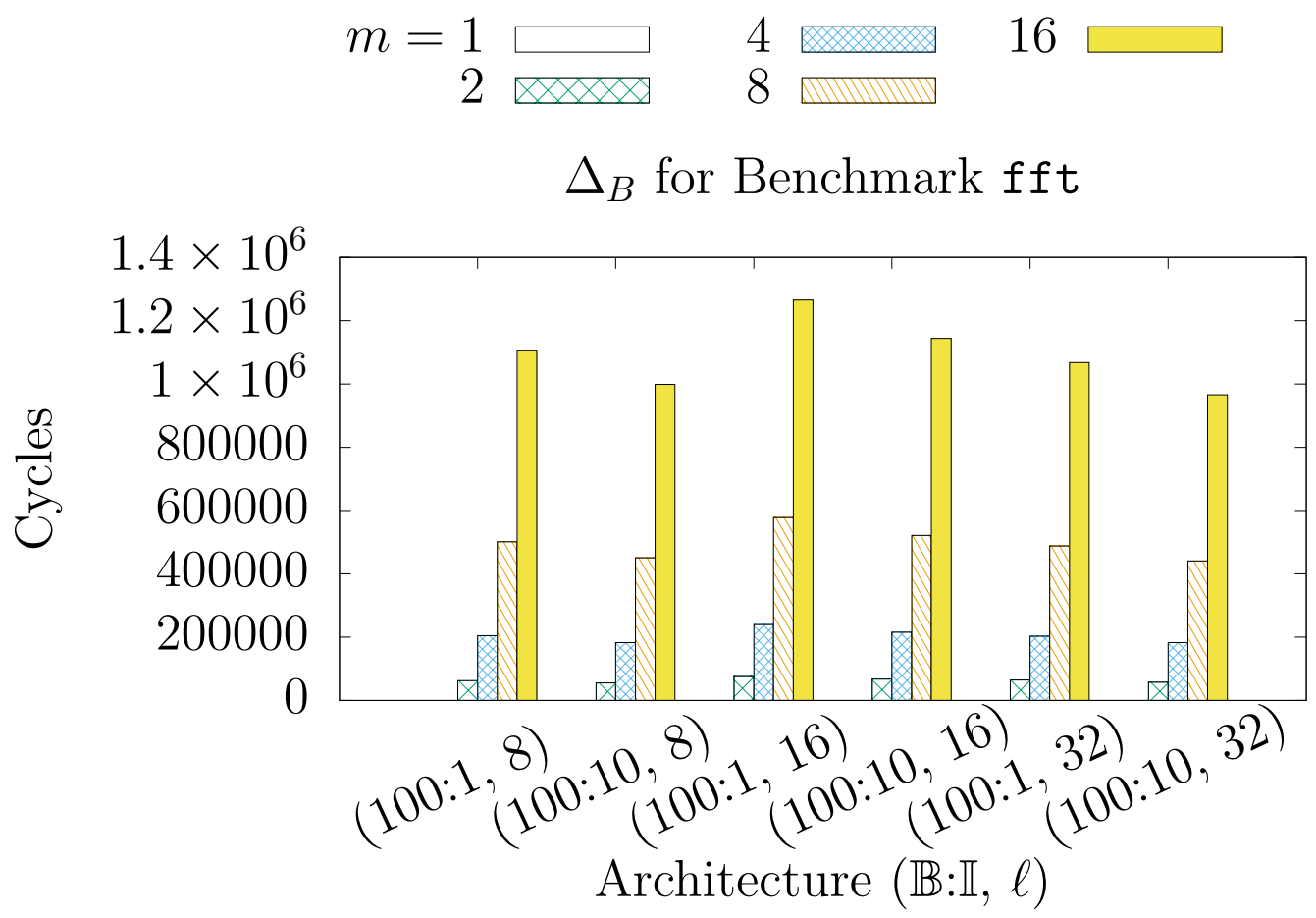
$m = 1$ 2 4 8 16
$\Delta_B$ for Benchmark fft
Cycles
$1.4 \times 10^6$
$1.2 \times 10^6$
$1 \times 10^6$
800000
600000
400000
200000
0
(100:1, 8) (100:10, 8) (100:1, 16) (100:10, 16) (100:1, 32) (100:10, 32)
Architecture ($\mathbb{B}$:$\mathbb{I}$, $\ell$)

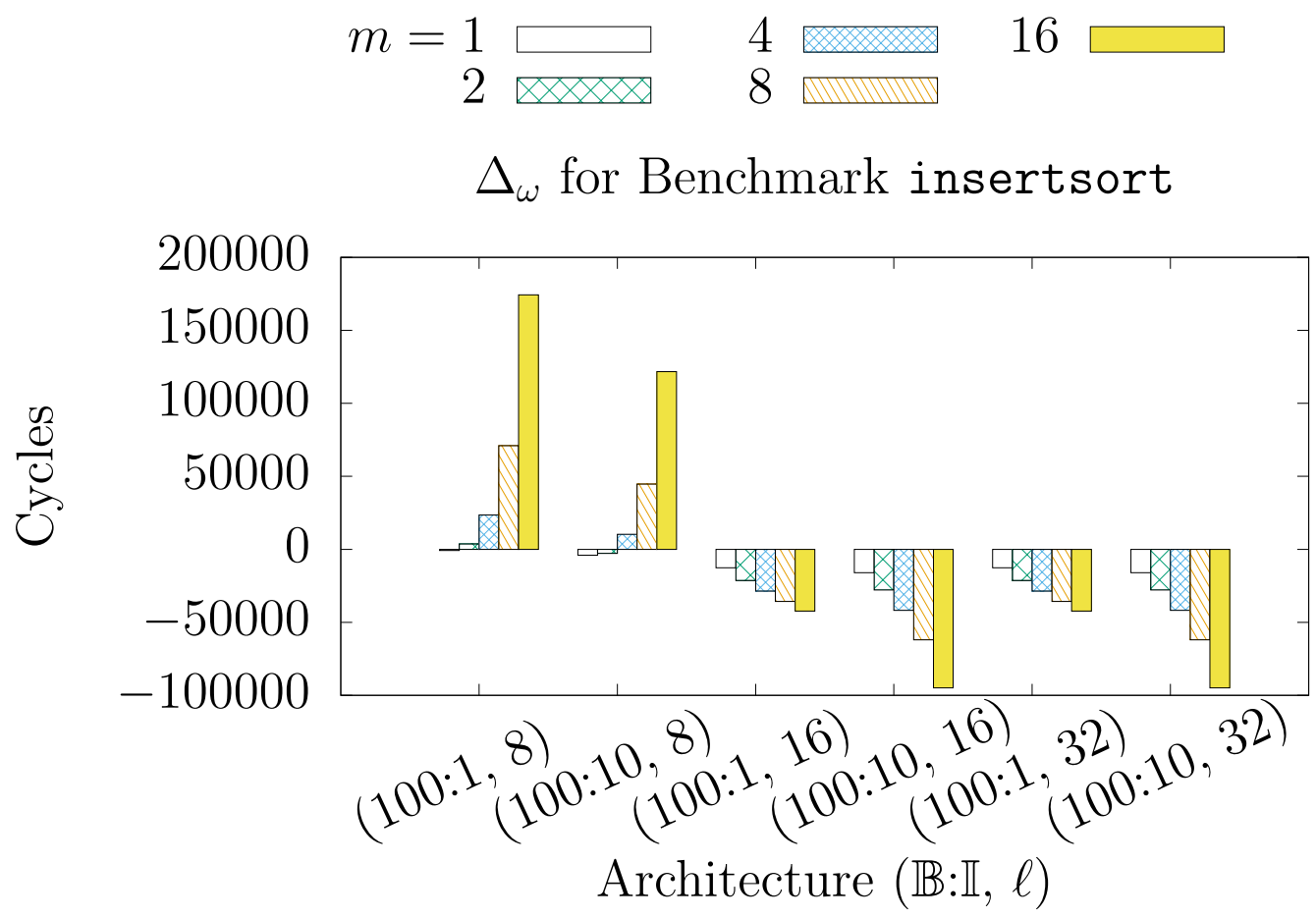
$m = 1$ 2 4 8 16
$\Delta_\omega$ for Benchmark insertsort
Cycles
200000
150000
100000
50000
0
−50000
−100000
(100:1, 8) (100:10, 8) (100:1, 16) (100:10, 16) (100:1, 32) (100:10, 32)
Architecture ($\mathbb{B}$:$\mathbb{I}$, $\ell$)

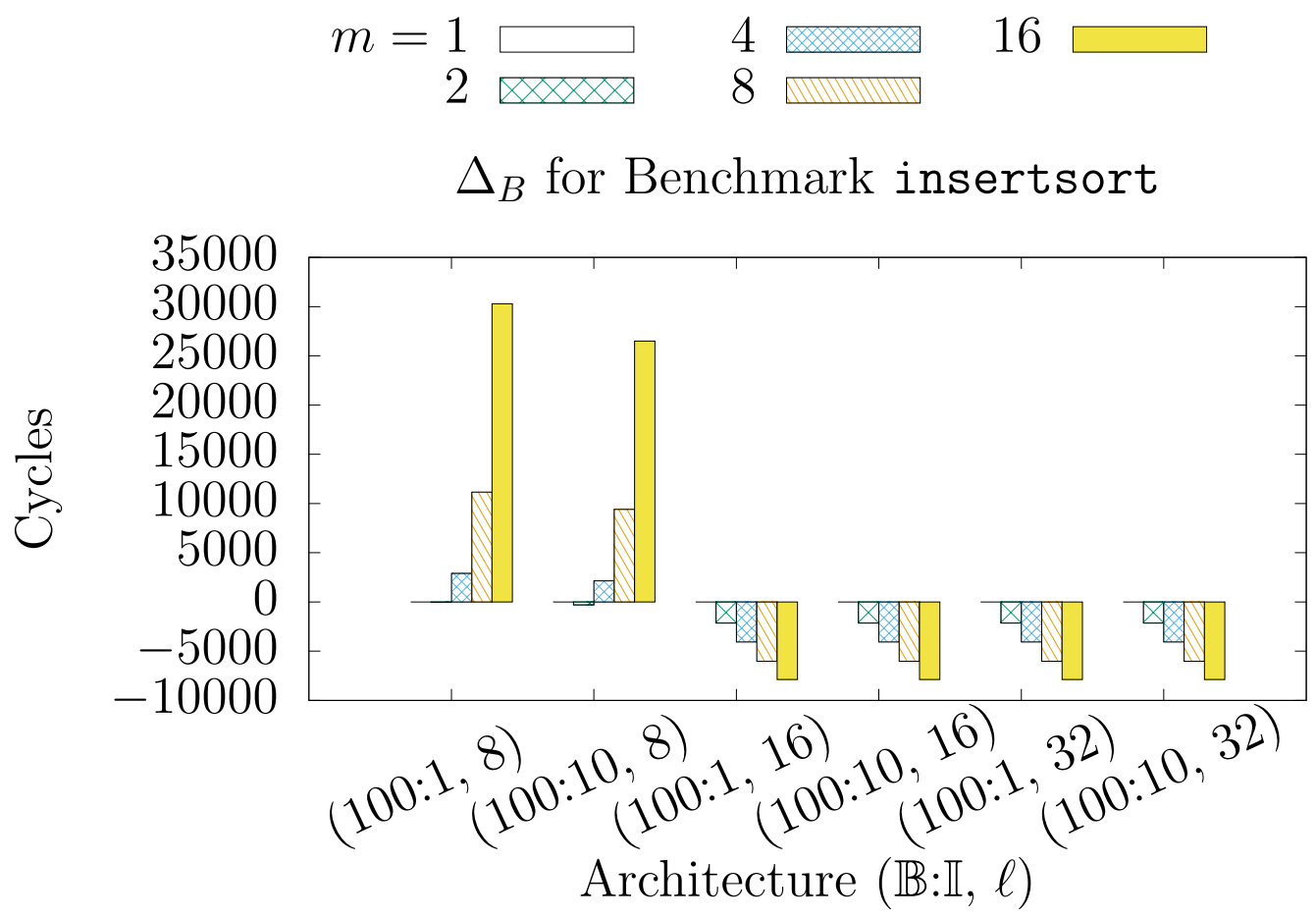
$m = 1$ 2 4 8 16
$\Delta_B$ for Benchmark insertsort
Cycles
35000
30000
25000
20000
15000
10000
5000
0
−5000
−10000
(100:1, 8) (100:10, 8) (100:1, 16) (100:10, 16) (100:1, 32) (100:10, 32)
Architecture ($\mathbb{B}$:$\mathbb{I}$, $\ell$)

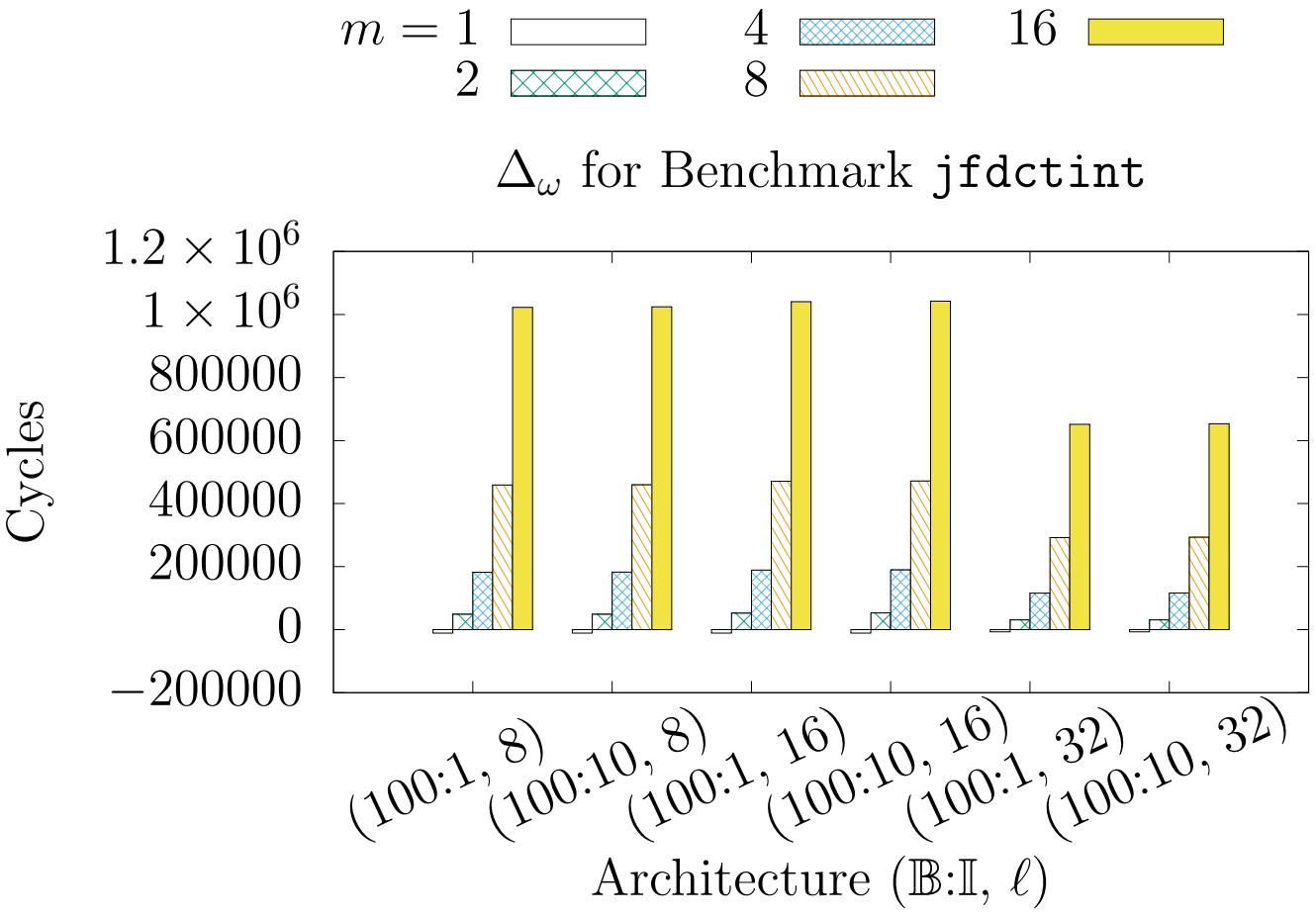

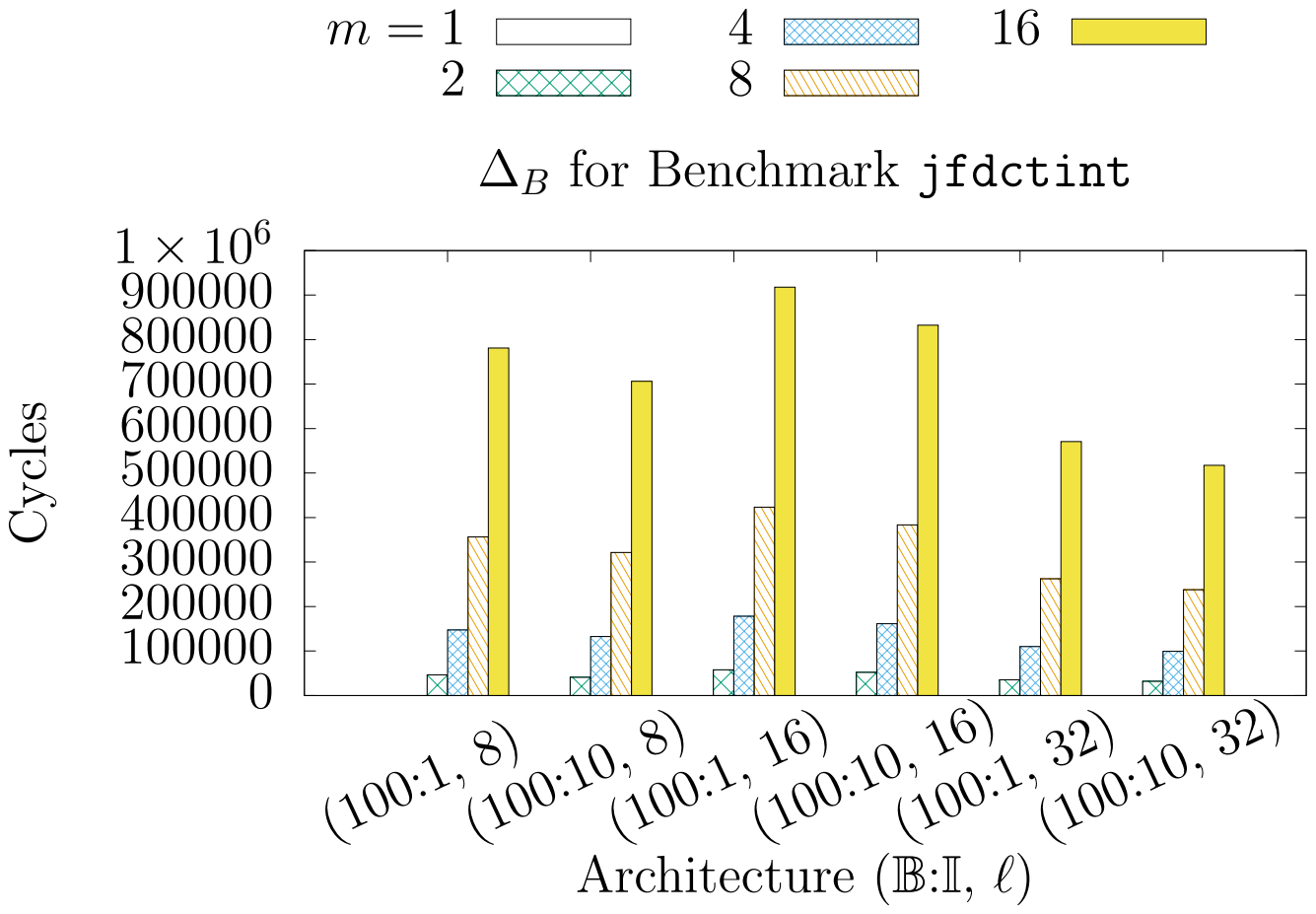

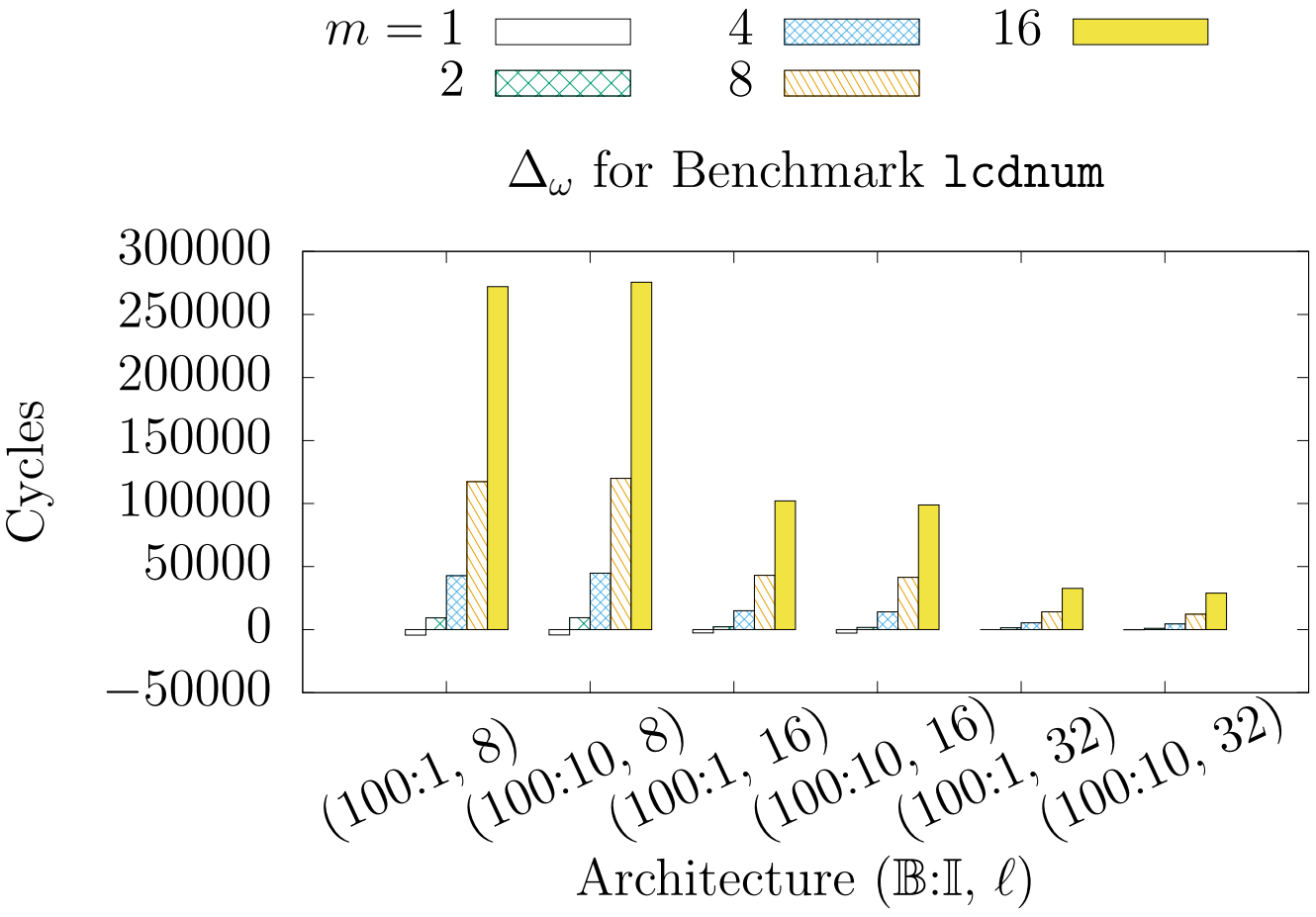

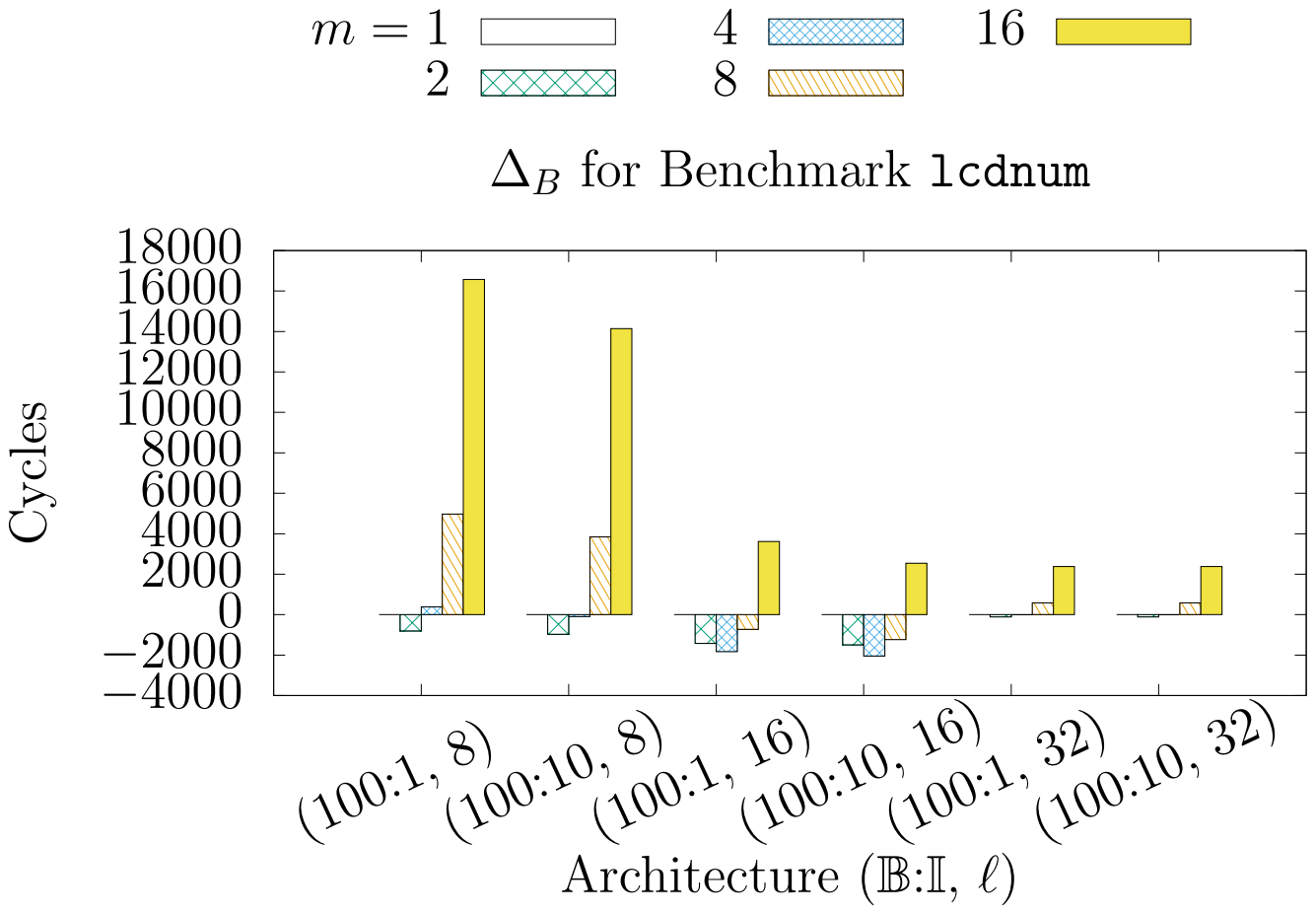

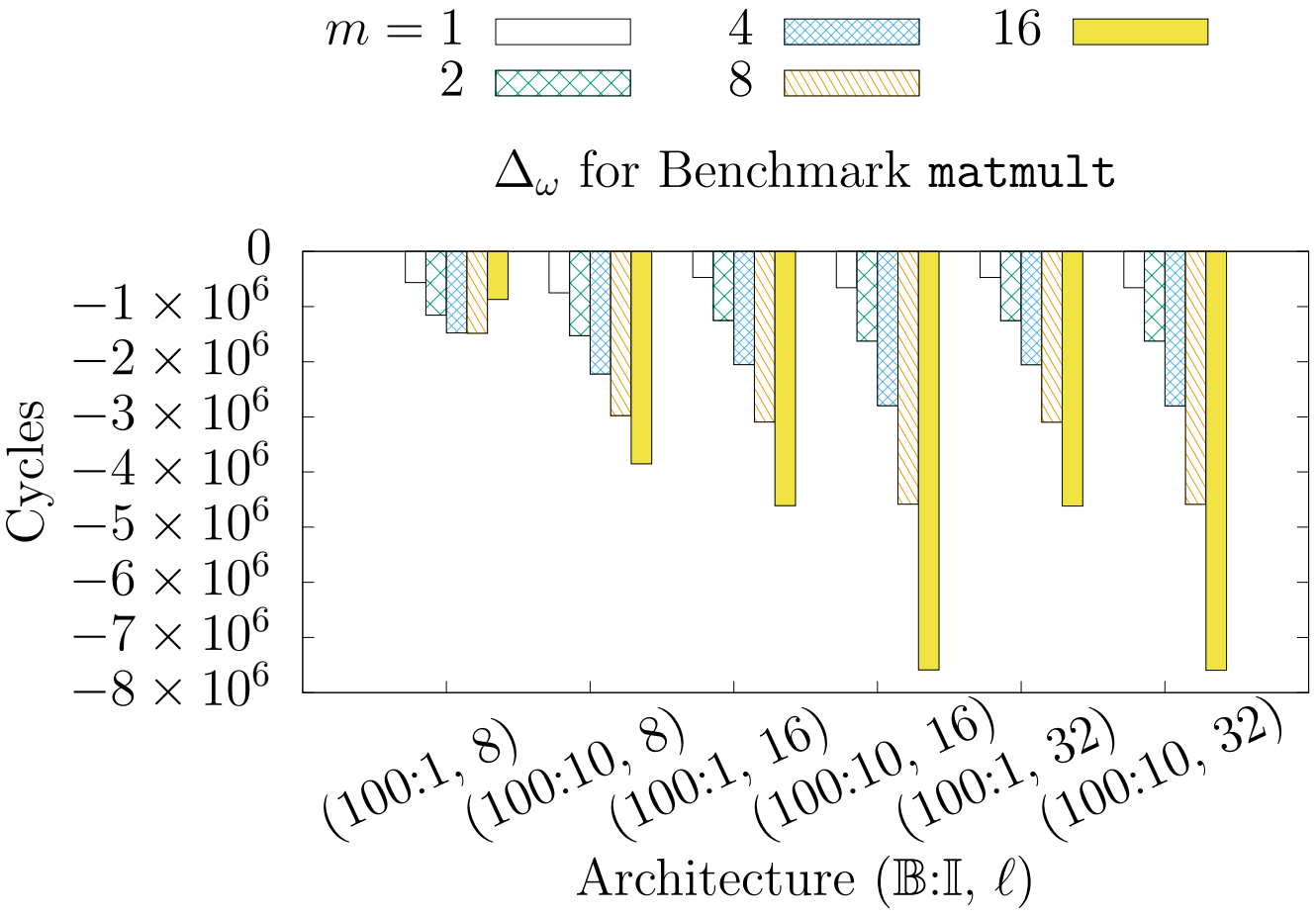

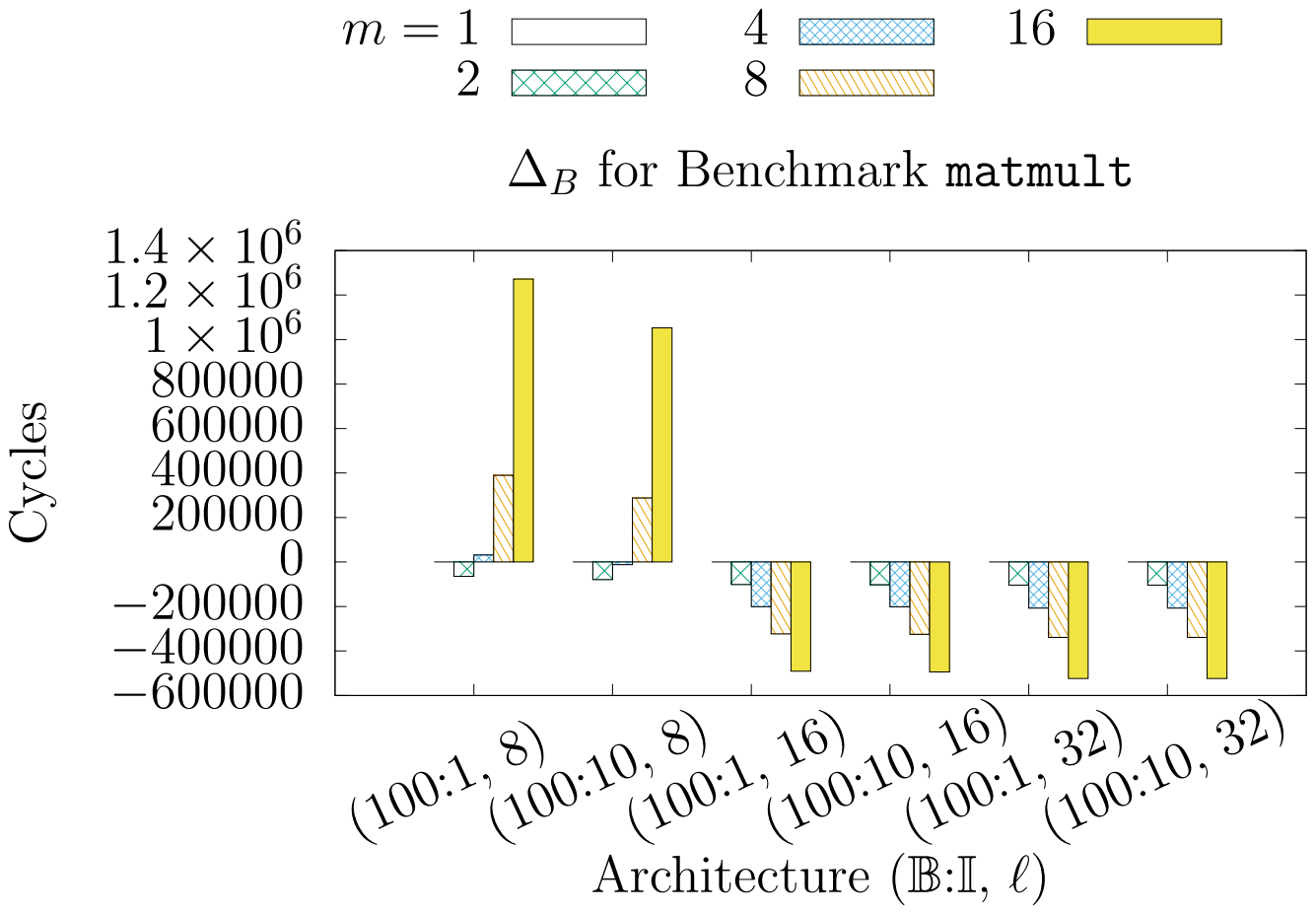

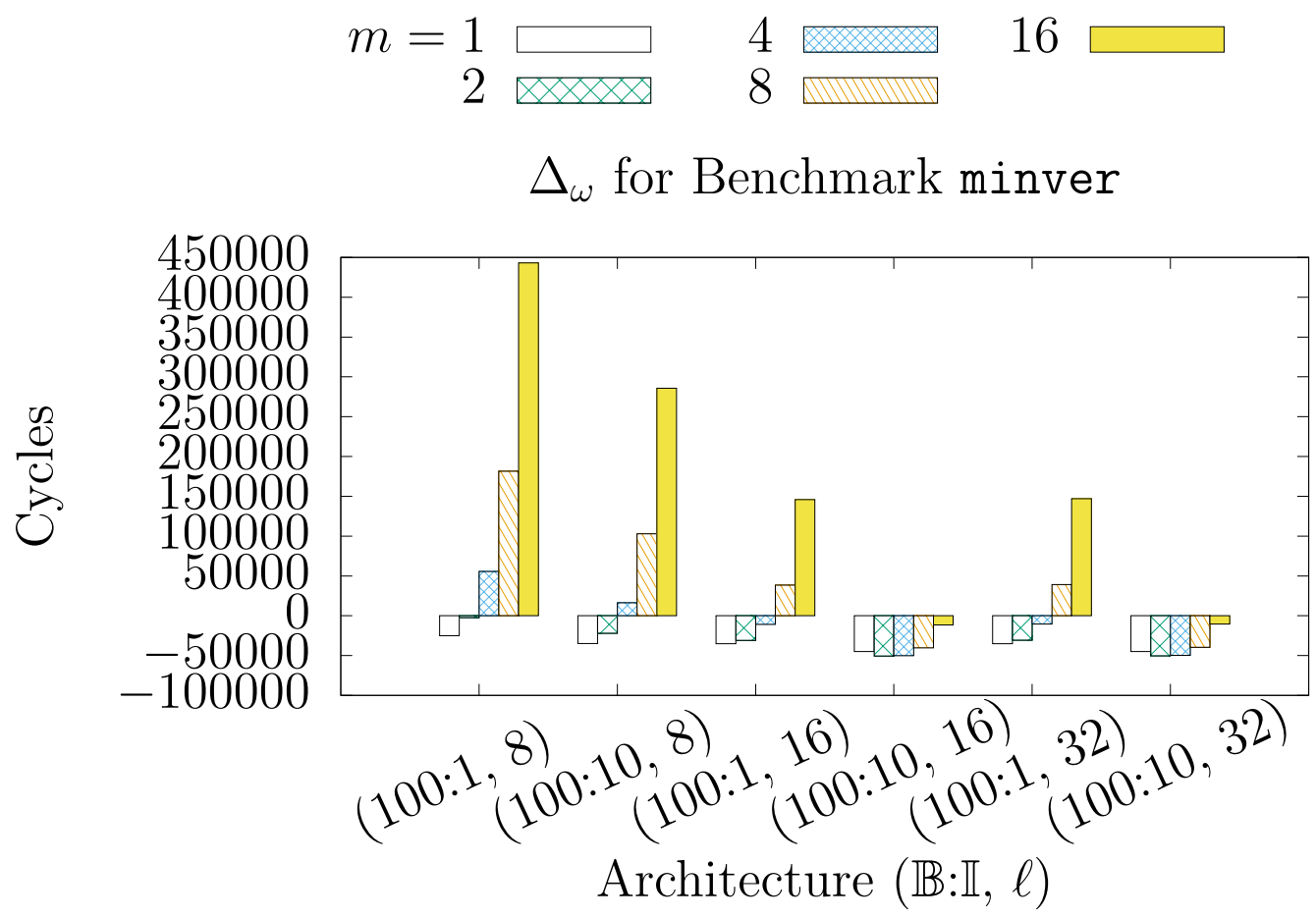

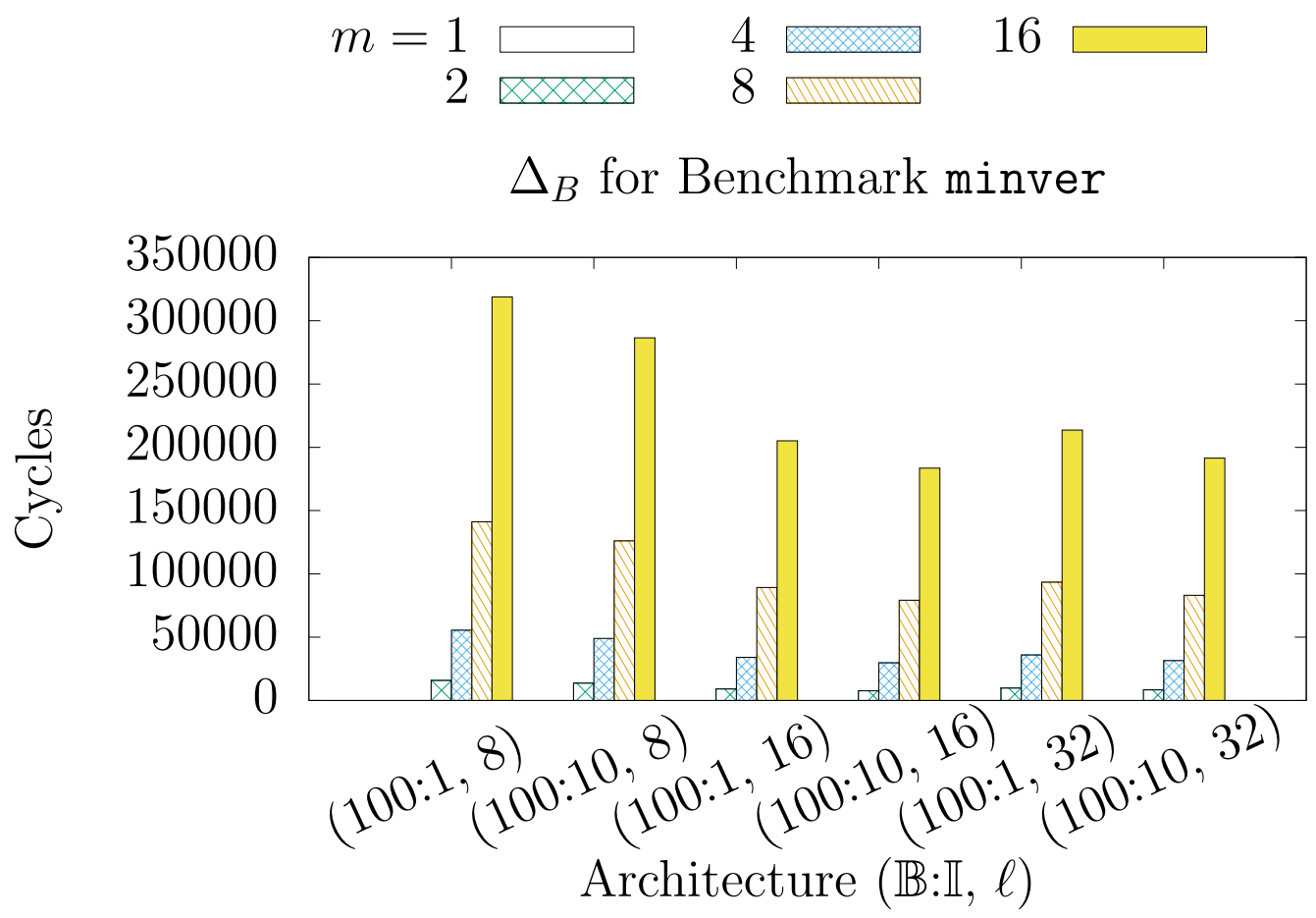

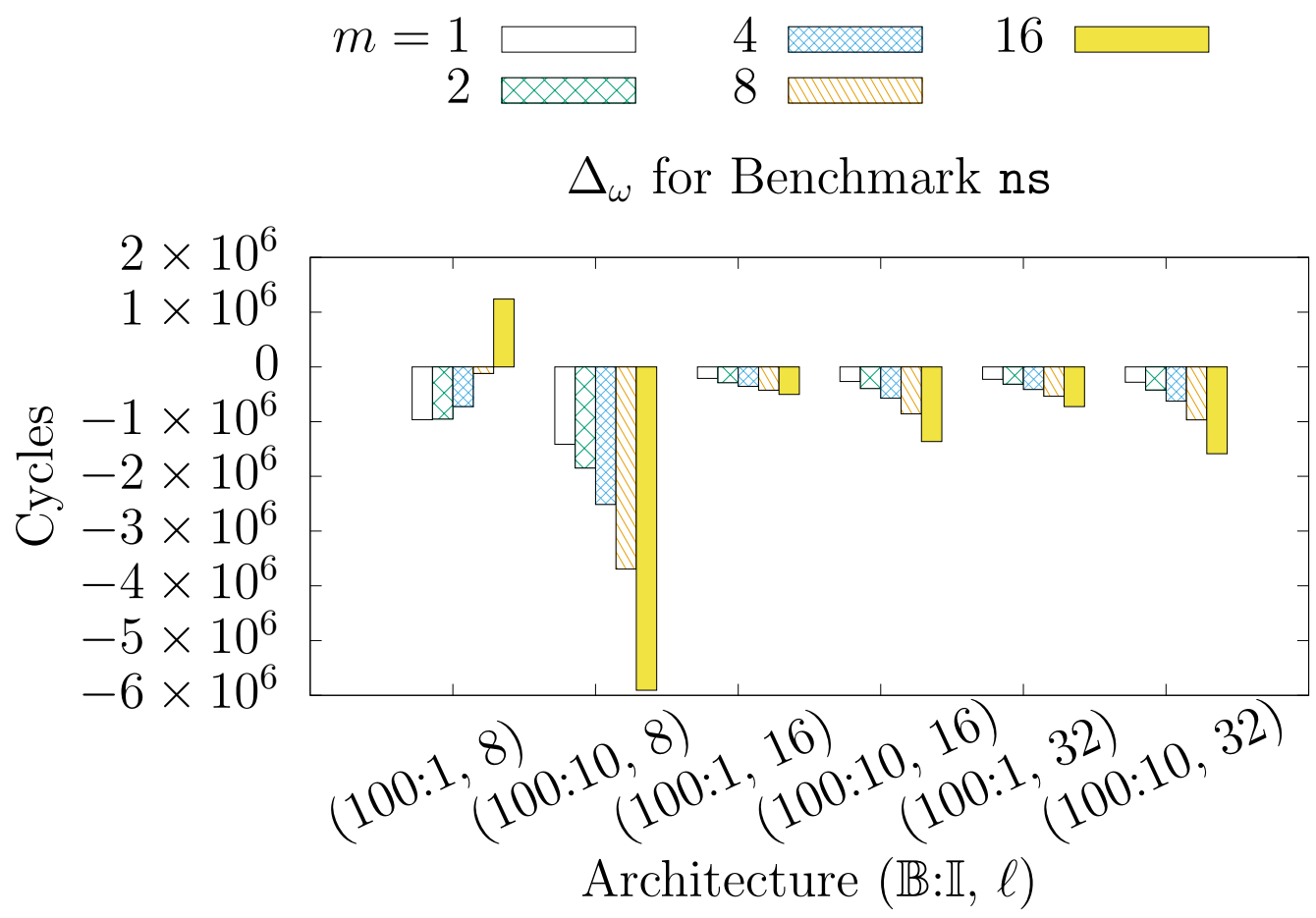

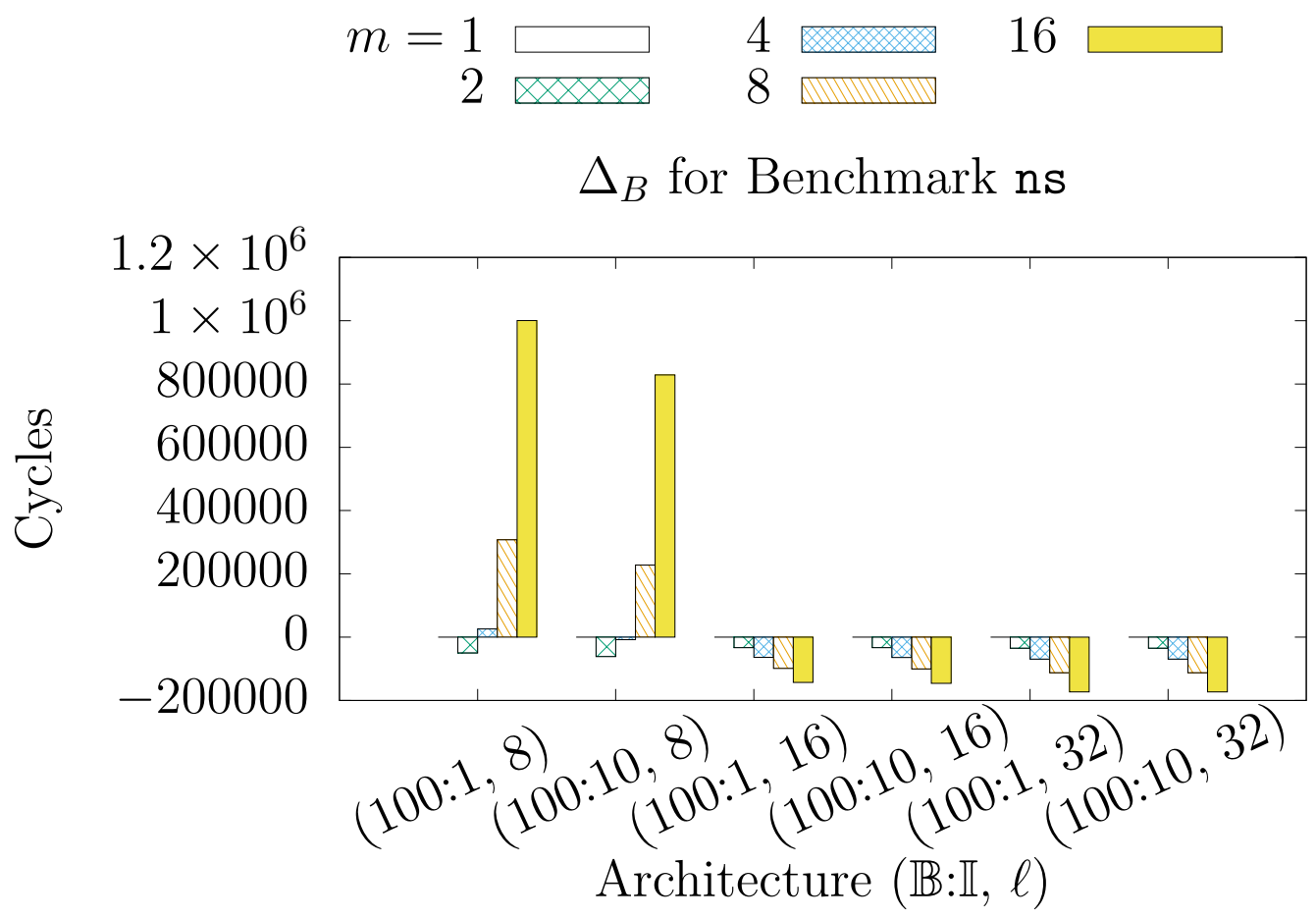

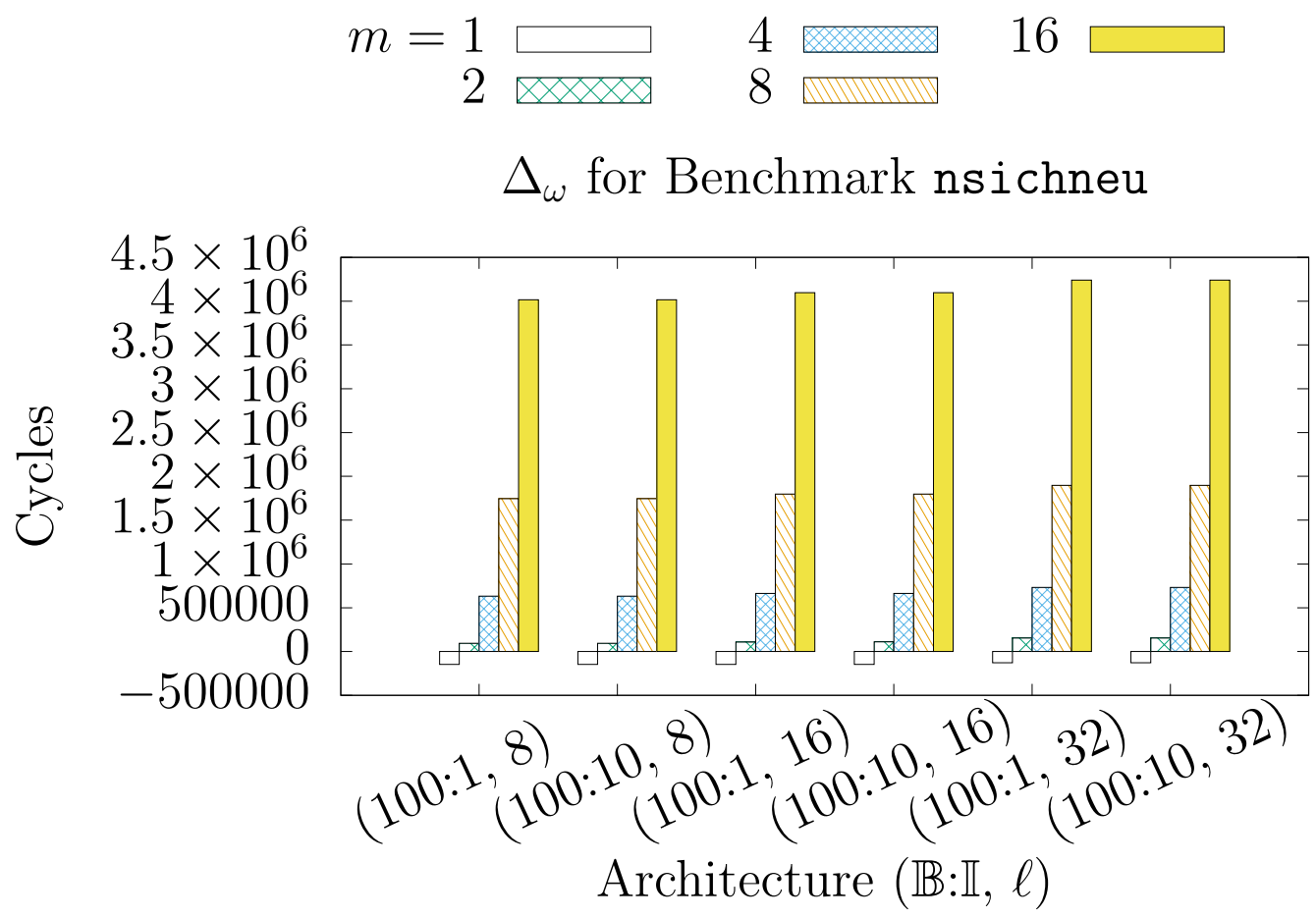

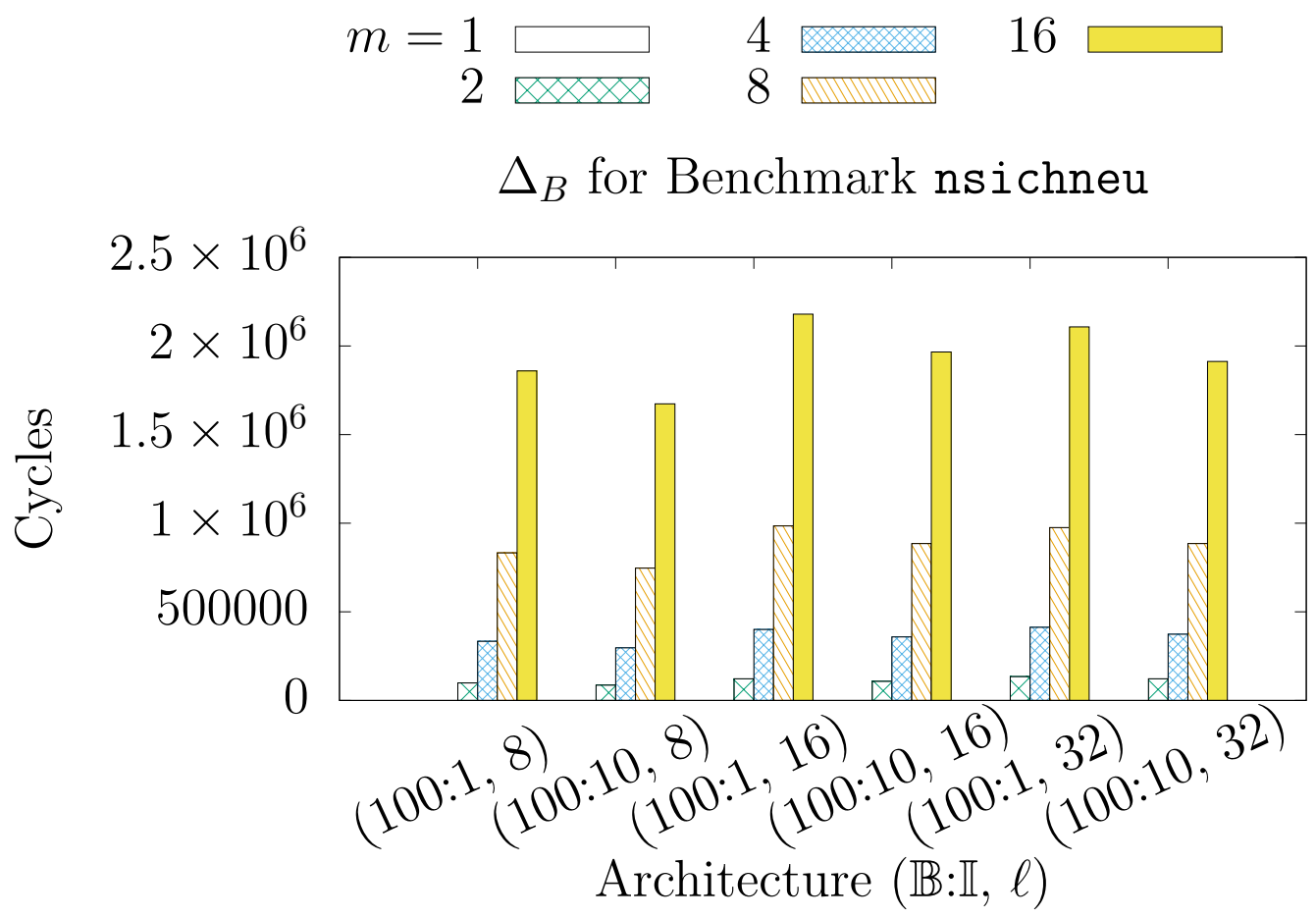

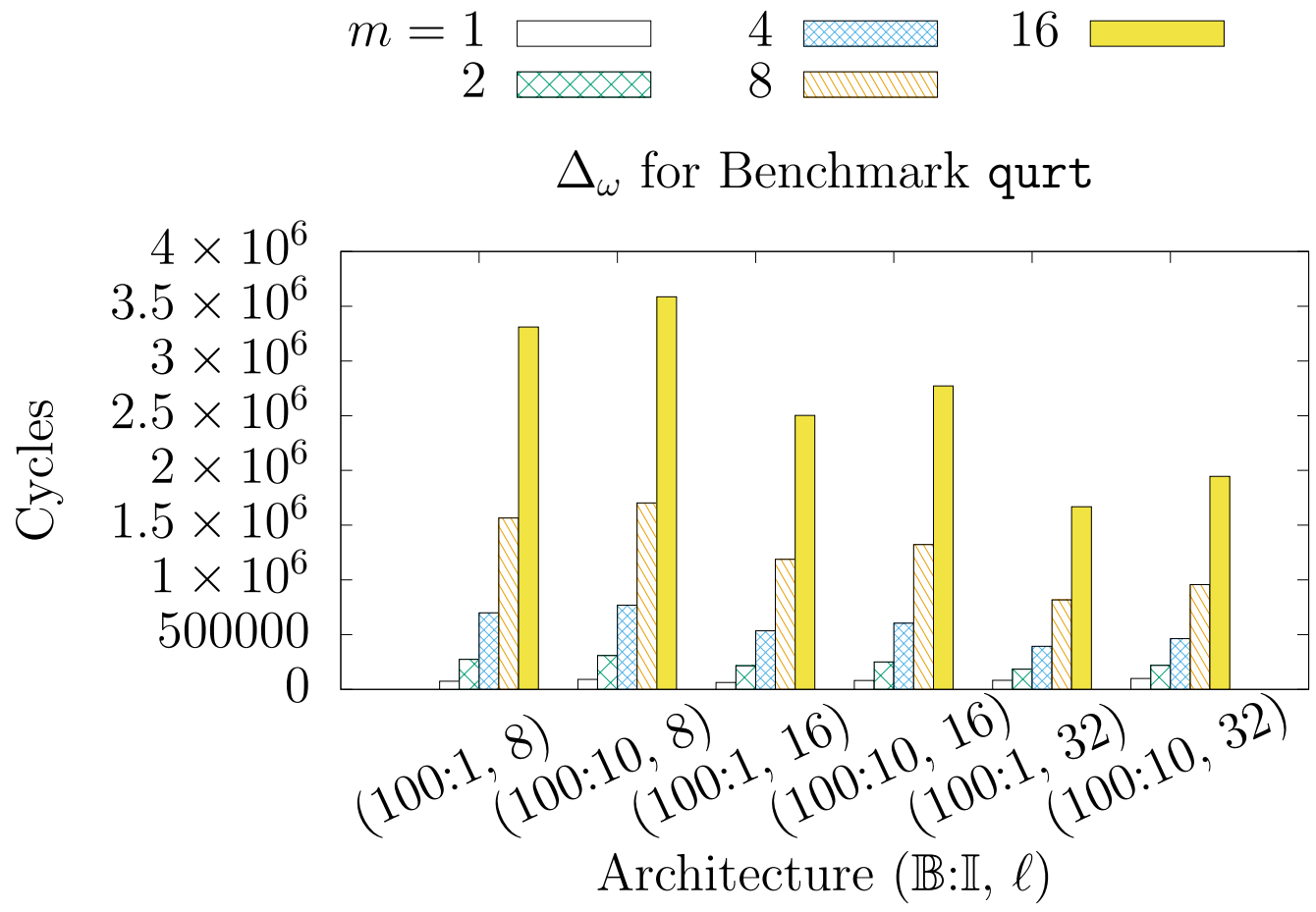

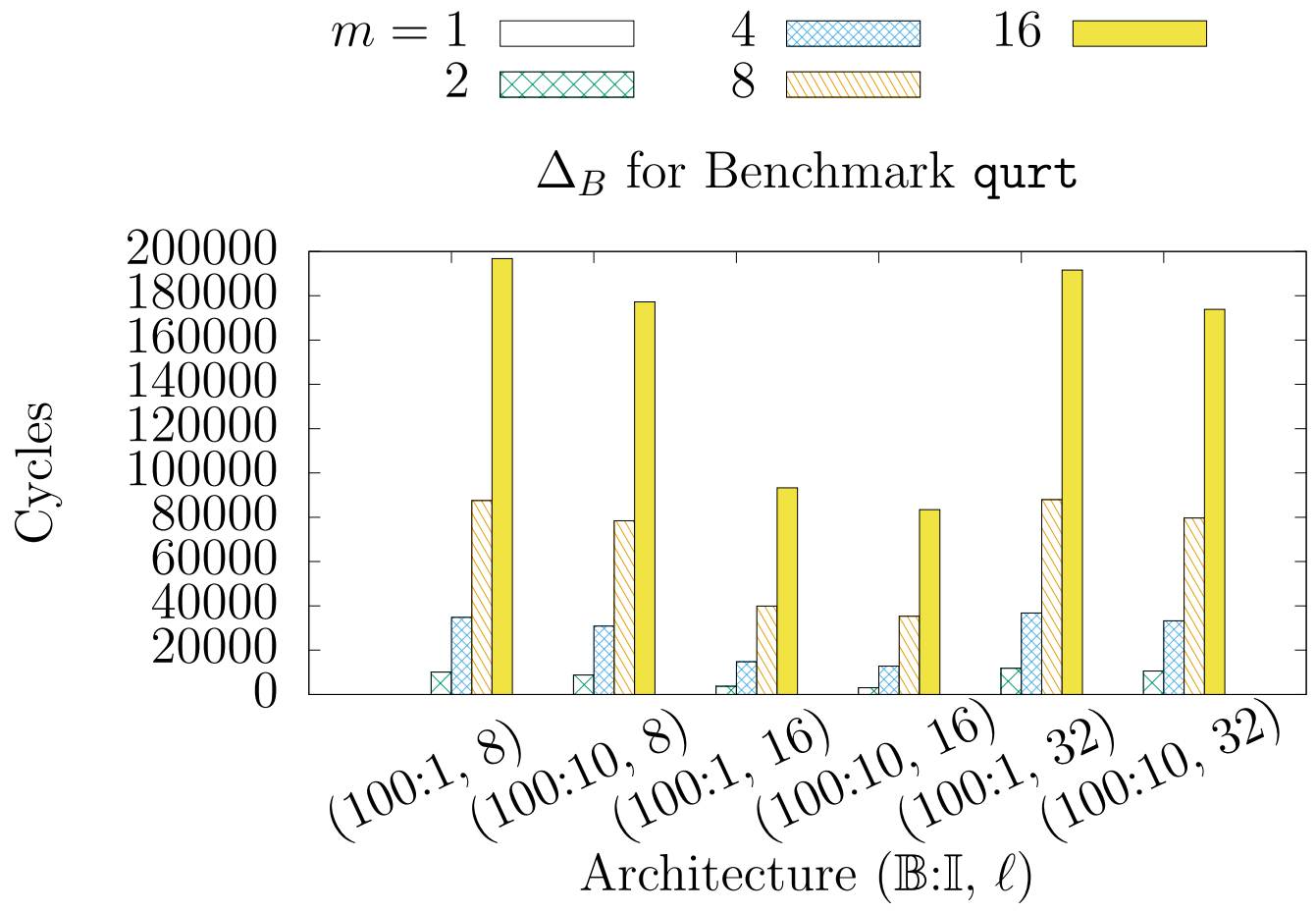

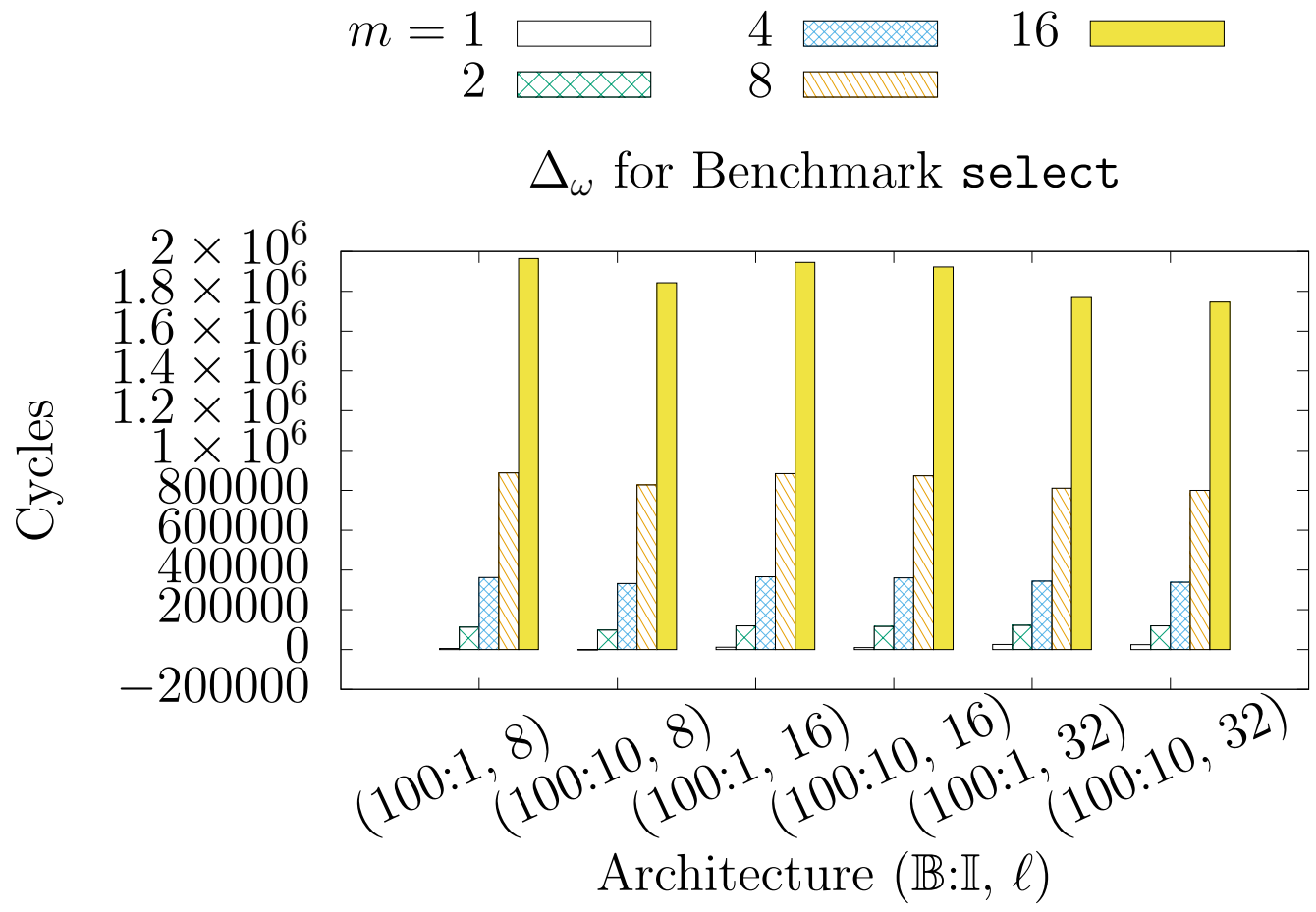

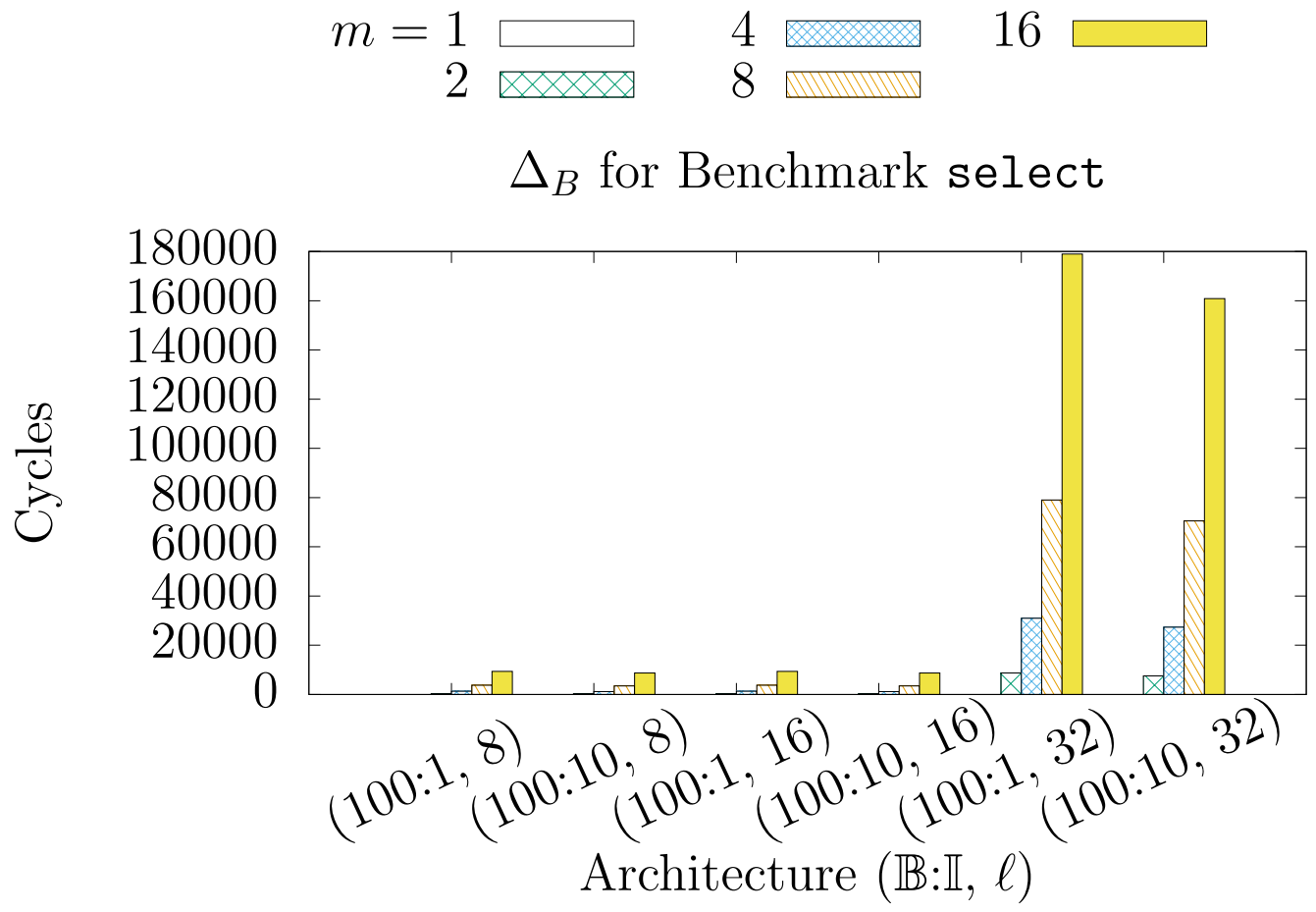

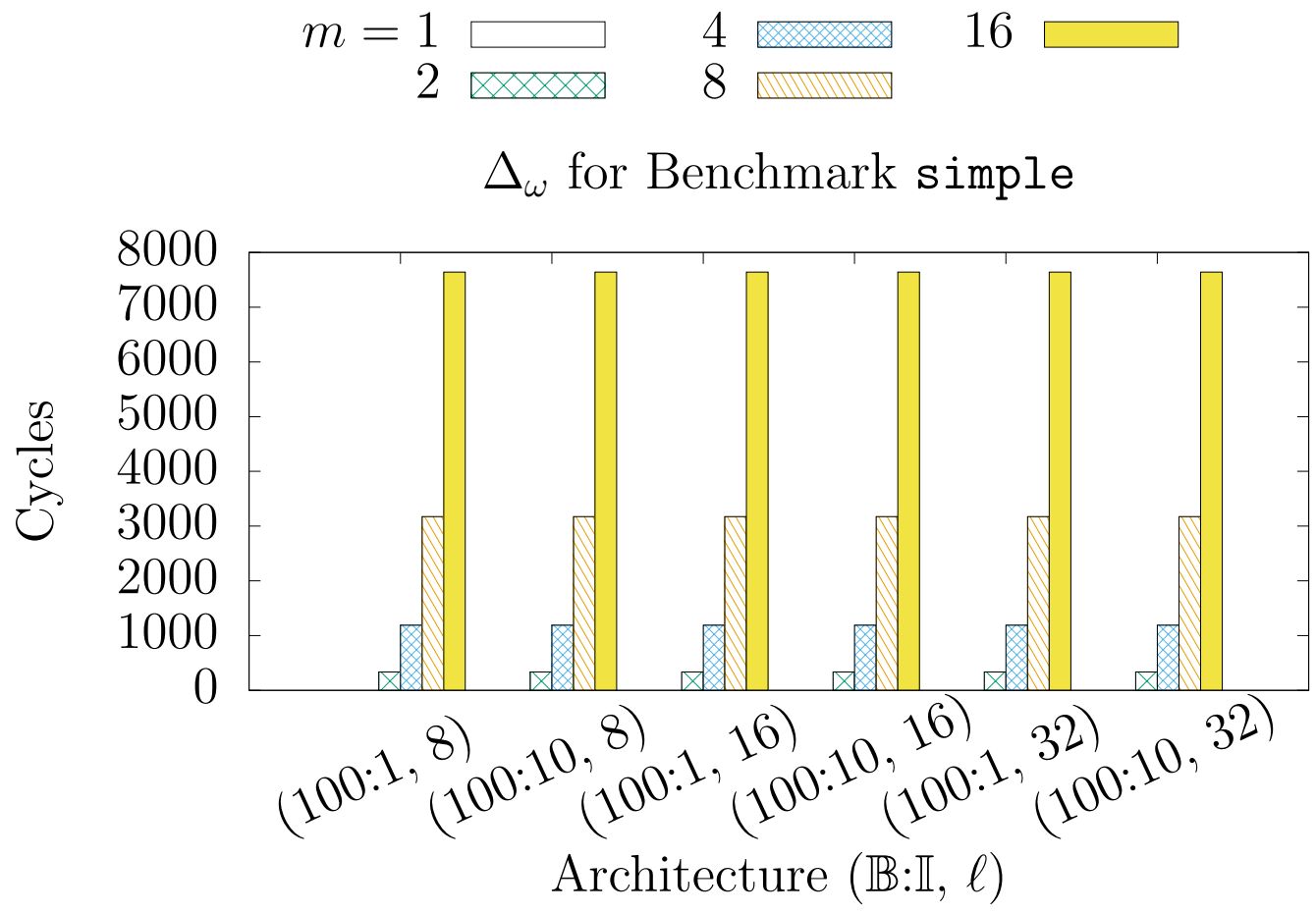

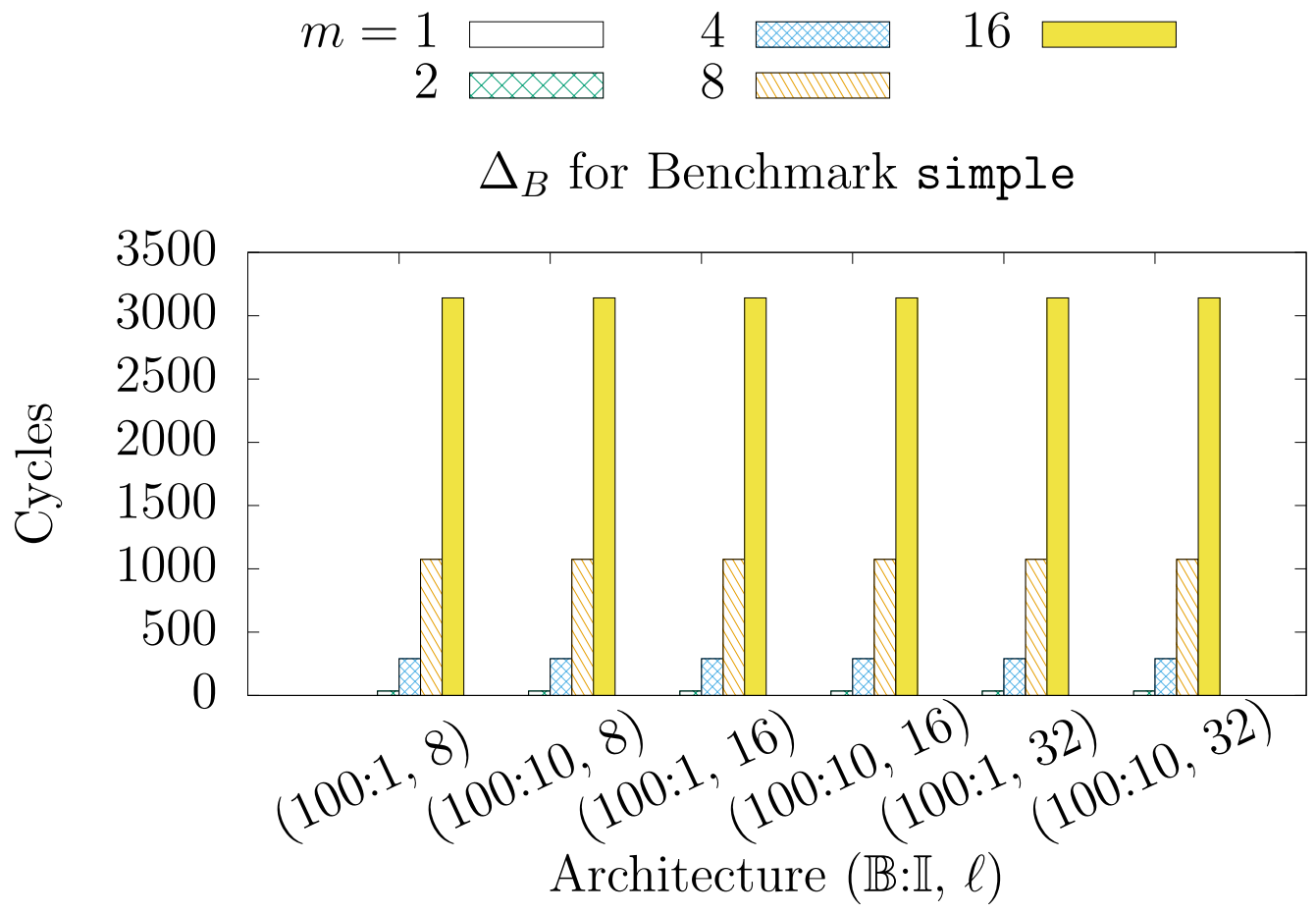

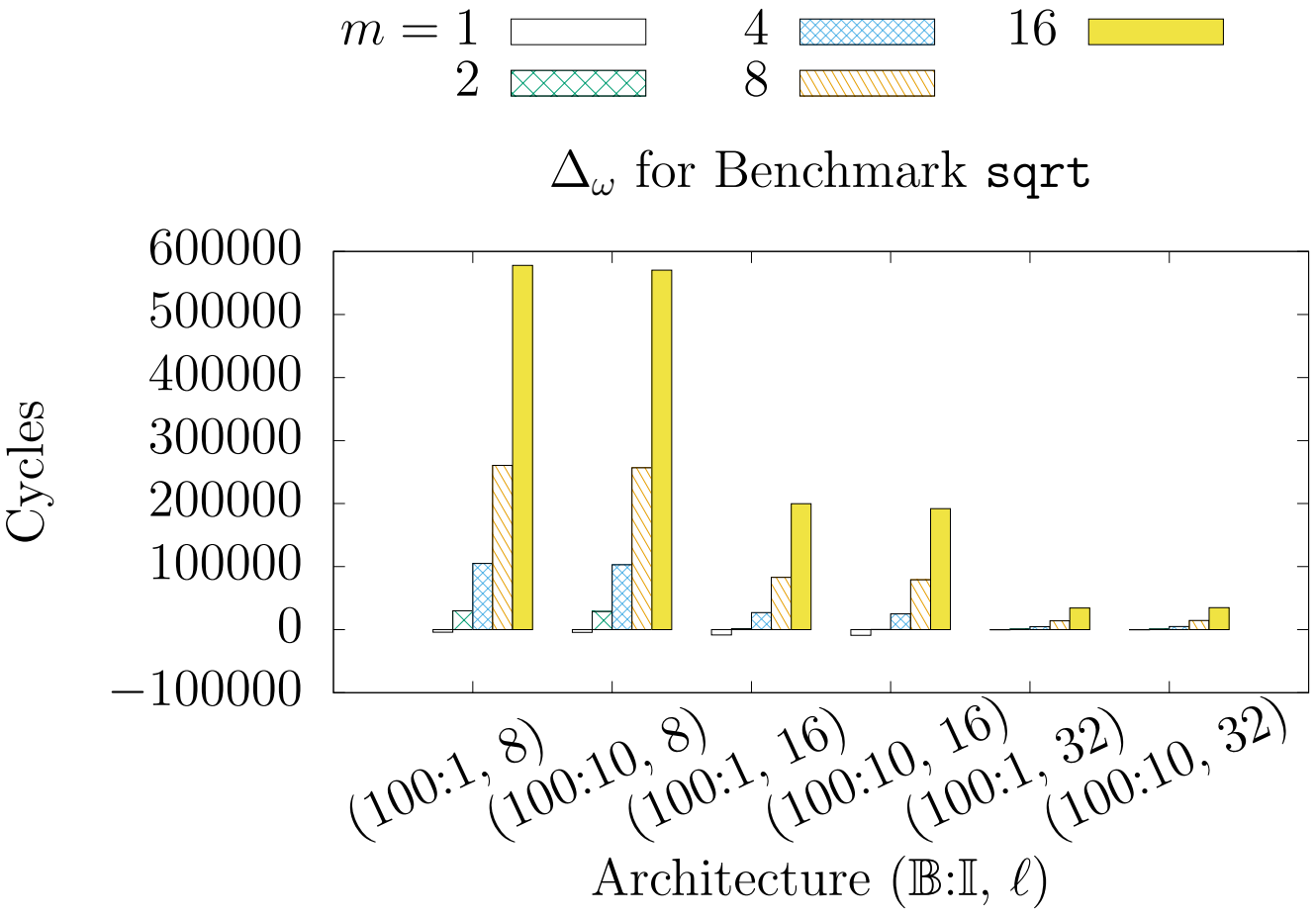

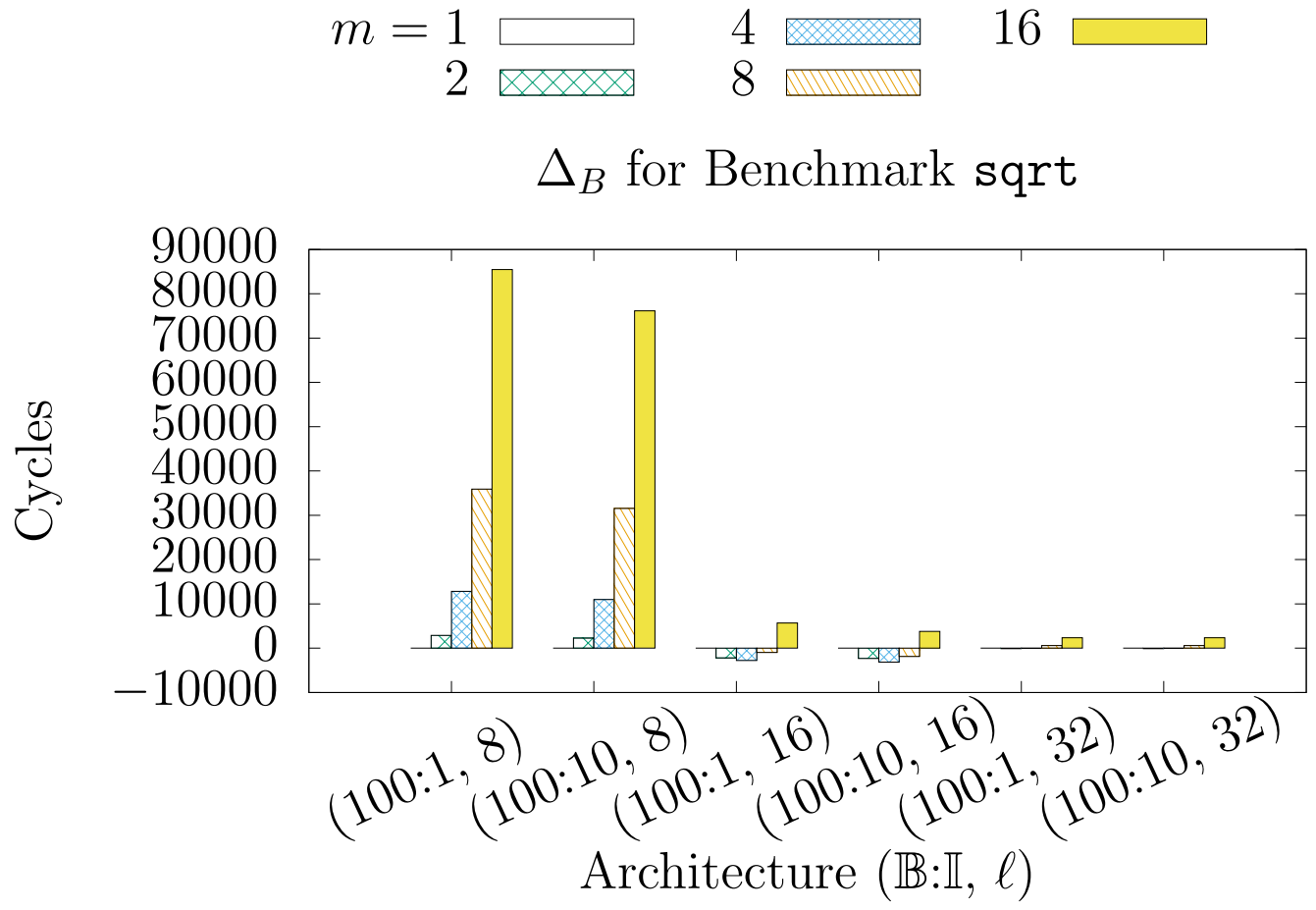

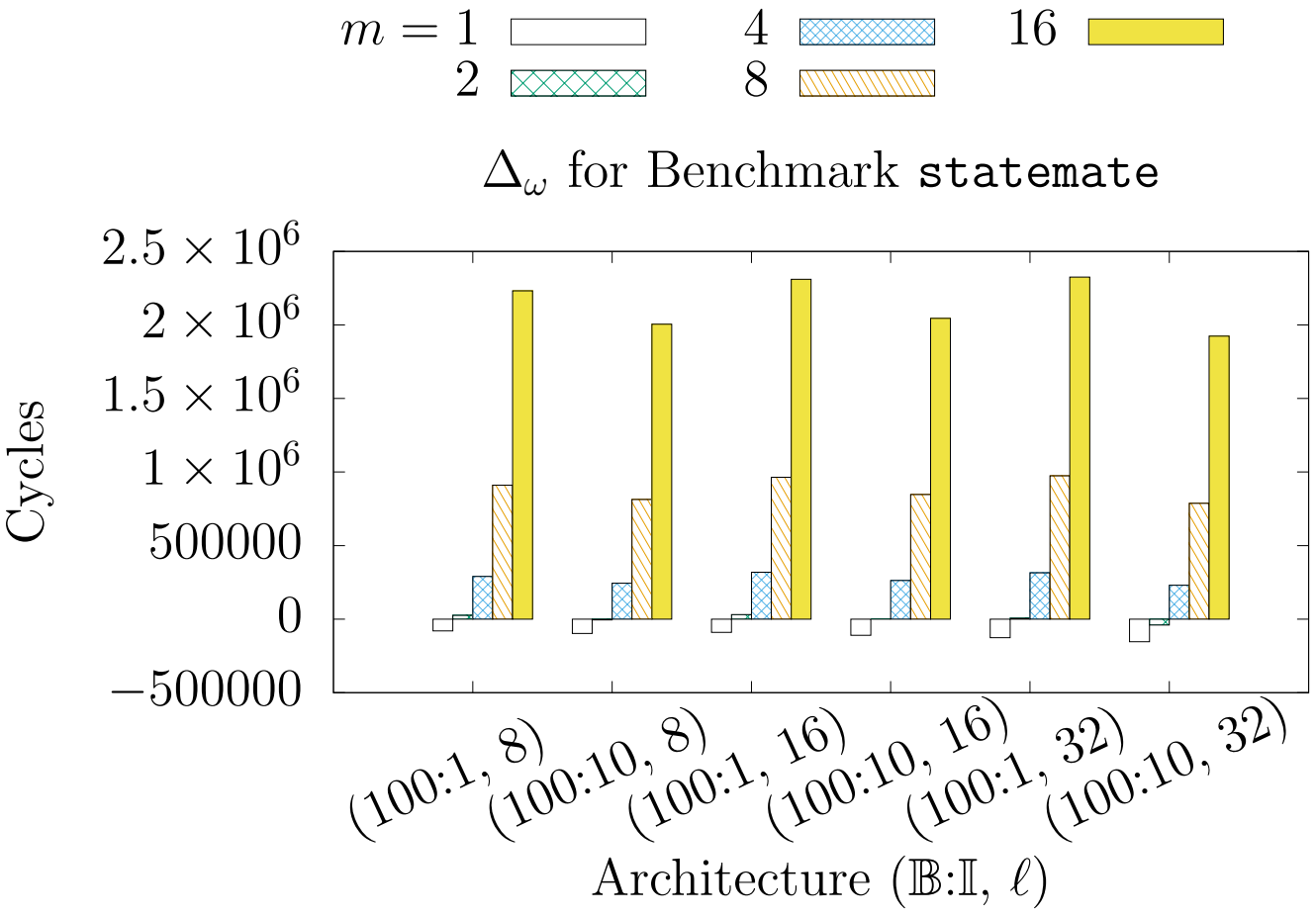

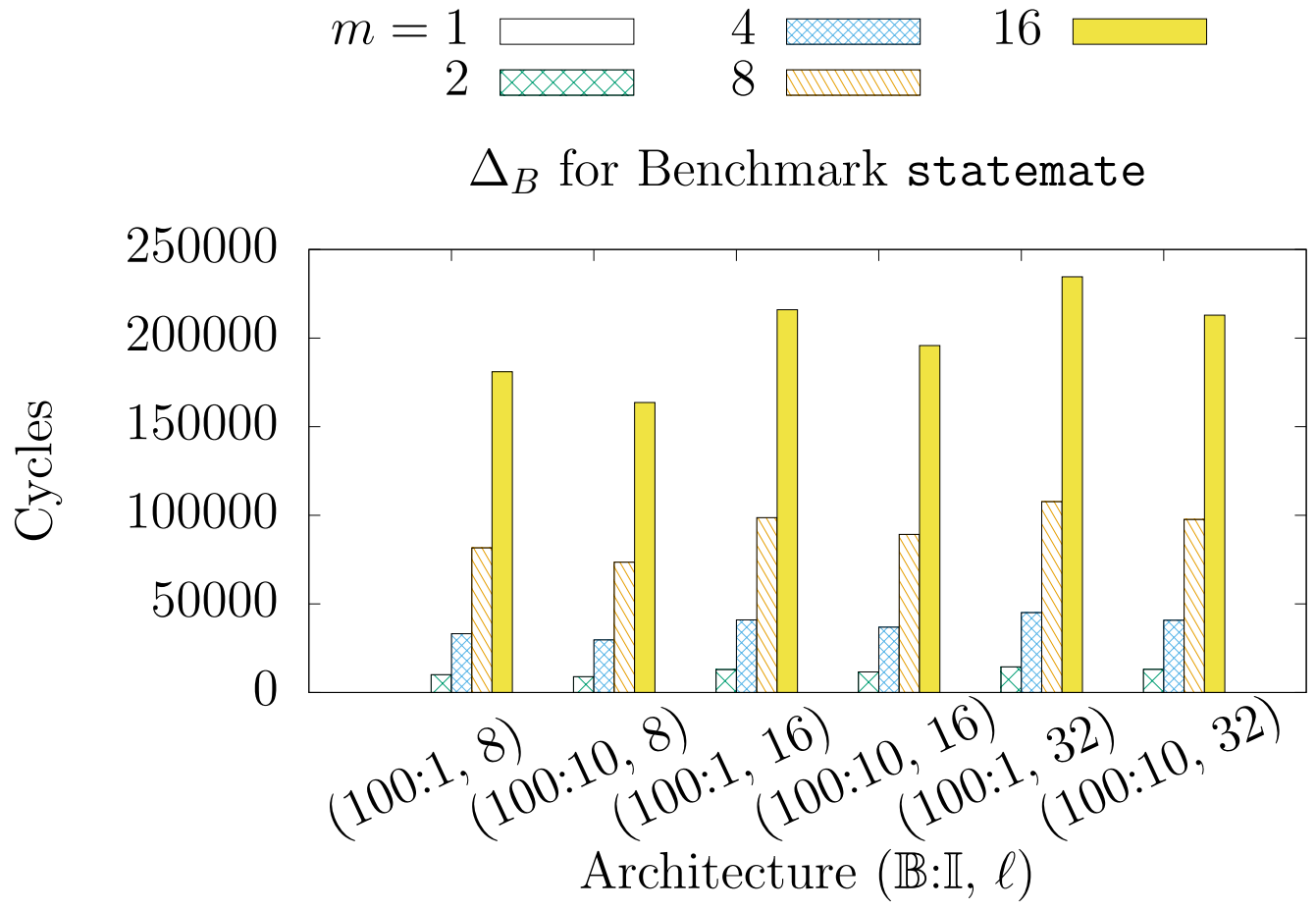

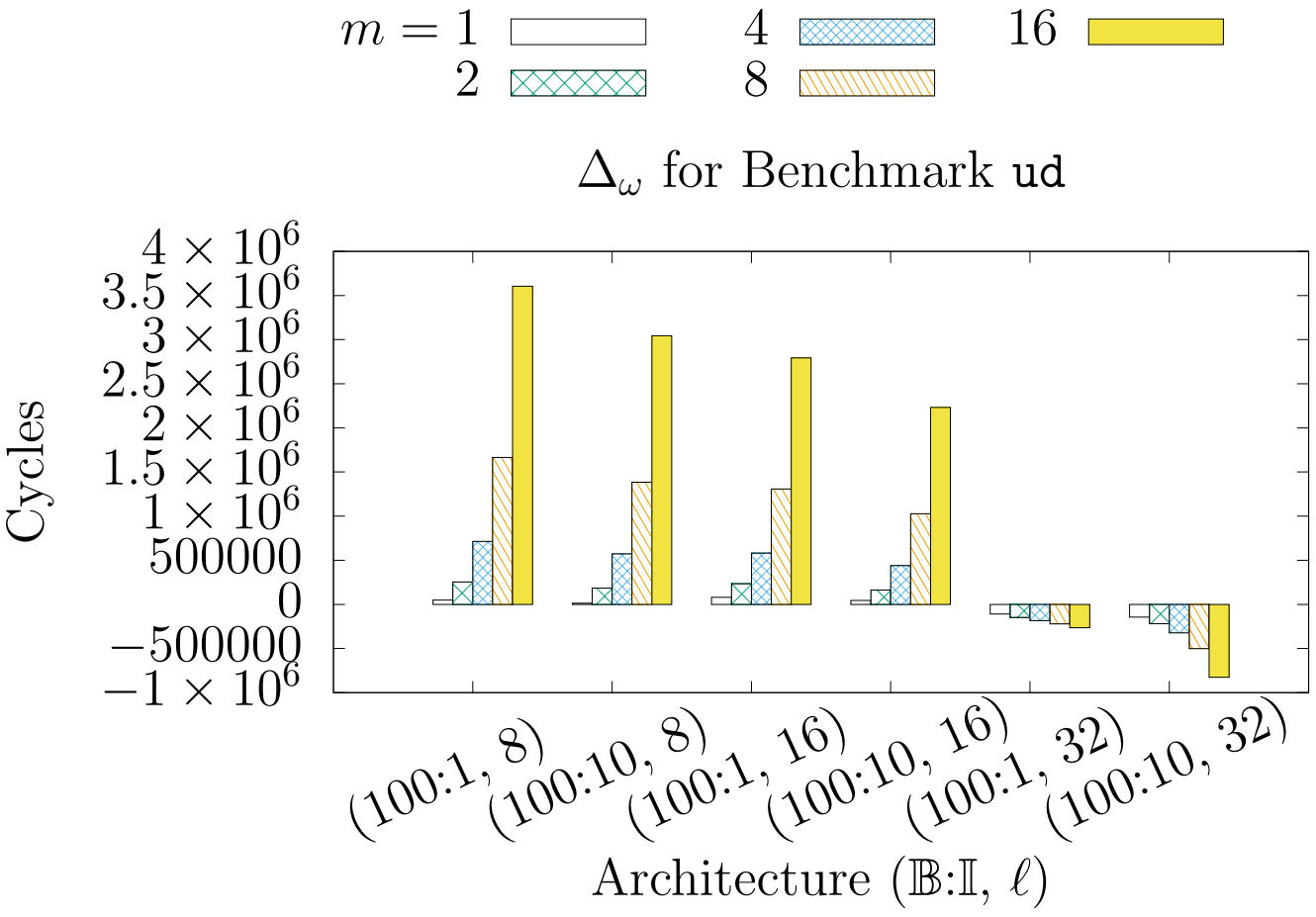

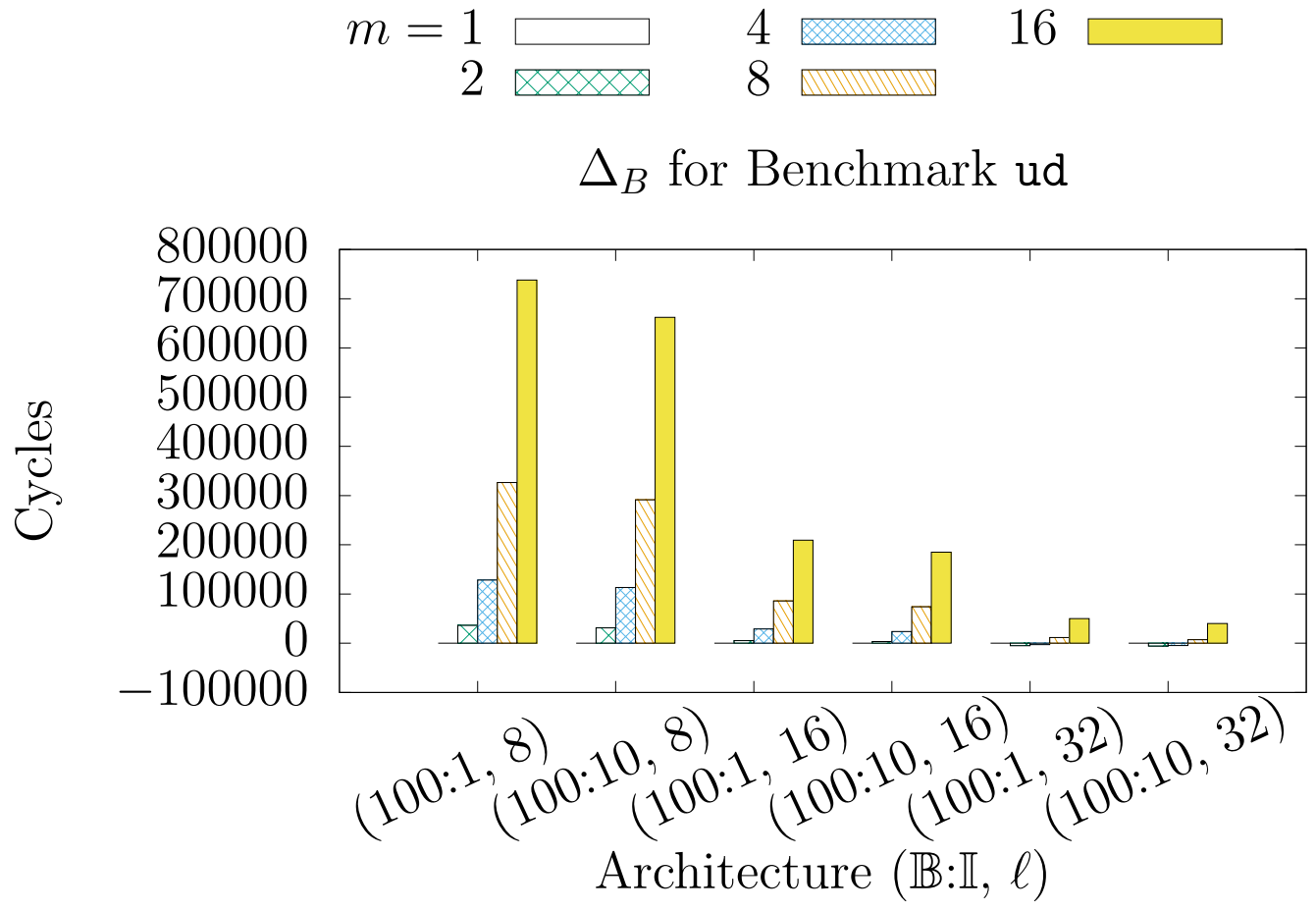